\documentclass[aps,prc,twocolumn,showpacs,preprintnumbers,
                          nofootinbib,float,longbibliography]{revtex4-1}
\usepackage{graphicx, fancybox}
\usepackage{amsmath,amssymb}
\usepackage[colorlinks=true, pdfstartview=FitV, linkcolor=red, citecolor=blue, urlcolor=blue]{hyperref}
\usepackage{color}
\usepackage{soul}
\usepackage{url}

\newcommand{\nobracket}{}

\newcommand{\tmmathbf}[1]{\ensuremath{\boldsymbol{#1}}}
\newcommand{\tmop}[1]{\ensuremath{\operatorname{#1}}}

\usepackage[normalem]{ulem}

\begin{document}

\title{Dynamical initial state model for relativistic heavy-ion collisions}

\author{Chun Shen}
\affiliation{Physics Department, Brookhaven National Laboratory, Upton, NY 11973, USA}

\author{Bj\"orn Schenke}
\affiliation{Physics Department, Brookhaven National Laboratory, Upton, NY 11973, USA}

\begin{abstract}
We present a fully three-dimensional model providing initial conditions for energy and net-baryon density distributions in heavy ion collisions at arbitrary collision energy. The model includes the dynamical deceleration of participating nucleons or valence quarks, depending on the implementation. 
The duration of the deceleration continues until the string spanned between colliding participants is assumed to thermalize, which is either after a fixed proper time, or a fluctuating time depending on sampled final rapidities. Energy is deposited in space-time along the string, which in general will span a range of space-time rapidities and proper times.
We study various observables obtained directly from the initial state model, including net-baryon rapidity distributions, 2-particle rapidity correlations, as well as the rapidity decorrelation of the transverse geometry. Their dependence on the model implementation and parameter values is investigated.
We also present the implementation of the model with 3+1 dimensional hydrodynamics, which involves the addition of source terms that deposit energy and net-baryon densities produced by the initial state model at proper times greater than the initial time for the hydrodynamic simulation.
\end{abstract}

{\maketitle}

\section{Introduction}

The Beam Energy Scan (BES) program at the Relativistic Heavy Ion Collider (RHIC) at Brookhaven National Laboratory \cite{Adamczyk:2013dal,Adamczyk:2014fia,Adare:2015aqk,Adamczyk:2017iwn} and the energy scans performed with  NA61/SHINE at CERN \cite{Mackowiak-Pawlowska:2017rcx} constitute systematic scans of heavy ion collisions over a range of beam energies. These programs provide the opportunity to explore the phase diagram of quantum chromo dynamics (QCD) by varying the typical temperature and baryon chemical potential of the produced matter. 

Precise measurements of the hadronic final state ought to allow for the extraction of transport properties of the Quark-Gluon Plasma (QGP) in a baryon rich environment as well as the determination of the QCD critical point \cite{Stephanov:1998dy,Stephanov:1999zu,Stephanov:2004wx}, should it exist in the accessible region of the phase diagram. In order to do so we require a reliable theoretical framework which can model the dynamical evolution of the collisions and all relevant sources of fluctuations.

Viscous relativistic hydrodynamics is a successful phenomenological model for heavy-ion collisions at high collision energies \cite{Heinz:2013th,Gale:2013da}. Its combination with a hadronic transport model, which provides a more detailed description of the dilute hadronic phase, creates a powerful hybrid framework for describing and predicting a wide range of observables in heavy-ion collisions at Large Hadron Collider (LHC) energies and top RHIC energies (for a review see \cite{Petersen:2014yqa}). 

Some complications arise at lower collision energies. While the hybrid framework reduces the theoretical uncertainties in the late stage of the evolution, a large uncertainty remains in the initial and early time pre-equilibrium stages. A particular problem for describing the early time evolution of the system for the lower BES energies arises because the Lorentz contraction factors for the two incoming nuclei can not be approximated by infinity. Thus, the nuclei have a finite size in the longitudinal (beam) direction and will consequently take a considerable time to pass through each other. One possibility is to start hydrodynamic simulations after the two nuclei have completely passed through each other. However, at $\sqrt{s_\mathrm{NN}} \sim \mathcal{O}(10)$ GeV, this may take approximately $\sim 2$ fm or more, leaving a large theoretical uncertainty for the early time dynamics of the system. 

To address this problem, in a previous work \cite{Karpenko:2015xea} the hadron transport approach UrQMD \cite{Bass:1998ca,Bleicher:1999xi} was used to describe the early stage of the collision. Then the switch to hydrodynamics was performed at a constant proper time, larger or equal to the passing time of the two nuclei. A similar model, using AMPT for the early stage was presented in \cite{Pang:2012he,Pang:2015zrq}, but it has not been applied to energies below top RHIC energies so far.

In this work, we aim to solve the problem by introducing a dynamical framework which interweaves the initial condition that produces three dimensional net-baryon and energy densities with the hydrodynamic evolution of the system. After a minimal thermalization time, which in general can be smaller than the time the two nuclei overlap, the hydrodynamic evolution is started. Collisions between nucleons that occur after this initial time will contribute additional energy and net-baryon number, which enter the hydrodynamic simulation via source terms. A similar idea using source terms in the hydrodynamic simulation but with different assumptions about the nature of these sources was presented in \cite{Okai:2017ofp}.

The time and location of the energy and net-baryon density deposition is determined from binary collisions of nucleons.
A Monte-Carlo Glauber model like prescription determines the position of the collisions in the transverse plane. To determine the longitudinal structure of the collision, a string is formed between colliding nucleons (or valence quarks within them, depending on the implementation) and they begin to decelerate. After a predetermined proper time the string will thermalize and deposit energy along its entire length and net-baryon density near its ends.

When using nucleon degrees of freedom, the only fluctuations will be those of the transverse structure and the longitudinal position of constant length strings. Introducing constituent quarks, whose longitudinal momentum fraction fluctuates, produces fluctuations of the string length. Additional fluctuations can be included by varying the time a string requires to thermalize. We study the effect of the latter by implementing varying final rapidities of each particle, sampled according to the distribution first introduced within the LEXUS model \cite{Jeon:1997bp,Monnai:2015sca}.

We focus on the analysis of net-baryon and multiplicity distributions in rapidity as well as measures of fluctuations, with emphasis on their dependence on different model assumptions.  We present Legendre coefficients of net baryon and energy density rapidity fluctuations, measures of decorrelations of the transverse geometry with rapidity, and cumulant ratios for net proton distributions obtained directly from the initial state.

The paper is organized as follows. In Section \ref{sec:model} we present the model, detailing the three-dimensional collision dynamics, string production and deceleration, as well as details like the possible choice of participants and rapidity fluctuations. We close Section \ref{sec:model} with a discussion of the form of the source terms for hydrodynamics. In Section \ref{sec:hydro} we lay out the form of the hydrodynamic equations with sources, and in Section \ref{sec:results} we present results of the numerical calculations. Conclusions are presented in Section \ref{sec:conc}.

\section{Three dimensional Monte-Carlo Glauber model}\label{sec:model}
There exist several models that provide fluctuating initial conditions in three spatial dimensions \cite{Pang:2012uw,Karpenko:2015xea,Bozek:2015bna,Broniowski:2015oif,Monnai:2015sca,Schenke:2016ksl}.
Here, we generalize the Monte Carlo Glauber model to three dimensions by introducing a prescription for the energy and net-baryon density deposition as a function of rapidity. We show that in general a (proper-) time dependent prescription for this deposition is required. This will be achieved by introducing source terms into the hydrodynamic simulation.

\subsection{Collision dynamics in 3D}
The time it takes two nuclei to pass through each other in a heavy ion collision depends on the collision energy. 
Given the nuclear radius $R$, the overlap time of two nuclei moving with opposite velocities $\pm v_z$ can be estimated in the laboratory frame as
\begin{equation}
  \tau_{\mathrm{overlap}} = \frac{2 R}{\gamma v_z}  = \frac{2 R}{\sinh(y_\mathrm{beam})},
\end{equation}
where $\gamma$ is the Lorentz factor and $y_\mathrm{beam} = \mathrm{arccosh}(\sqrt{s_\mathrm{NN}}/(2m_p))$ is the beam rapidity. Here $\sqrt{s_\mathrm{NN}}$ is the collision energy per nucleon pair and $m_p = 0.938$ GeV is the mass of a proton.\footnote{We approximate the neutron mass $m_n\approx m_p$.}
%
\begin{figure}[ht!]
  \centering
  \begin{tabular}{c}
  \includegraphics[width=0.9\linewidth]{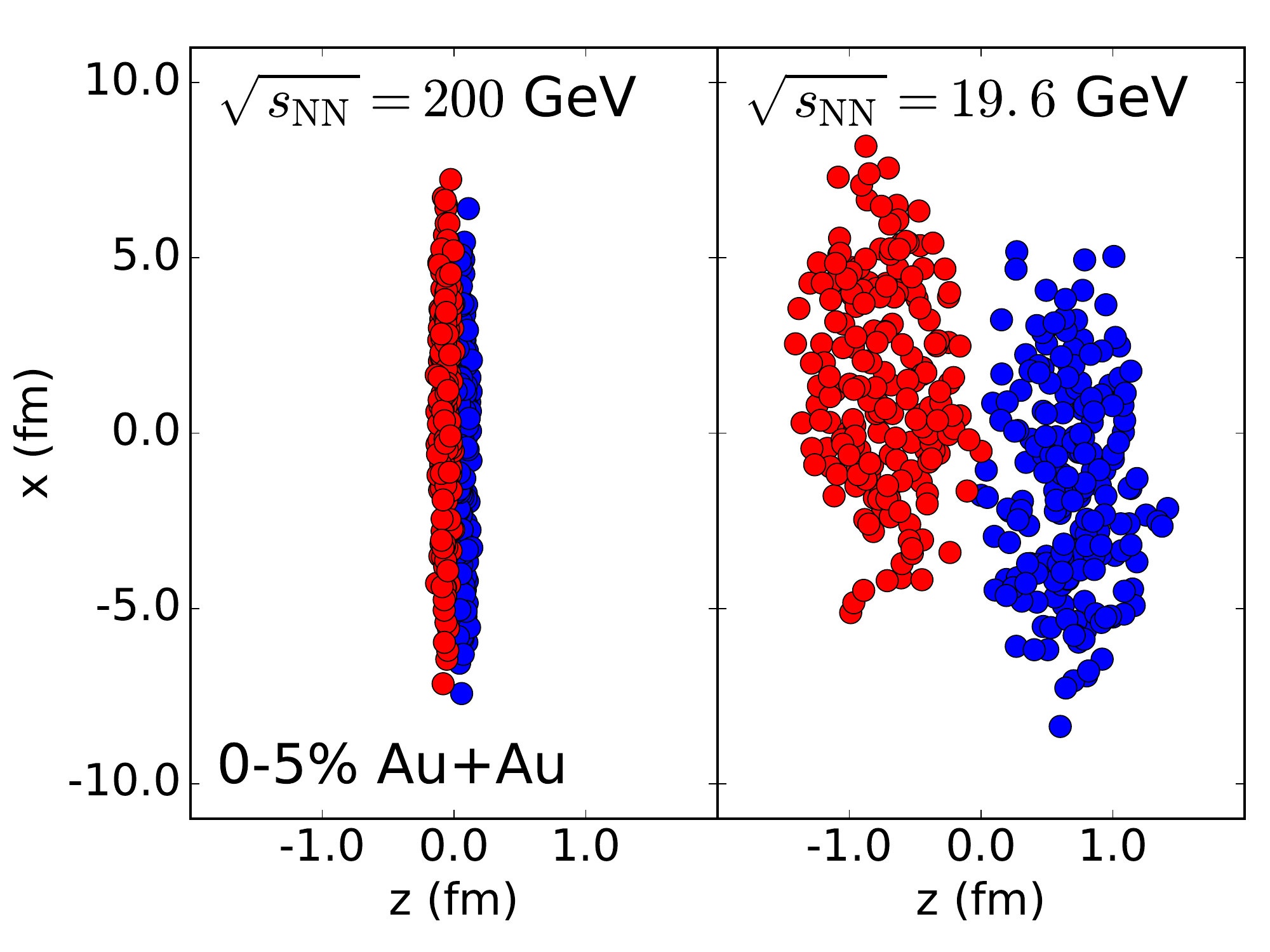}
   \end{tabular}
  \caption{Nucleon positions as a function of one transverse ($x$) and the longitudinal direction ($z$) for to different collision energies.}
  \label{fig:positions}
\end{figure}
%
For two example Au+Au events Fig.\,\ref{fig:positions} shows the distribution of nucleons in the laboratory frame at the time of the first NN-collision. Nucleon positions $(x^i_{P}, y^i_{P}, z^i_{P})$ and $(x^j_{T}, y^j_{T}, z^j_{T})$, where $i$ and $j$ run over all projectile ($P$) and target ($T$) nucleons, respectively, were sampled from a three dimensional isotropic Woods-Saxon distribution, then Lorentz contraction in the $z$-direction was applied according to the collision energy. This illustrates that it will take a finite time for the two nuclei to pass through one another and that nucleon-nucleon collisions will occur at different positions $z$ and over an extended range in time $t$.

%
%
\begin{figure}[ht!]
  \centering
  \begin{tabular}{c}
   \includegraphics[width=0.9\linewidth]{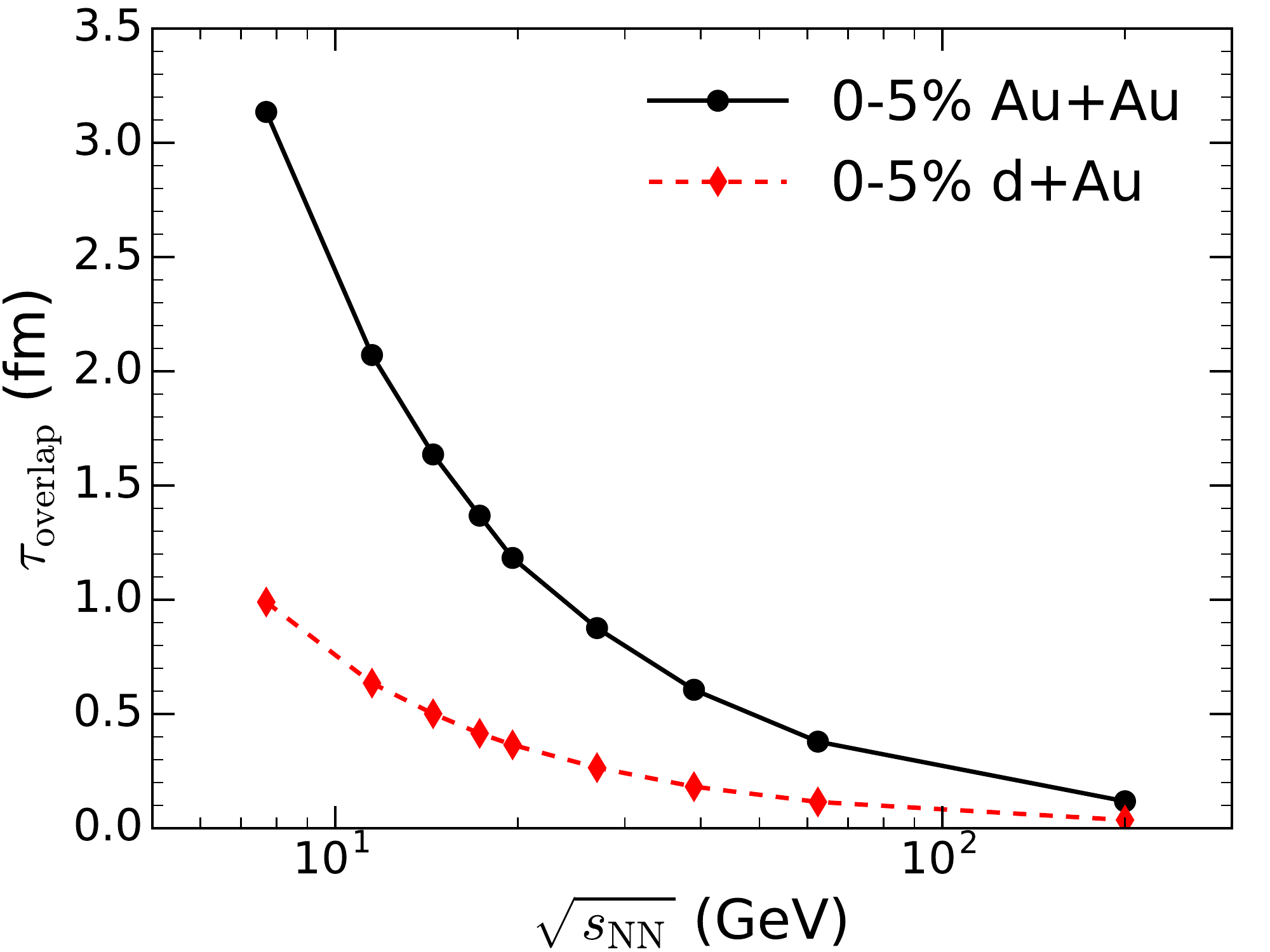}
   \end{tabular}
  \caption{The nuclear overlapping time of 0-5\% central d+Au and Au+Au collisions as a function of the collision energy at the RHIC BES program.}
  \label{fig1}
\end{figure}
%
%
Fig.~\ref{fig1} shows the overlapping time $\tau_\mathrm{overlap}$ as a function of the collision energy $\sqrt{s_\mathrm{NN}}$ for Au+Au and d+Au collision systems. Due to the finite $\tau_{\mathrm{overlap}}$, binary collisions of the
nucleons cannot be approximated to all occur at the origin of the light cone $t = z = 0$.
In order to get the collision time and position for every binary
collision, we perform a simple transport simulation, described in the following. We
assume straight line trajectories for the colliding nucleons - a binary collision does
not change the direction of the colliding nucleons but only slows them down. This assumption
can be relaxed by adding small random transverse kicks to nucleons at every binary
collision in the future.

Before the collision, all projectile and target nucleons are assigned the velocities
\begin{equation}
  \tmmathbf{v}_{P(T)} = (0, 0, \tanh(\pm y_{\mathrm{beam}}))\,,
\end{equation}
respectively.
At time $t = 0$, we set $\mathrm{max}\{z^P_i\} = \mathrm{min} \{z^T_j\}= 0$.
This ensures that all binary collisions occur inside the forward light cone. The space-time positions of individual binary collisions are determined using a simple collision driven transport scheme. Whether collisions between nucleons $i$ and $j$ occur is determined with a standard Monte-Carlo Glauber model using the geometric interpretation of the nucleon-nucleon cross section and the transverse position of the nucleons $(x^i_P, y^i_P)$ and $(x^j_T, y^j_T)$ \cite{Miller:2007ri}. 
Collision positions are computed as
\begin{equation}
  x_c^{ij} = (x^i_P + x^j_T) / 2, \quad y_c^{ij} = (y^i_P + y^j_T)/2 \,,
\end{equation}
and
\begin{equation}
  z_c^{ij} = z_P^i + \Delta t_{ij} \tmmathbf{v}_P^i,
\end{equation}
where the $\Delta t_{ij} = (z^j_T - z^i_P)/(\tmmathbf{v}_P - \tmmathbf{v}_T)$ are the collision times.
Note that with $|\tmmathbf{v}_P - \tmmathbf{v}_T|$ being defined in the laboratory frame, there is no problem with this difference being greater than the speed of light $c$. It is defined as the rate of change of the distance between the two approaching nucleons.

\subsection{String production and deceleration}
Having determined the positions and times of all binary collisions, we need to develop a prescription for how produced energy and net-baryon number are distributed.
In order to do so we assume that strings are produced between all colliding nucleons.\footnote{We will discuss below how to modify the model to use constituent quark degrees of freedom.} For details on how strings are connected between participating nucleons we refer the reader to the appendix.

Before a string breaks and thermalizes with the system (i.e. contributes its energy to the medium), its end points decelerate according to \cite{Bialas:2016epd}
\begin{equation} \label{eq:deceleration}
  \frac{d E}{d z} = - \sigma ~\tmop{and}~ \frac{d p_z}{d t} = - \sigma\,,
\end{equation}
where $\sigma$ is the string tension. A similar prescription with the constant string tension replaced by space and time dependent components of the energy momentum tensor of the Glasma was introduced in \cite{Li:2016wzh}. That framework is however likely constrained to high (i.e. top RHIC and higher) energies.

We note here that we do not assign a specific color structure to each string and the string tension $\sigma$ can be understood as an effective parameter characterizing the average force between interacting components (either nucleons or what we will call valence quarks) of the two nuclei. We make this choice for simplicity but note that in the future detailed color information can be added to the model, and its effect on the longitudinal structure and fluctuations studied. The implementation could for example follow the method employed in HIJING \cite{Gyulassy:1994ew}, where a nucleon-nucleon collision produces two color neutral strings, each between a quark and a di-quark \cite{Capella:1992yb}.

An early version of this model \cite{Shen:2017ruz} assumed an instantaneous energy loss at the time of the binary collision followed by a free-streaming propagation for the strings. In that case, the produced string length along the longitudinal direction was anti-correlated with the amount of energy lost in the collisions. This leads to the unphysical situation that a collision without energy loss would produce a string that feeds the most energy to the hydrodynamic medium. Adopting the deceleration dynamics cures this shortcoming. The energy of the produced string from Eqs.\,(\ref{eq:deceleration}) is proportional to the energy lost during the collision.

From the Eqs.\,(\ref{eq:deceleration}) we find that the rapidity of a string end is decelerated to
\begin{equation}
  \tilde{y}(\Delta \tau_f) = \tilde{y}_i \pm \tmop{arccosh} \left( \frac{\Delta \tau_f^2 \sigma^2}{2 m^2} + 1 \right)\,,\label{eq:yofdtau}
\end{equation}
where we always take the solution with $| \tilde{y}(\Delta \tau_f) | < |\tilde{y}_i | \nobracket$, the absolute value of the initial rapidity of the string's endpoints (which is the same for both endpoints in the rest frame of the string).
If one nucleon is connected to multiple strings, its incoming rapidity $\tilde{y}_i$ starts with the beam rapidity in the earliest collision and is decelerated sequentially for a time $\Delta \tau_f$ for each string it is connected to. 
The sign in Eq.\,(\ref{eq:yofdtau}) depends on the direction the endpoint is moving. The final rapidity of the string end point is $\tilde{y}_f=\tilde{y}(\Delta\tau_f)$, however, when the string end point comes to a halt and Eq.\,(\ref{eq:yofdtau}) would lead to an acceleration, evolution of the string is stopped. The maximum deceleration time at which this happens is
\begin{equation}
\Delta \tau_\mathrm{max} = \frac{m}{\sigma} \sqrt{2 (\cosh(\tilde{y}_i) - 1)}.
\end{equation}
To obtain equations for the space-time coordinates of the string ends in the lab frame, 
we need to apply a boost with the center of mass rapidity of the string
\begin{equation}
  Y = \frac{1}{2} (y^i_l + y^i_r)\,,
\end{equation}
where $y^i_l$ and $y^i_r$ are the initial lab frame rapidities of the left ($l$) moving and right ($r$) moving string ends, respectively.
Using Eq.~(\ref{eq:yofdtau})
we determine the positions of the left and right moving end points at the time of string breaking as
\begin{align}
  t_{\tmop{lab}}^{l/r} = t_c + \Delta \tau_f \Big( &-\frac{\Delta \tau_f \sigma_{l/r}}{2 m}
  \sinh (y_{l/r}^i)  \notag \\ & + \sqrt{\frac{\Delta \tau_f^2 \sigma_{l/r}^2}{4 m^2} + 1} \cosh (y_{l/r}^i) \Big)\,, \label{eq9}
\end{align}
\vspace{-0.4cm}
\begin{align}
  z_{\tmop{lab}}^{l/r} = z_c + \Delta \tau_f \Big( &-\frac{\Delta \tau_f \sigma_{l/r}}{2 m}
  \cosh (y_{l/r}^i) \notag \\ & + \sqrt{\frac{\Delta \tau_f^2 \sigma_{l/r}^2}{\text{} 4 m^2} + 1}
  \sinh (y_{l/r}^i) \Big) \,, \label{eq10}
\end{align}
with the initial lab frame rapidities of the end points, $y_{l/r}^i$. They are obtained from the rapidities in the rest frame of the collision as $y_{l/r}^i=\tilde{y}_{l/r}^i+Y$. For the left moving end, because $y_l < 0$, we use $\sigma_l = -\sigma$ and for the right moving end, because $y_r > 0$, we use $\sigma_r = \sigma$.

\begin{figure}[ht!]
  \centering
  \begin{tabular}{c}
  \includegraphics[width=0.95\linewidth]{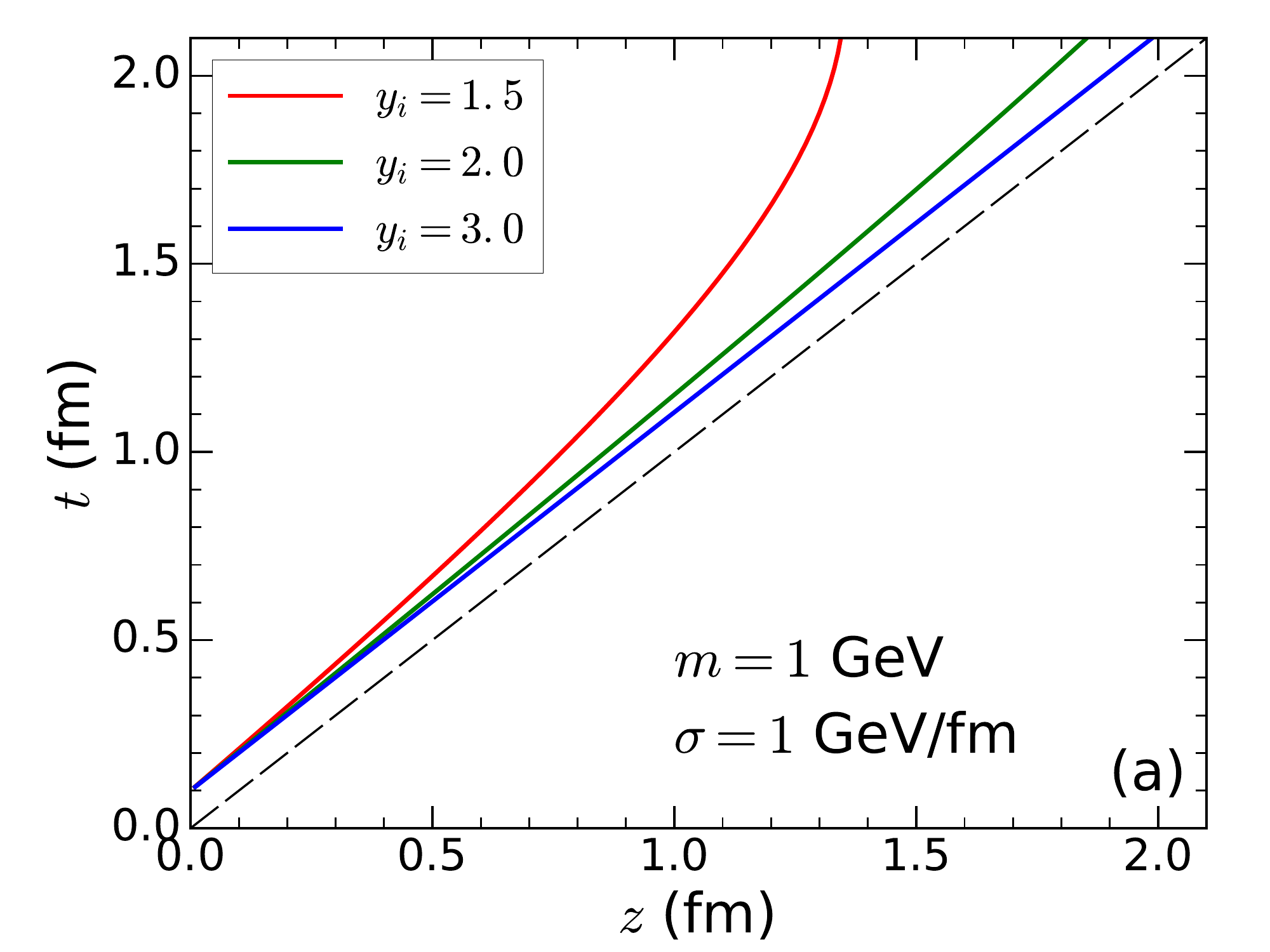} \\
   \includegraphics[width=0.95\linewidth]{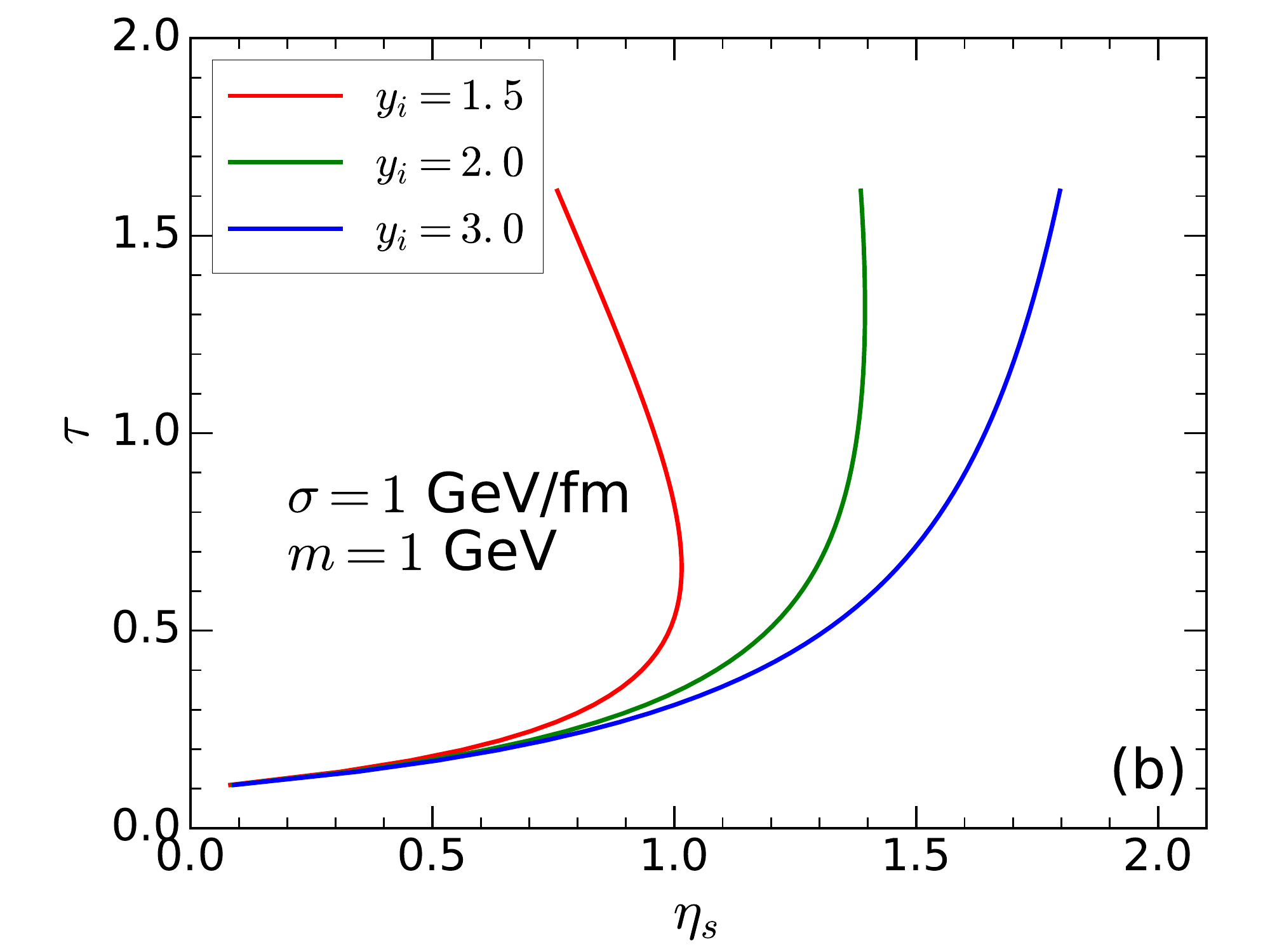}
   \end{tabular}
  \caption{The trajectories of the end points of the decelerating strings with different initial rapidities in $t-z$ (a) and $\tau-\eta_s$ (b) coordinates.}
  \label{fig2}
\end{figure}

Fig.~\ref{fig2} shows the trajectories of the right moving end of a string during its deceleration with constant $\sigma=1\,{\rm GeV/fm}$ and $m = 1$ GeV for different initial rapidities.

Using the definition of the space-time rapidity
\begin{equation}
  \eta_s = \frac{1}{2} \ln \left( \frac{t + z}{t - z} \right) , \label{eq11}
\end{equation}
Eqs. (\ref{eq9}) and (\ref{eq10}) provide the space-time rapidity of the left and right end of the string, $\eta_{s, l}$ and
$\eta_{s, r}$. 

%
%

%
\begin{figure*}[ht!]
  \centering
  \begin{tabular}{cc}
   \includegraphics[width=0.45\linewidth]{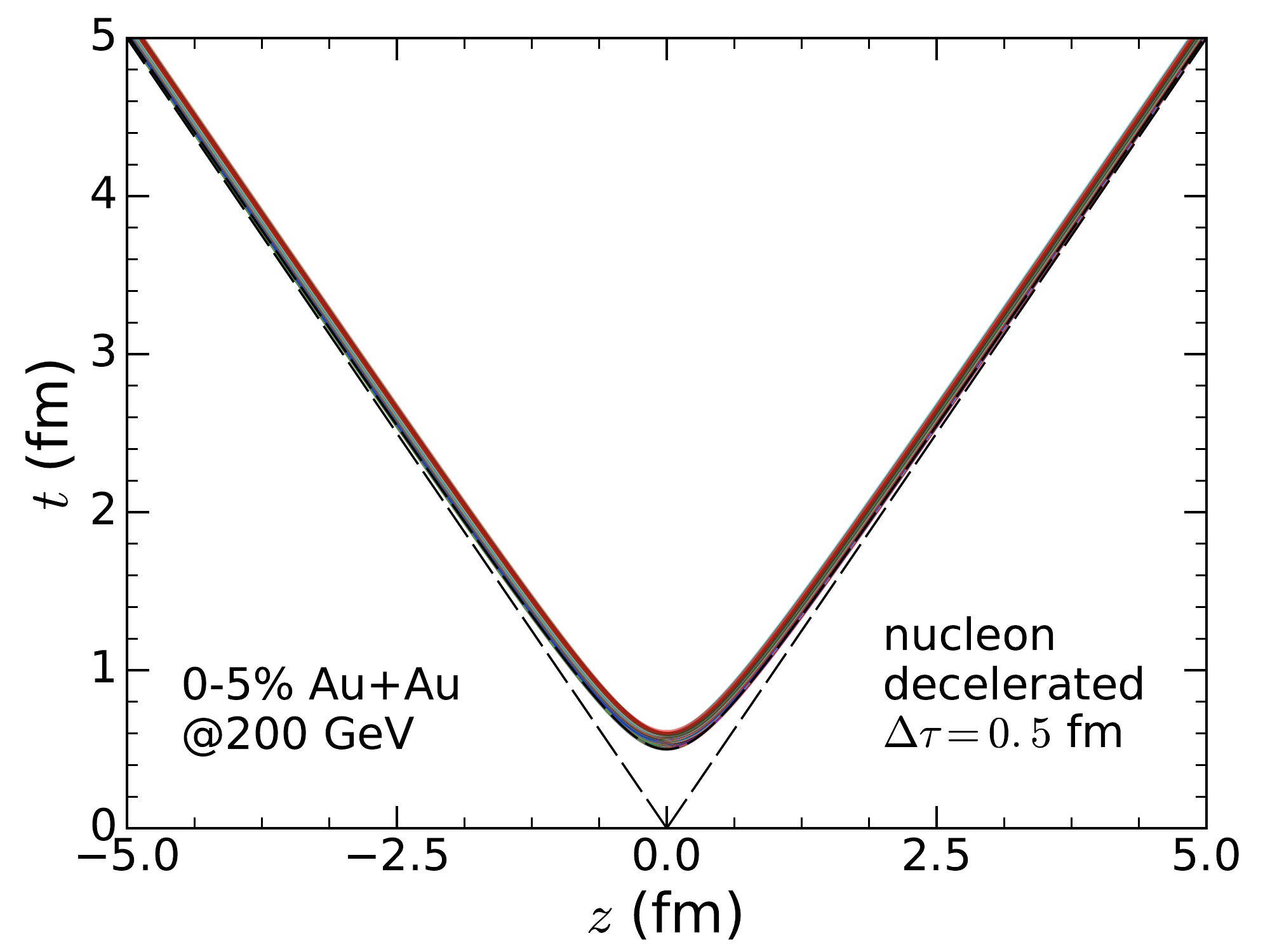} & 
   \includegraphics[width=0.45\linewidth]{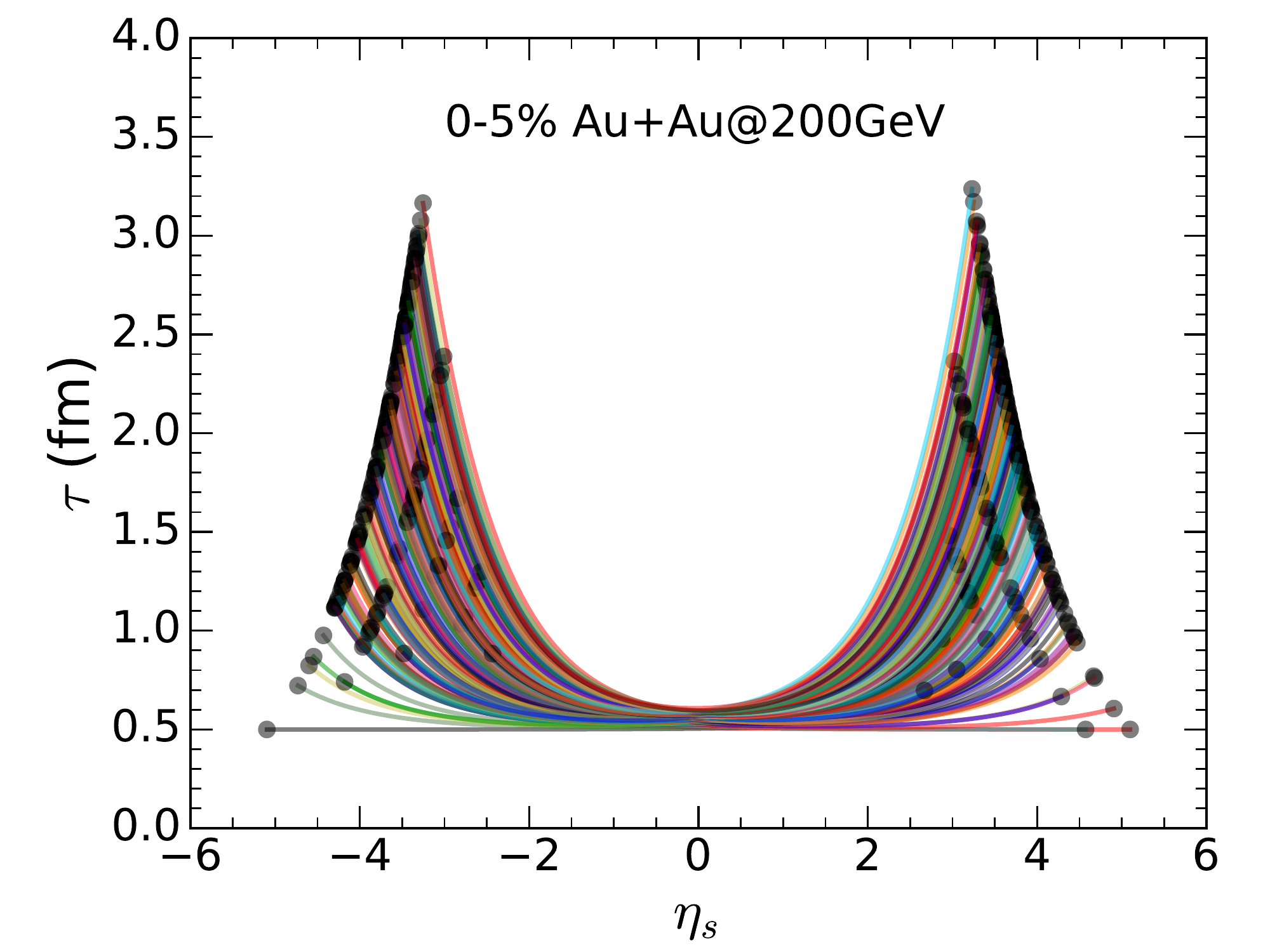} \\
  \includegraphics[width=0.45\linewidth]{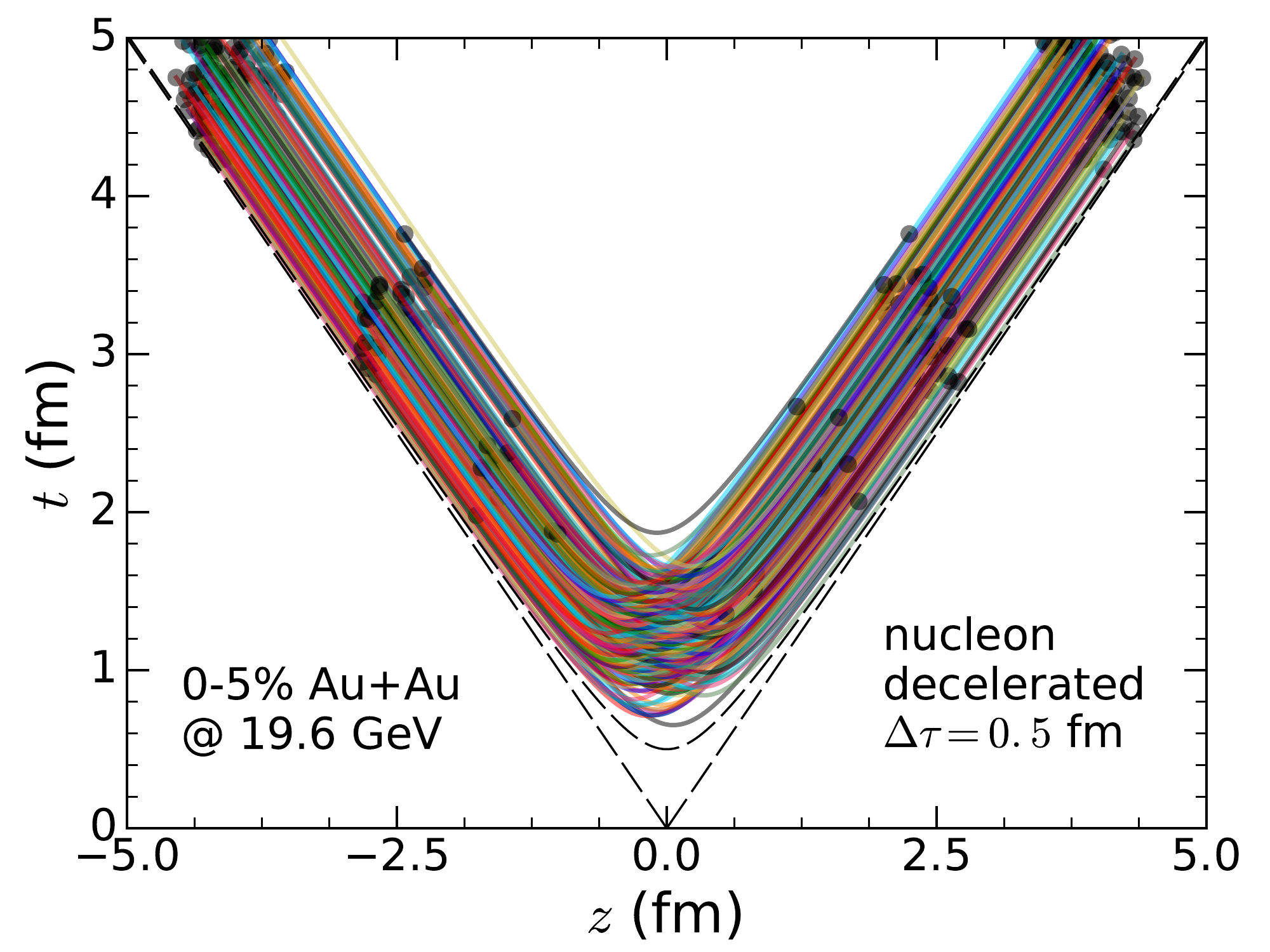} & 
   \includegraphics[width=0.45\linewidth]{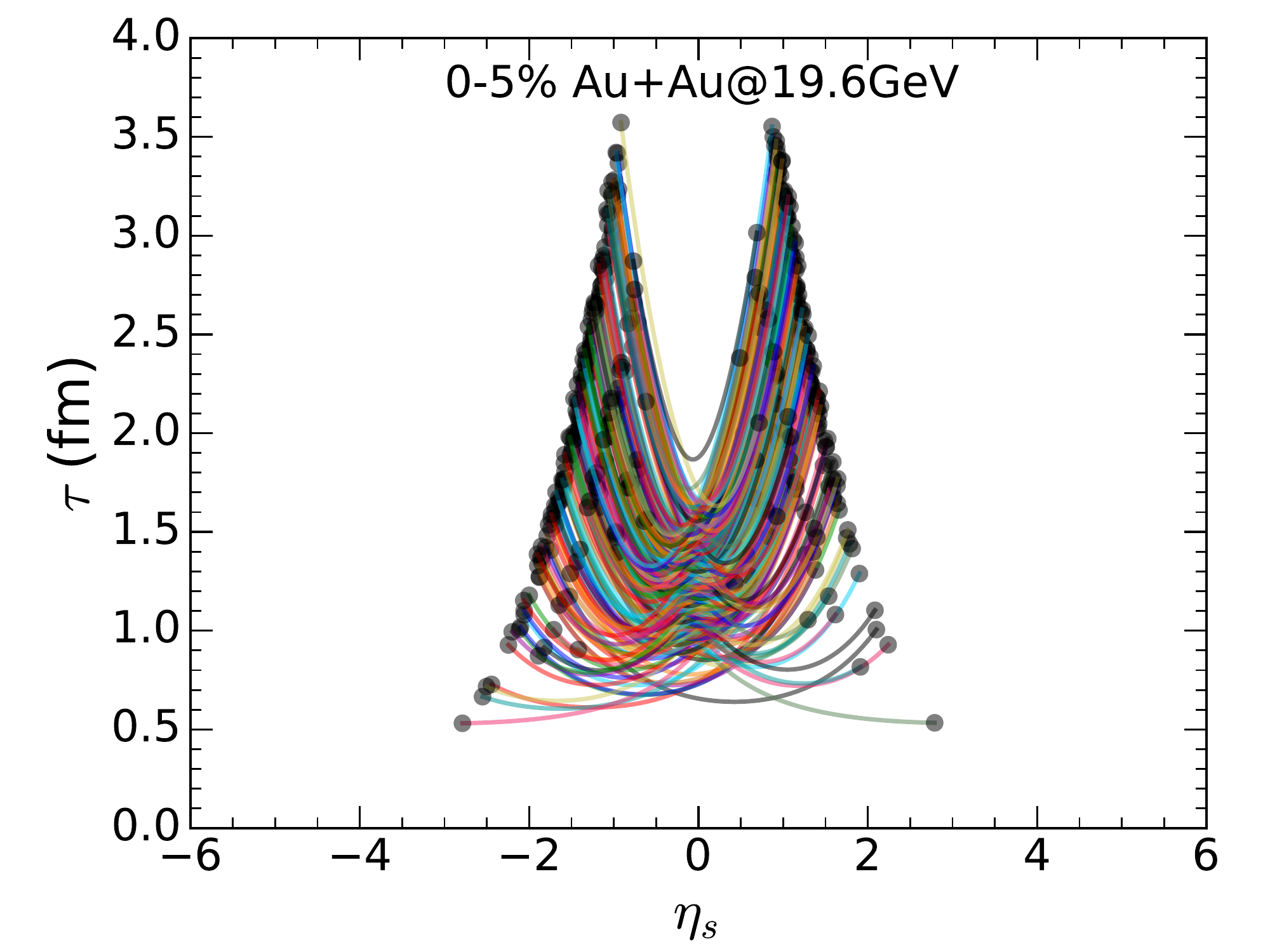}
   \end{tabular}
  \caption{The space-time distribution of the strings at their thermalization in a Au+Au collision at 200 GeV and 19.6 GeV in $t-z$ and $\tau-\eta_s$ coordinates. The black dots indicate the space-time position of the net baryon charges.}
  \label{fig:AuAuStrings}
\end{figure*}
%

%
\begin{figure*}[ht!]
  \centering
  \begin{tabular}{cc}
   \includegraphics[width=0.45\linewidth]{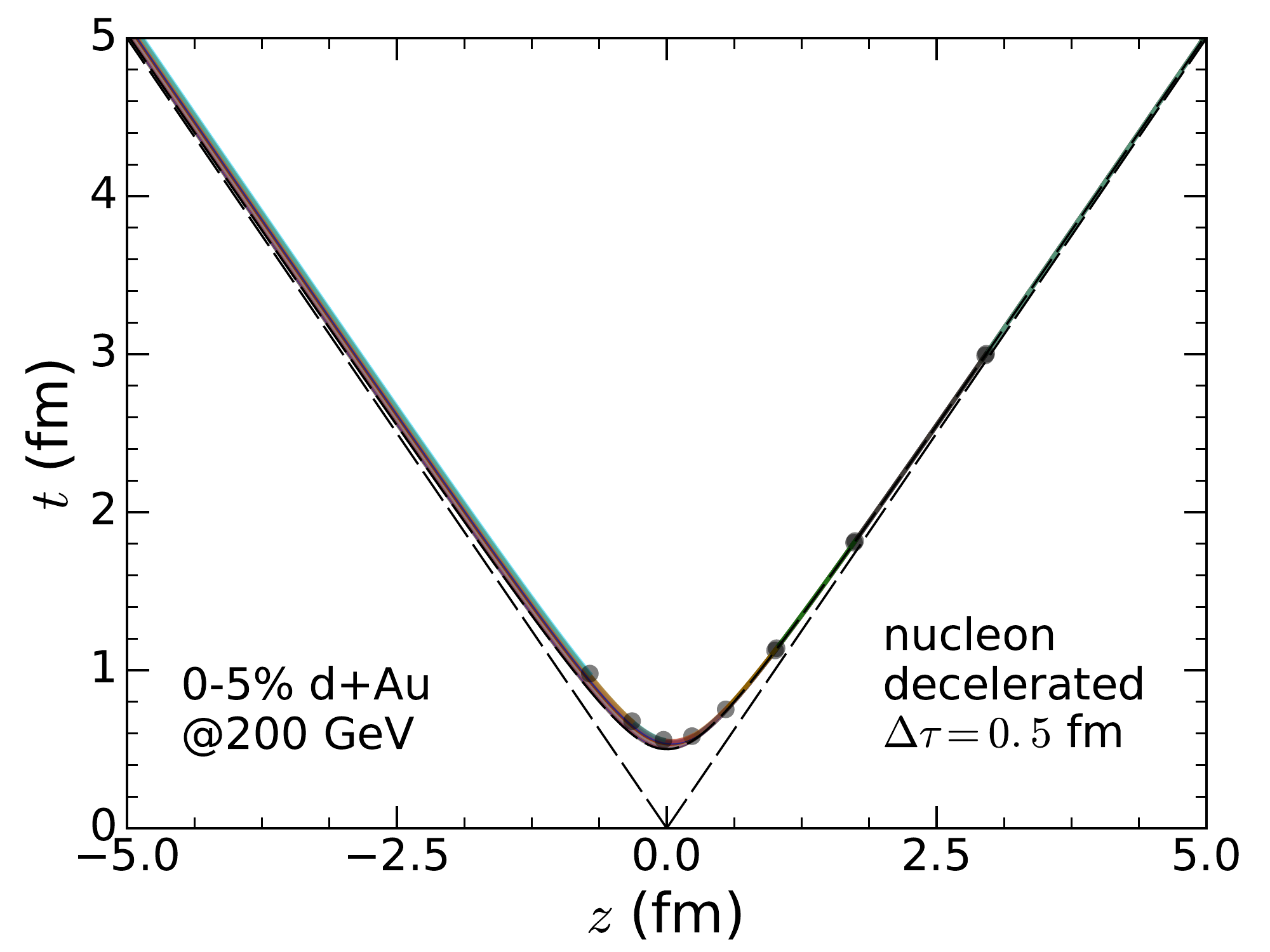} & 
   \includegraphics[width=0.45\linewidth]{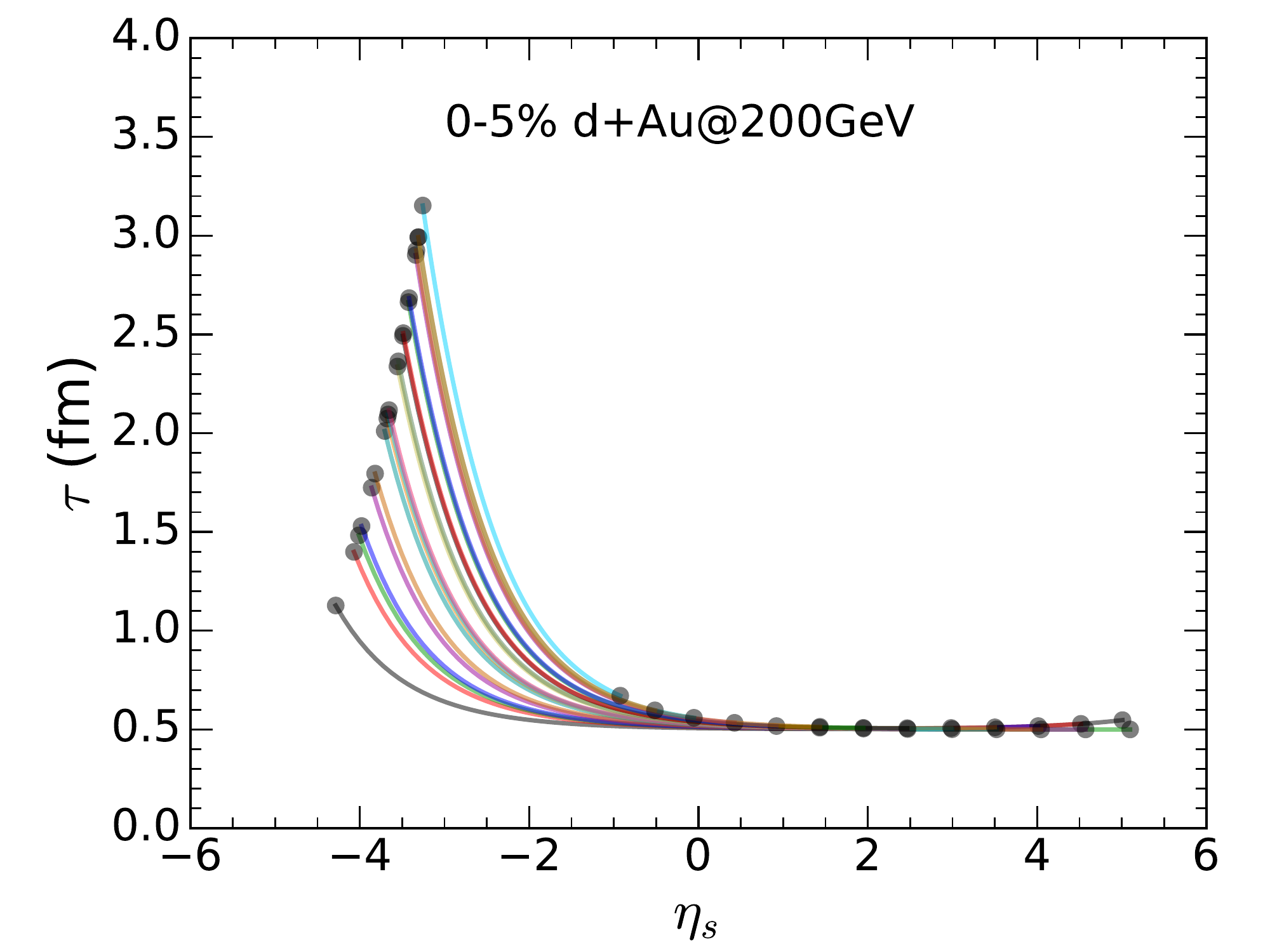} \\
  \includegraphics[width=0.45\linewidth]{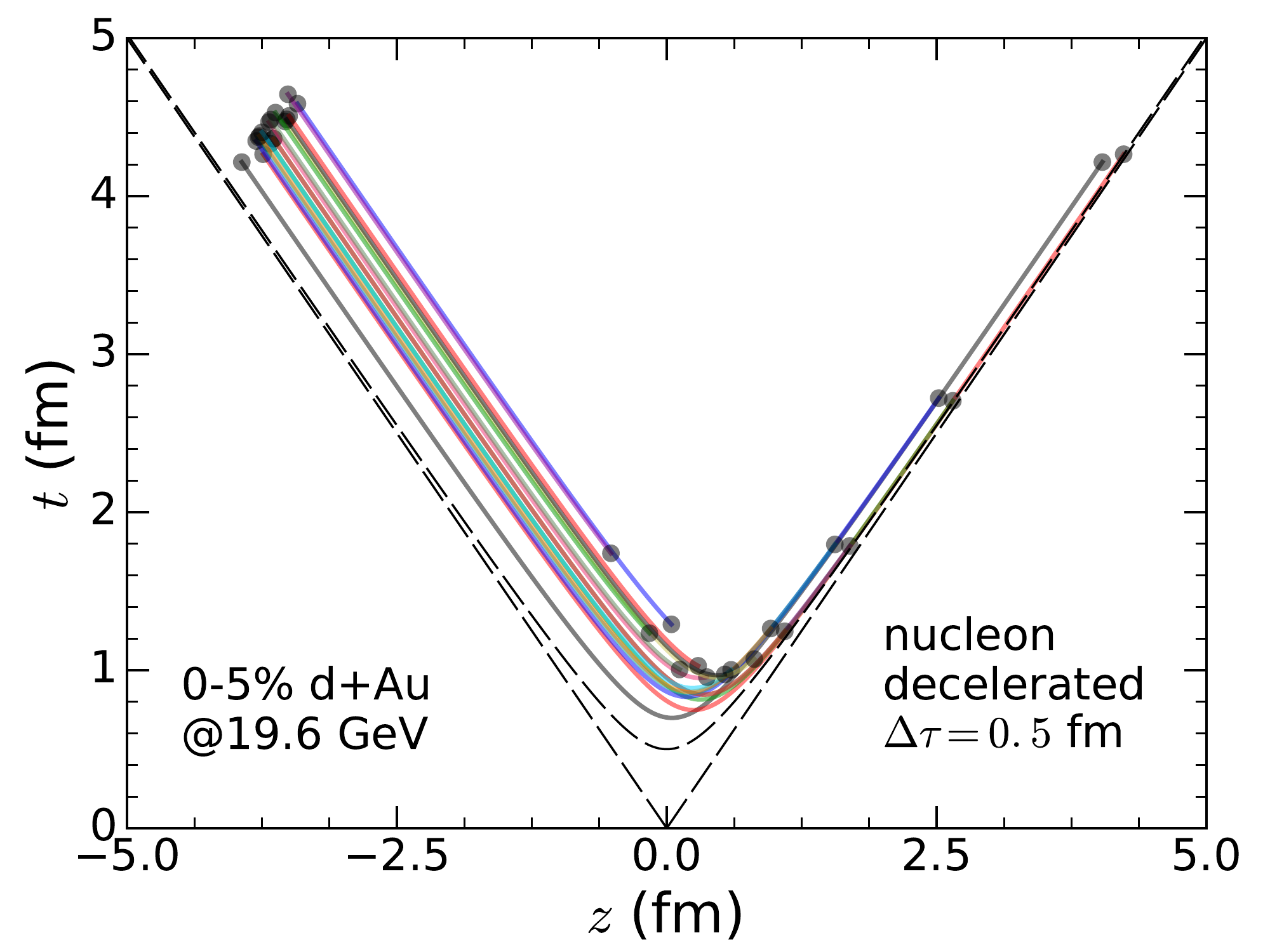} & 
   \includegraphics[width=0.45\linewidth]{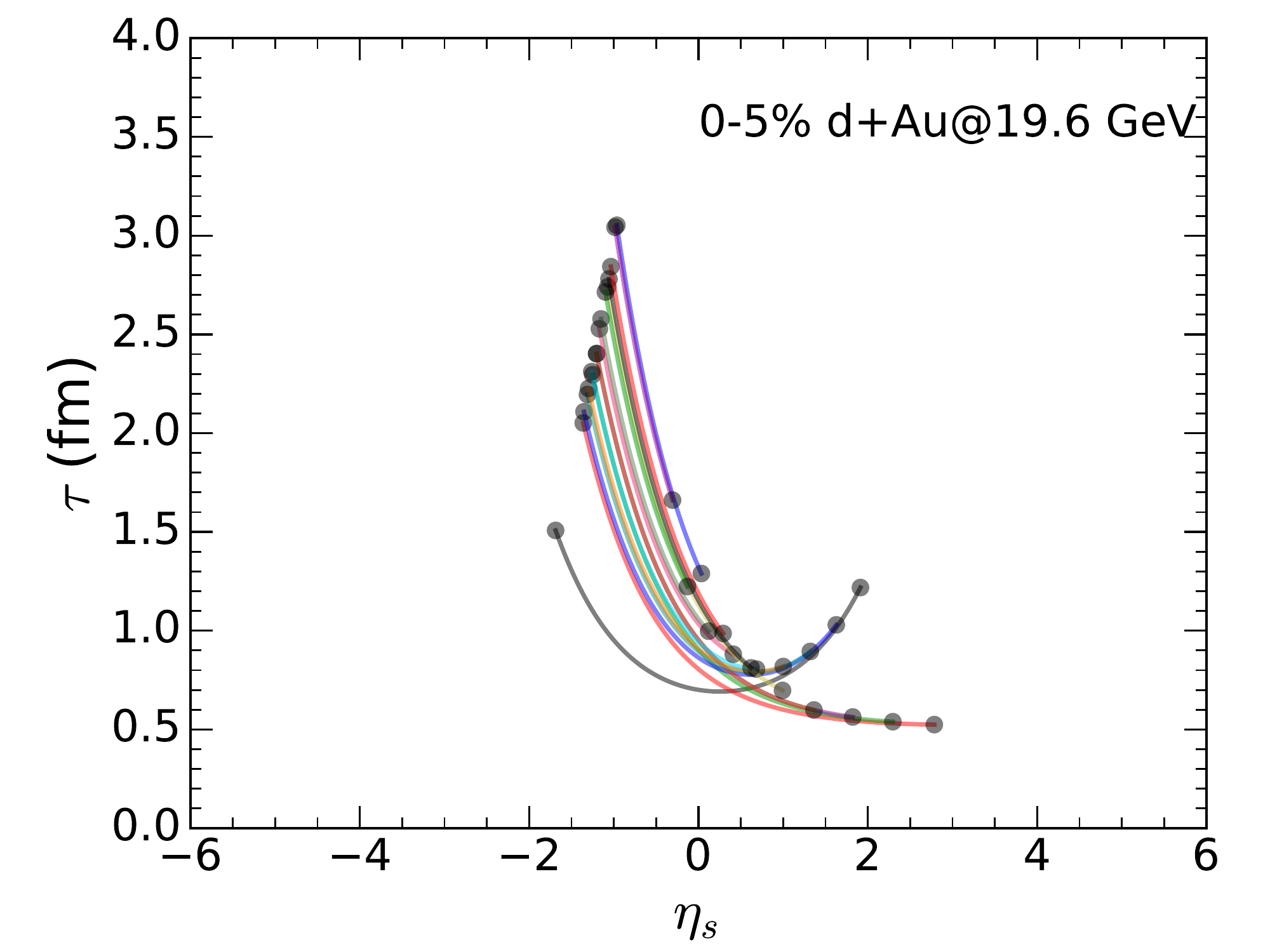}
   \end{tabular}
  \caption{Same as for Fig.~\ref{fig:AuAuStrings} but for a d+Au collision at 200 GeV and 19.6 GeV.}
  \label{fig:dAuStrings}
\end{figure*}
%

To fully construct source terms of energy and net baryon number density for hydrodynamic simulations we need the following information for each string
\begin{equation}
  \tau_c, \eta_{s, c}, x_c, y_c, \Delta \tau_f, \eta_{s, l}, \eta_{s, r}, y_l,
  y_r\,,
\end{equation}
where $\eta_{s, c}$ is the space time rapidity and $\tau_c$ the proper time of the collision, obtained from $t_c$ and $z_c$.

The space-time position of the entire string in the lab frame after evolving for $\Delta \tau$ is given by the equation
\begin{equation}
  (t - t_c)^2 - (z - z_c)^2 = \Delta \tau^2\,.
\end{equation}
The string will cross a given constant proper time $\tau$ surface at
\begin{equation}
  \eta_s = \eta_{s, c} \pm \tmop{arccosh} \left( \frac{\tau^2 + \tau_c^2 -
  \Delta \tau^2}{2 \tau \tau_c} \right) . \label{eq:etas}
\end{equation}
or
\begin{eqnarray}
  \tau &=& \tau_c \cosh (\eta_s - \eta_{s, c}) \notag \\
         && + \sqrt{\tau_c^2 \cosh^2 (\eta_s - \eta_{s, c}) - (\tau_c^2 - \Delta \tau^2)} .\label{eq:tau}
\end{eqnarray}
Eq.\,(\ref{eq:etas}) only has solutions for $\tau > \tau_c + \Delta \tau$, with the smallest value of $\tau$ realized at $\eta_s=\eta_{s,c}$. In practice, we will use above equations with $\Delta \tau = \Delta \tau_f$ to determine every string's space time position when it thermalizes and becomes part of the hydrodynamic medium. 

We know a string's extension in space-time rapidity, $[\eta_{s, l},
\eta_{s, r}]$, so this string will contribute as a source term during the
proper time interval determined by varying $\eta_s$ in Eq.\,(\ref{eq:tau}) between $\eta_{s,
l}$ and $\eta_{s, r}$.

Examples of the space time distribution of thermalizing strings, and thus the positions of sources for the hydrodynamic simulation, which will be detailed in the next section, are shown for Au+Au collisions in Fig.\,\ref{fig:AuAuStrings} and for d+Au collisions in Fig.\,\ref{fig:dAuStrings}. At the highest considered energy, $\sqrt{s}=200\,{\rm GeV}$, the $\tau$-range occupied by strings is rather limited around space-time rapidity $\eta_s=0$, however, at larger $|\eta_s|$ thermalizing strings are present for up to $\tau\approx 3\,{\rm fm}$. At the lower considered energy, $\sqrt{s}=19.6\,{\rm GeV}$, strings (sources) are spread over a wide range in $\tau$ for all space-time rapidities $\eta_s$. This clearly demonstrates the necessity to initialize the hydrodynamic simulation dynamically at energies below $200\,{\rm GeV}$. Furthermore, if one is interested in the dynamics away from mid-rapidity, considering source terms for up to $\tau\approx 3\,{\rm fm}$ may be necessary.

Let us note here that while we describe the deceleration of string ends dynamically, for the sake of simplicity we do not model string breaking or the emerging substructure of strings explicitly. Including string fragmentation as done for example in \cite{Andersson:1983ia,Werner:1994tw,Ivanyi:1999bv} can be added in future extensions of the model and will likely affect the detailed longitudinal structure of the deposited energy and net baryon numbers. Comparison to experimental observables discussed in Section \ref{sec:results} can then be used to constrain the detailed mechanisms of string breaking and deceleration.

\subsection{Choice of participants}
So far we have assumed that strings are connected to participant nucleons and disregarded any possible nucleonic substructure. In that case
the initial rapidities are fixed to $y^i_l = - y_\mathrm{beam}$ and $y^i_r = y_\mathrm{beam}$. Alternatively, for every wounded nucleon valence quarks can be sampled from the parton distribution function (PDF) and strings are spanned between two such quarks. Their initial rapidity can be estimated from the following formula,
\begin{equation}
y_q = \mathrm{arcsinh} \left(x_q \sqrt{\frac{s}{4 m_q^2} - 1}\right).
\label{eq.valenceQuark}
\end{equation}
In the high energy limit, $s \rightarrow \infty$, Eq.~(\ref{eq.valenceQuark}) reduces to the often used expression $y_q = \log \left(x_q \sqrt{s}/m_q \right)$, which would however lead to negative $y_q$ when $x_q \sqrt{s} < m_q$. 

Sampling initial rapidities for the constituent quark participants leads to fluctuations of the strings' lengths. 
The transverse position of the valence quarks are sampled from a 2D Gaussian distribution \cite{Monnai:2015sca}.

\subsection{Rapidity loss fluctuations}
%
\begin{figure}[ht!]
  \centering
   \includegraphics[width=0.9\linewidth]{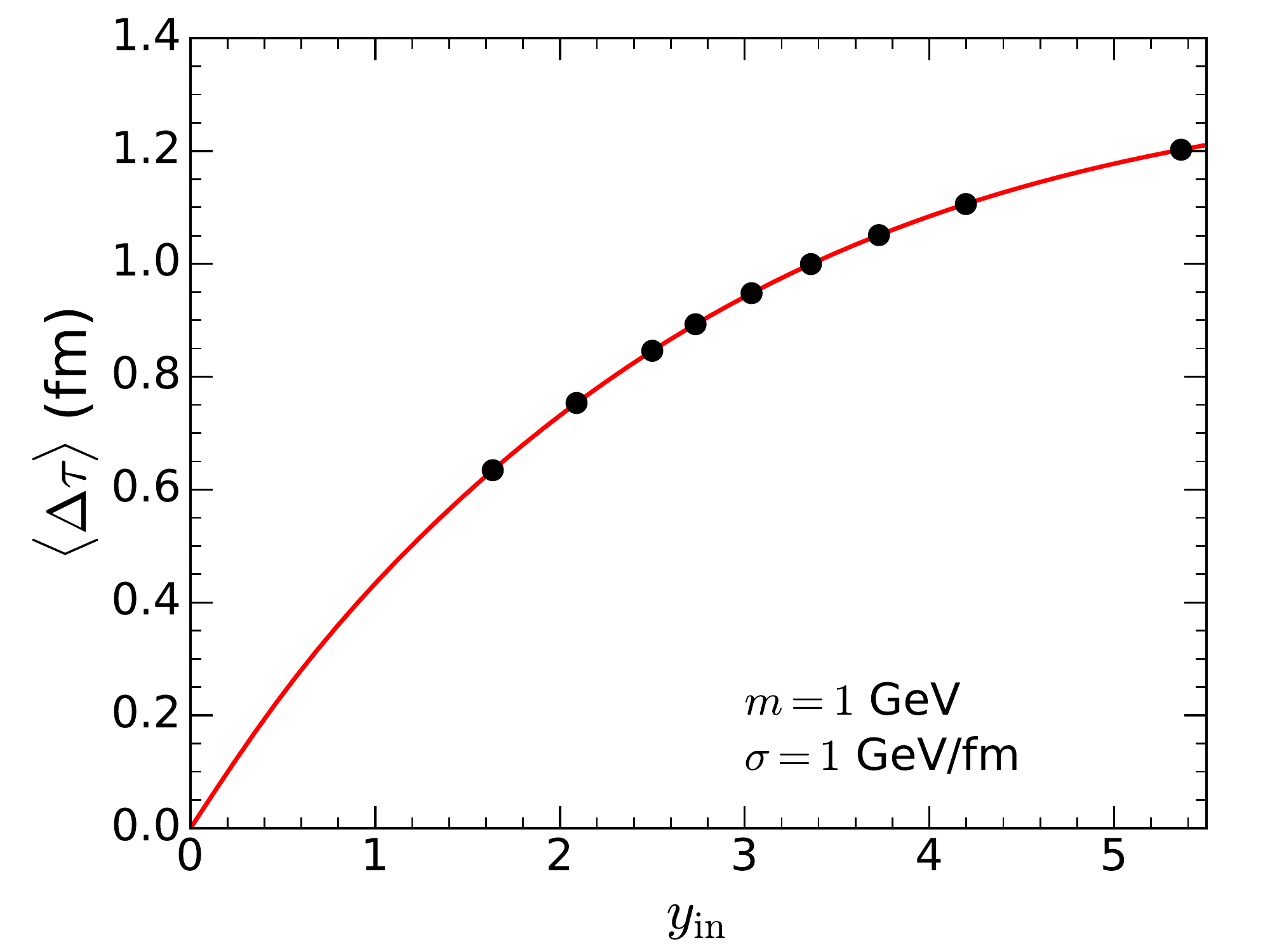}
   \caption{The average string deceleration time $\Delta \tau$ as a function of the initial string rapidity in the LEXUS model. The black points indicate the RHIC BES collision energies, 5, 7.7, 11.5, 14.5, 19.6, 27, 39, 62.4, and 200 GeV.}
  \label{fig:meanDeltatau}
\end{figure}
%

Alternatively to using a constant thermalization time for every string, in every nucleon-nucleon collision the rapidity loss of the incoming nucleons in their pair rest frame can be sampled from a probability distribution
\begin{eqnarray}
  P(y_\mathrm{loss}) = \frac{\cosh(2y^\mathrm{in}_\mathrm{lrf} - y_\mathrm{loss})}{\sinh(2y^\mathrm{in}_\mathrm{lrf}) - \sinh(y^\mathrm{in}_\mathrm{lrf})},
  \label{eq17}
\end{eqnarray}
where $y^\mathrm{in}_\mathrm{lrf}$ denotes the absolute value of the incoming nucleons' rapidity in their pair rest frame.
This distribution was first introduced within the LEXUS model \cite{Jeon:1997bp}.

The rapidity loss $y_\mathrm{loss}$ fluctuates in the range $[0, y^\mathrm{in}_\mathrm{lrf}]$, introducing fluctuations of the string thermalization time 
\begin{equation}
\Delta \tau_f= \frac{m}{\sigma}\sqrt{2(\cosh(y_\mathrm{loss}) - 1)}.
\end{equation}
The mean rapidity loss and average string deceleration time are
\begin{equation}
\langle y_\mathrm{loss} \rangle = \frac{\cosh(2  y^\mathrm{in}_\mathrm{lrf}) - \cosh( y^\mathrm{in}_\mathrm{lrf}) -  y^\mathrm{in}_\mathrm{lrf}\sinh( y^\mathrm{in}_\mathrm{lrf})}{\sinh(2 y^\mathrm{in}_\mathrm{lrf}) - \sinh(y^\mathrm{in}_\mathrm{lrf})}
\end{equation}
and
\begin{equation}
\langle \Delta \tau \rangle = \int_0^{y^\mathrm{in}_\mathrm{lrf}} dy_\mathrm{loss} \frac{m}{\sigma}\sqrt{2(\cosh(y_\mathrm{loss}) - 1)} P(y_\mathrm{loss}).
\end{equation}
For $ y^\mathrm{in}_\mathrm{lrf} \rightarrow \infty$, the mean rapidity loss $\langle y_\mathrm{loss} \rangle \rightarrow 1$. The average string deceleration time $\langle \Delta \tau \rangle$ is shown in Fig.~\ref{fig:meanDeltatau} for a string tension $\sigma = 1$ GeV/fm and particle mass $m = 1$ GeV. 
We see that even on average it takes a significant amount of time $\langle \Delta \tau \rangle = \mathcal{O}(1 {\rm fm})$ for a string to thermalize and deposit energy into the medium.

We will demonstrate the effect of these additional rapidity fluctuations on observables sensitive to longitudinal fluctuations in Section\,\ref{sec:results}.

\subsection{Sources for hydrodynamic fields}\label{sec:sources}
Individual strings are not thermalized at a fixed proper time and depending on the collision energy, many of them will thermalize after the hydrodynamic simulation has already started.\footnote{Once the deceleration of a string's endpoints is completed, the string is assumed to be thermalized with the medium and its energy and net baryon numbers are added to the hydrodynamic medium. We do not perform any explicit thermalization procedure, whose details are not known.} This is why we need to treat the energy and net-baryon number deposition dynamically
and introduce source terms to the hydrodynamic simulation. 
The velocity profile of the string is
\begin{align}
  u_\mathrm{string}^{\mu} (\tau, \eta_s) = \Big(\cosh &[y_\mathrm{string}(\eta_s) - \eta_s], 0, 0, \notag\\
    &\frac{1}{\tau}\sinh [y_\mathrm{string}(\eta_s) - \eta_s] \Big).
\end{align}
where $y_\mathrm{string}(\eta_s)$ is the momentum rapidity along the string defined at the time of string breaking $\tau_{\rm break}=\tau_c+\Delta\tau_f$.
We define the momentum rapidity profile via a linear interpolation
\begin{equation}
  y_\mathrm{string} (\eta_s) = y_l + \frac{y_r - y_l}{\eta_{s, r} - \eta_{s, l}} (\eta_s -
  \eta_{s, l}) . \label{eq:rapProfile}
\end{equation}
In order to vary how much of the source longitudinal velocity will be affected by the flow velocity from the medium at its thermalization we introduce a quenching factor $\alpha$
\begin{equation}
y^\mathrm{source}_{L, \alpha}(\eta_s) =  y_\mathrm{string} (\eta_s) - \alpha (y_\mathrm{string} (\eta_s) - y^\mathrm{flow}_L).
\end{equation}
The rapidity corresponding to the source's transverse velocity, which is entirely due to the motion of the medium, is given by
\begin{equation}
y^\mathrm{source}_{\perp, \alpha}(\eta_s) =  \alpha y^\mathrm{flow}_\perp\,,
\end{equation}
where $y^\mathrm{flow}_\perp$ is the transverse flow rapidity.
The four velocity of the source can be written as
\begin{eqnarray}
&& \!\!\!\!\!\!\!\!\!\!\!\!\! u^\mu_\mathrm{source}(\tau, \eta_s) = \notag \\
&&  \bigg( \cosh[y^\mathrm{source}_{L, \alpha}( \eta_s) - \eta_s] \cosh[y^\mathrm{source}_{\perp,\alpha}(\eta_s)], \notag \\
&&  \quad \sinh[y^\mathrm{source}_{\perp,\alpha}(\eta_s)] \cos(\phi),  \notag \\
&&  \quad \sinh[y^\mathrm{source}_{\perp,\alpha}(\eta_s)] \sin(\phi),  \notag \\
&&  \quad \sinh[y^\mathrm{source}_{L, \alpha}( \eta_s) - \eta_s] \cosh[y^\mathrm{source}_{\perp,\alpha}(\eta_s)] \bigg).
\end{eqnarray}
For $\alpha = 0$ the string is not affected by the background medium such that $y^\mathrm{source}_{L,\alpha}(\eta_s) = y_\mathrm{string} (\eta_s)$ and $y^\mathrm{source}_{\perp, \alpha} = 0$. When $\alpha = 1$ the string is fully `stopped' to the medium flow velocity $y^\mathrm{source}_{L,\alpha}(\eta_s) = y^\mathrm{flow}_L$ and $y^\mathrm{source}_{\perp, \alpha} = y_\perp^\mathrm{flow}$. 

The energy-momentum current $J^\mu_\mathrm{source}$ at a given space-time point, $x^\alpha$, is defined as
\begin{equation}
J^\mu_\mathrm{source} (x^\alpha) = \sum_{i \in \mathrm{\{strings\}}} e_i u_{i, \mathrm{source}}^{\mu} f_\mathrm{smear}(x^\alpha; x_i^\alpha).
\end{equation}
Here $e_i$ is the local energy density of the string $i$. The spatial smearing function $f_\mathrm{smear}(x^\alpha; x_i^\alpha)$ takes the form
\begin{equation}
f_\mathrm{smear}(x^\alpha; x_i^\alpha) = \frac{\delta(\tau - \tau_i)}{\tau} f_{\perp} (x, y; x_i, y_i) f_{\eta_s} (\eta_s; \eta_{s, i}).
\end{equation}
We use Gaussian smearing profiles in the transverse and longitudinal directions,
\begin{equation}
  f_{\perp} (x, y; x_i, y_i) = \frac{1}{\pi \sigma_{\perp}^2} \exp \left[ - \frac{(x - x_i)^2 - (y - y_i)^2}{\sigma_{\perp}^2} \right]\,,
\end{equation}
and
\begin{equation}\label{eq:fLong}
  f_{\eta_s} (\eta_s; \eta_{s, i}) = \frac{1}{\sqrt{\pi} \sigma_{\eta_s}} \exp \left[ - \frac{(\eta_s - \eta_{s, i})^2}{\sigma_{\eta_s}^2} \right] .
\end{equation}
with $\sigma_\perp = 0.5$ fm and $\sigma_{\eta_s} = 0.2$ as the size of the hot spot for a source at $(x_i, y_i, \eta_{s,i})$.

In this work we consider net-baryon charge, but below description could also be extended to other conserved quantities. 
To get the source term for the conserved charge density $\rho_{\tmop{source}}$, we first consider the charge number current,
\begin{equation}
N^\mu_Q(x^\alpha) = \sum_{i \in \mathrm{\{participants\}}} Q_i \frac{P_i^\mu}{P_i^\tau} f_\mathrm{smear}(x^\alpha, x_i^\alpha),
\label{eq18}
\end{equation}
where $P^\mu$ is the momentum of the baryon charge.
Here the index $i$ sums over all the participants in the collision. $Q_i$ stands for the quantum charge of participant $i$. The source term $\rho_\mathrm{source}$ in one fluid cell can be computed as
\begin{eqnarray}
\!\!\!\!\!\!\!\!\!\! \rho_\mathrm{source}(x^\alpha) &=& u^\mathrm{flow}_\mu N_Q^\mu(x^\alpha) \notag \\
&=& \sum_{i \in \mathrm{\{participants\}}} Q_i \frac{u^\mathrm{flow}_\mu P_i^\mu}{P_i^\tau} f_\mathrm{smear}(x^\alpha, x_i^\alpha).
\end{eqnarray}

The energy density distribution, obtained from the sum over all strings should exhibit a plateau in the $\eta_s$ direction.
Since in the hydrodynamic simulation, we propagate the system in proper time $\tau$, we need to include the Jacobian in
the Gaussian profile to take into account the difference in $d \tau$ and $d \eta_s$ when we integrate over the positions of all Gaussians 
\begin{eqnarray}
  \tmop{plateau} &=& \int_{\eta_{\min}}^{\eta_{\max}} d \eta_G e^{- (\eta -
  \eta_G)^2 / \sigma_{\eta}^2} \notag \\
   &= &\int d \tau \frac{d \eta_G}{d \tau} e^{-
  (\eta - \eta_G)^2 / \sigma_{\eta}^2}
\end{eqnarray}
with
\begin{equation}
  \frac{d \eta_G}{d \tau} = \pm \frac{1}{\tau} \frac{\tau^2 - \tau_0^2 + \Delta
  \tau^2}{\sqrt{(\tau^2 + \tau_0^2 - \Delta \tau^2)^2 - 4 \tau^2 \tau_0^2}},
\end{equation}
where $d\eta_G/d\tau > 0$ for $\eta_s > \eta_{s,0}$ and $d\eta_G/d\tau < 0$ for $\eta_s < \eta_{s,0}$.

\section{Initialize hydrodynamic fields with source terms} \label{sec:hydro}

The hydrodynamic equations with source terms \cite{Kajantie:1982jt,Kajantie:1982nh} can be written as
\begin{equation}
  \partial_{\mu} T^{\mu \nu} = J^{\nu}_{\tmop{source}}
  \label{eq3.1}
\end{equation}
and
\begin{equation}
  \partial_{\mu} J^{\mu} = \rho_{\tmop{source}}.
  \label{eq3.2}
\end{equation}
To understand the effect of the energy momentum source current, we can consider the ideal part of the hydrodynamic equations in the local rest frame. By projecting Eq.~(\ref{eq3.1}) with $u^{\mu}$ and $\Delta^{\mu \nu}$, we have
\begin{equation}
  D e = - (e + P) \theta + u_{\nu} J^{\nu}_{\tmop{source}}
  \label{eq3.3}
\end{equation}
and
\begin{equation}
  D u^{\mu} = \frac{\nabla^{\mu} P + \Delta^{\mu \nu} J_{\nu,
  \tmop{source}}}{e + P}\,,
  \label{eq3.4}
\end{equation}
where $D=u_\mu \partial^\mu$ and $\theta=\partial_\mu u^\mu$.
The component of $J^{\nu}_{\tmop{source}}$ that is parallel to $u^\nu$ in Eq.~(\ref{eq3.3}) feeds energy to the local hydrodynamic medium. We denote it as $\delta e = u_{\nu} J^{\nu}_{\tmop{source}}$.
In Eq.~(\ref{eq3.4}), the orthogonal component of $J^{\nu}_{\tmop{source}}$ acts as an acceleration force on the fluid cell. We denote the acceleration vector as
\begin{equation}
  \delta u^{\mu} = \frac{\Delta^{\mu \nu} J_{\tmop{source}, \nu}}{e + P} .
\end{equation}
So the energy momentum source vector can be decomposed as
\begin{equation}
  J^{\mu}_{\tmop{source}} = \delta e u^{\mu} + (e + P) \delta u^{\mu} .
\end{equation}
The charge current conservation equation Eq.~(\ref{eq3.2}) can be written as,
\begin{equation}
  D \rho = - \rho \theta + \rho_{\tmop{source}}.
\end{equation}
The $\rho_{\tmop{source}}$ is understood as a source term that contributes to the
local conserved charges. In general, the velocity of the baryon
charge may not be exactly equal to the flow velocity of the medium. In this case, the conserved charge current $N_Q^\mu$ in Eq.~(\ref{eq18}) contributes to a charge diffusion current
\begin{equation}
  q_Q^{\mu} = \Delta^{\mu \nu} N_{Q,\nu} \,.
\end{equation}
The size of the diffusion current can be quantified using the inverse Reynolds number for the diffusion current
\begin{equation}
  R_q^{-1} = \sqrt{ - \frac{q_{Q, \mu} q^{\mu}_Q}{\rho^2_\mathrm{source}}} = \sqrt{ 1 - \frac{N_{Q,\mu} N_Q^\mu}{(u_\mu N_Q^\mu)^2}}.
\end{equation}
With $N_Q^\mu$ in Eq. (\ref{eq18}), we find $0 \le R_q^{-1} \le 1$. 

\section{Results}\label{sec:results}

In this section we present results from four variations of the initial state model introduced above. They differ by the choice of participants (nucleons or constituent quarks) and by whether we decelerate participants for a constant proper time $\Delta \tau$ or a deceleration time that fluctuates together with the rapidity loss. The valence quarks' $x$ values are sampled from the CT10NNLO PDFs \cite{Gao:2013xoa}. For Au and Pb nuclei, nuclear many-body effects are included by using the EPS09 nuclear PDF \cite{Eskola:2009uj}.

\subsection{Baryon stopping}

%
\begin{figure}[h!]
  \centering
  \includegraphics[width=0.9\linewidth]{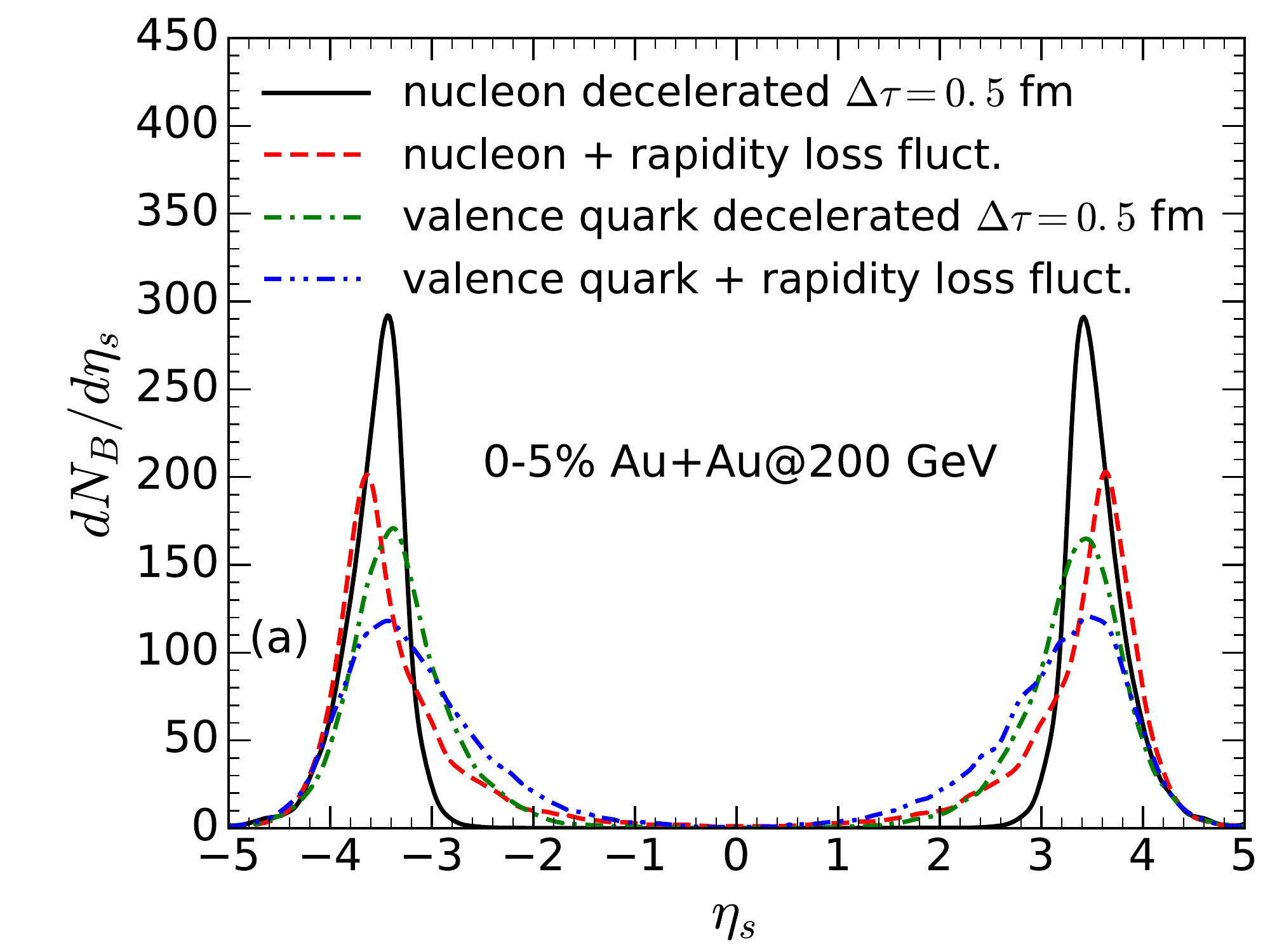} \\
  \includegraphics[width=0.9\linewidth]{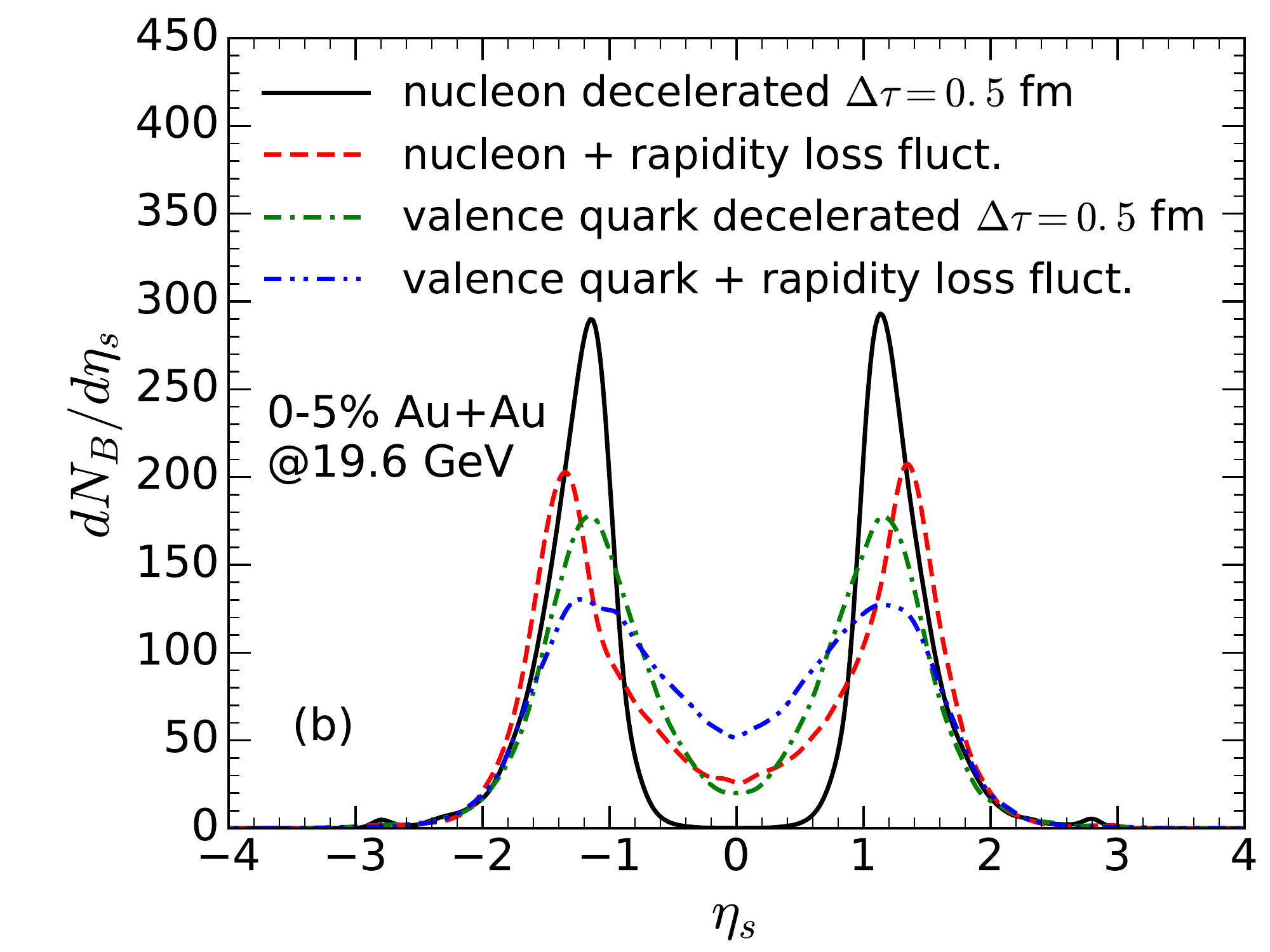}
  \caption{The space-time rapidity distributions of charged hadrons and net baryon numbers from the four initial conditions for 0-5\% Au+Au collisions at 200 GeV (a) and 19.6 GeV (b).}
  \label{fig9}
\end{figure}
%
Fig.~\ref{fig9} shows the comparison of the space-time rapidity distribution of the net baryon number from the four initial state models at two collision energies. The peak positions of the net baryon distributions are approximately the same in all four models. The largest differences are visible around midrapidity. The model using nucleon degrees of freedom and a constant deceleration time results in the smallest, almost zero, baryon density at midrapidity for both energies. This is easy to understand, because the incoming nucleons are all assigned the same beam rapidity and the constant deceleration time corresponds to a constant rapidity loss for every binary collision that produced a string. Because  there are very few nucleons connected with multiple strings with our numerical method described in the Appendix, the model for nucleons decelerated with a constant $\Delta \tau = 0.5\,$ fm predominantly produces a shift of the initial peaks around the beam rapidity, which for the considered energies is not large enough to move baryons to mid-rapidity. 

Incorporating fluctuations of the outgoing rapidities inspired by the LEXUS model in Eq.~(\ref{eq17}) allows for a large rapidity loss in a single collision. Hence there is a finite probability for a nucleon to be stopped near mid-rapidity. 

When considering valence quark degrees of freedom, additional rapidity fluctuations appear because their $x$ values are sampled from the PDF. In addition, this leads to shifts towards smaller rapidities and quark participants are more likely to be stopped at mid-rapidity. Finally, the valence quark model with additional rapidity fluctuations includes both effects mentioned above. Consequently it produces the highest baryon density around mid-rapidity.

We note that including a string breaking mechanism in the model will likely modify the baryon density distribution along the string. This is because the formation of baryons away from the original string ends would be possible. It will be interesting to study the effect on average baryon production and its fluctuations.

%
\begin{figure}[h!]
  \centering
  \includegraphics[width=0.9\linewidth]{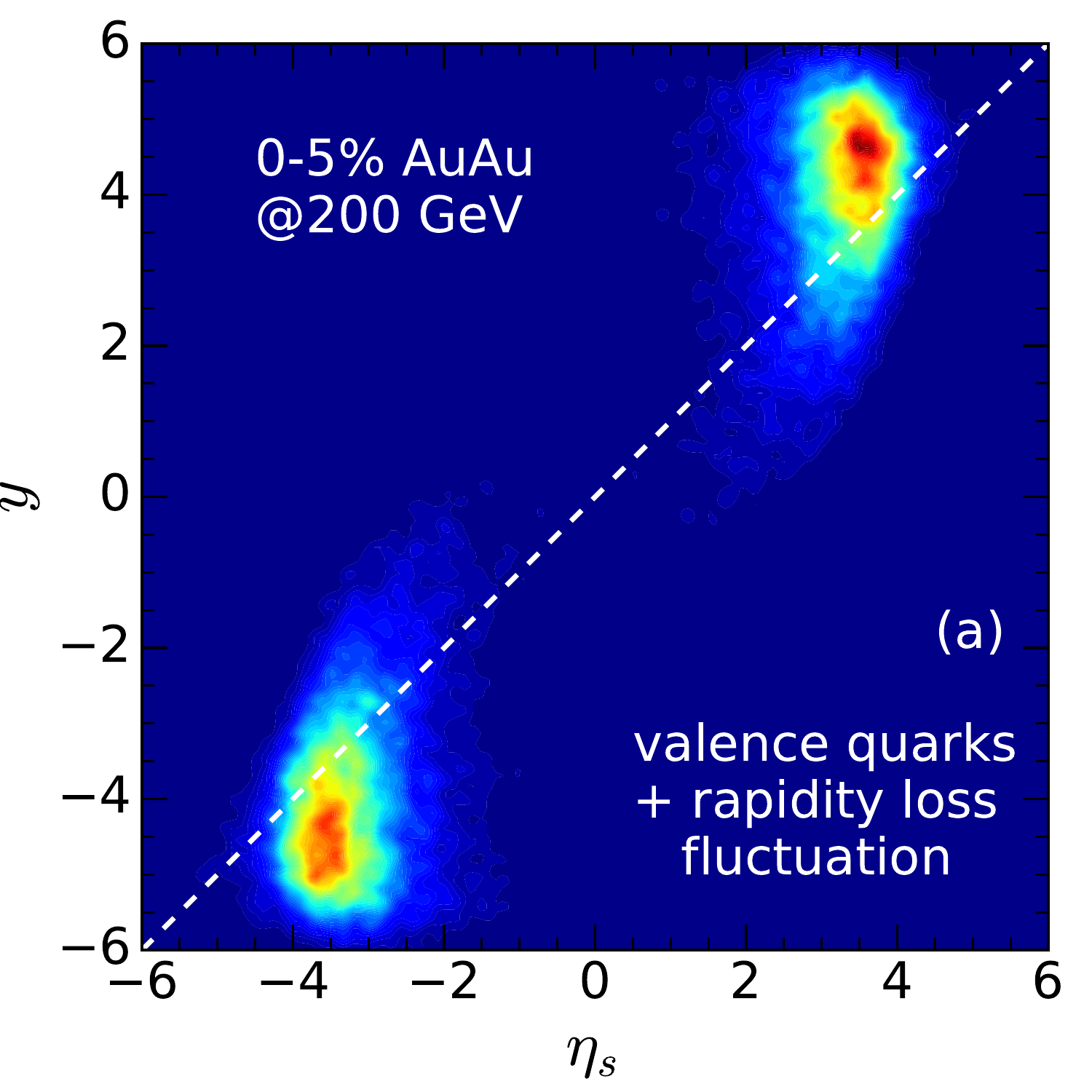} \\
  \includegraphics[width=0.9\linewidth]{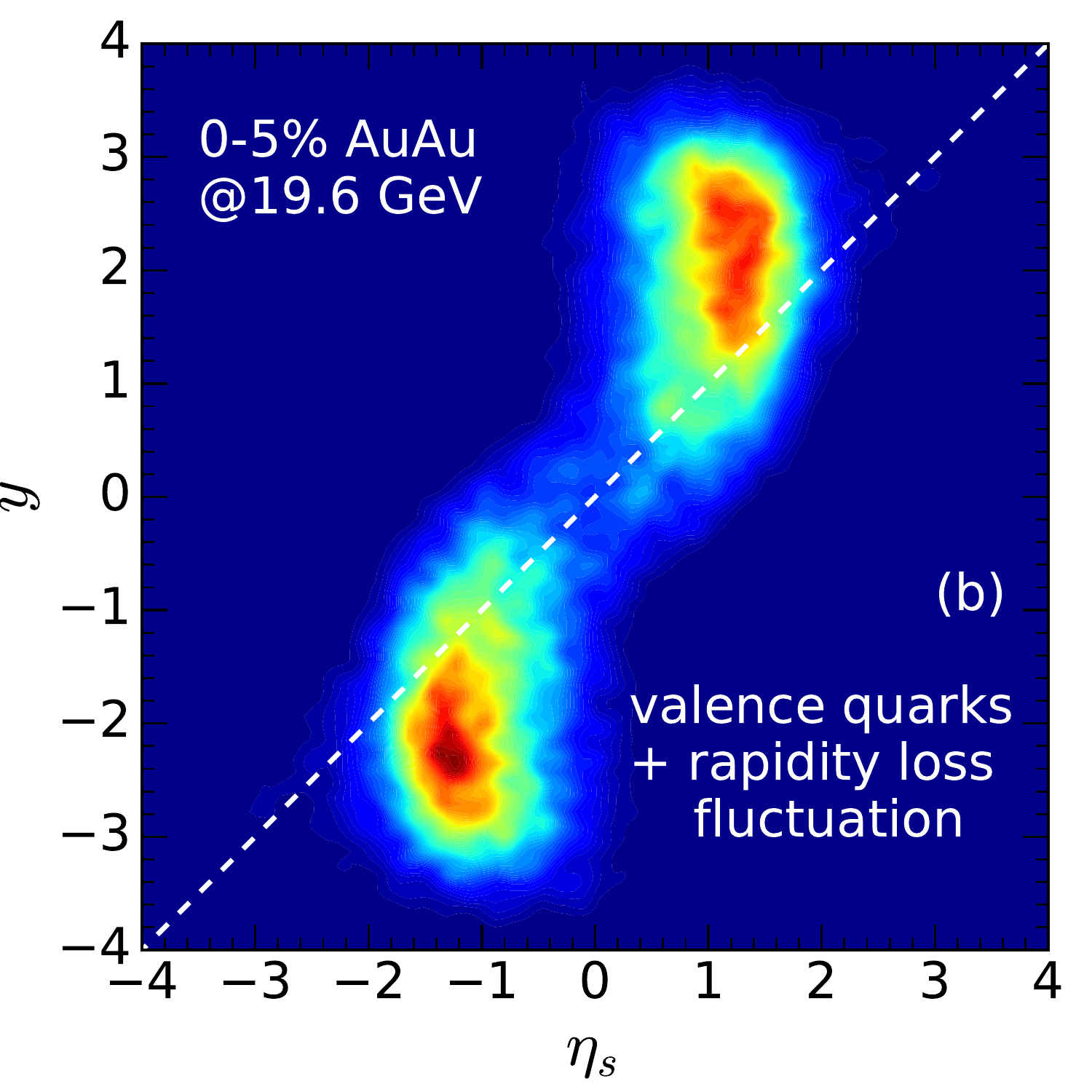}
  \caption{Contour plots for the correlation of the net baryon rapidity in momentum space to its space-time rapidity $\eta_s$ from the valence quark + rapidity loss fluctuation model for 0-5\% Au+Au collisions at 200 GeV (a) and 19.6 GeV (b).}
  \label{fig10}
\end{figure}
%

Our model simulates the space-time evolution of the participant nucleons or quarks dynamically assuming deceleration caused by a constant string tension. Consequently, the resulting momentum rapidities of the baryons are not equal to their space-time rapidities. Fig.~\ref{fig10} shows the correlation between the net baryons' momentum space rapidities $y$ with their space-time rapidities $\eta_s$ after the deceleration dynamics. At forward rapidities, the net baryons' rapidities $y$ are typically approximately 1-2 units larger than their space-time rapidities $\eta_s$. This can be understood because the deceleration time is finite (also see Fig.\,\ref{fig2}) and the binary collision points are not at the origin of the collision system $t = 0, z = 0$.  

%
\begin{figure}[ht!]
  \centering
  \includegraphics[width=0.9\linewidth]{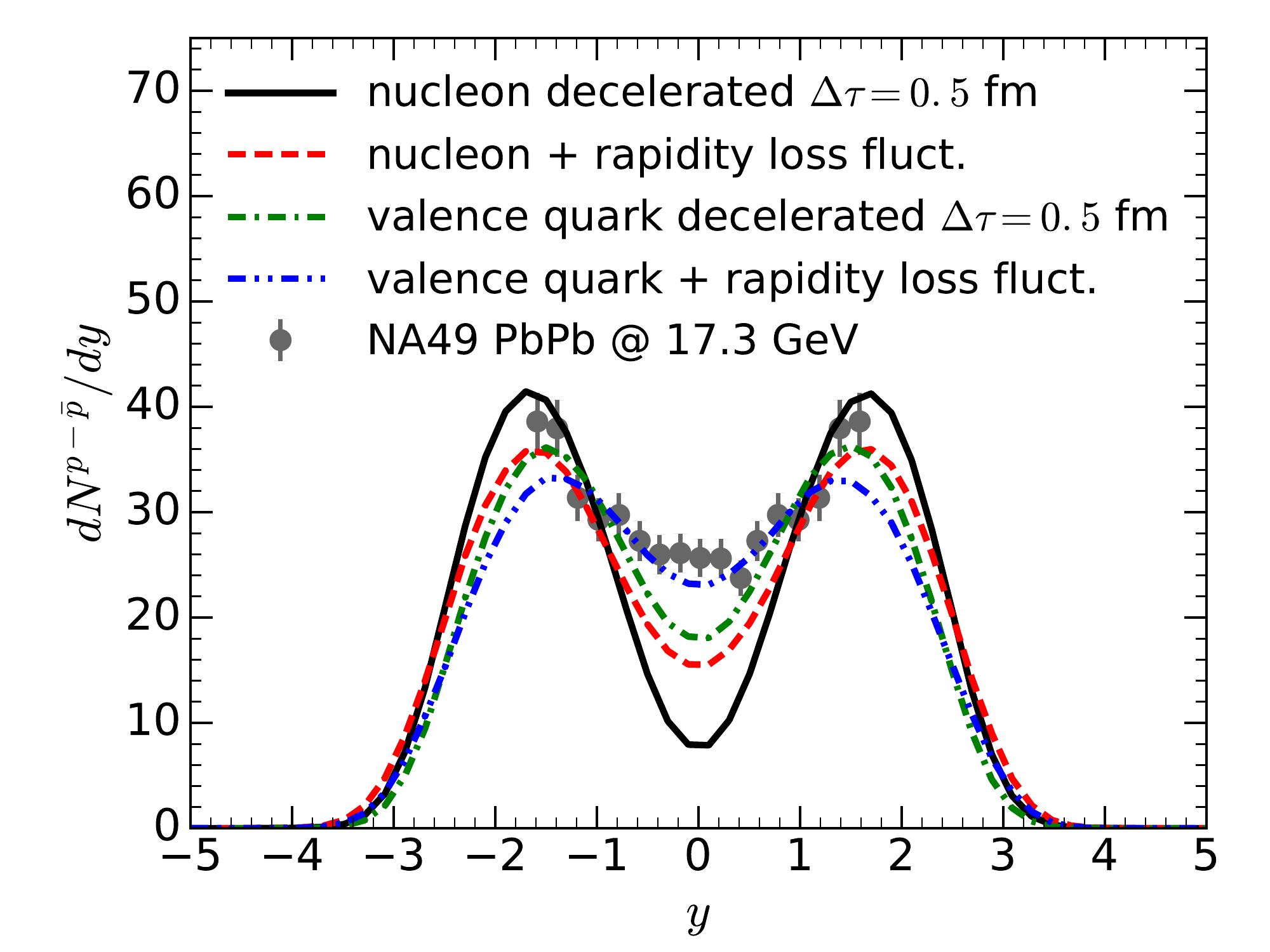}
  \caption{The rapidity distribution of net protons from the four different initial state models coupled with hydrodynamics + hadronic cascade simulations in central Pb+Pb collisions at the top SPS energy \cite{Anticic:2010mp}.}
  \label{fig11.1}
\end{figure}
%
%
%
\begin{figure}[ht!]
  \centering
  \includegraphics[width=0.9\linewidth]{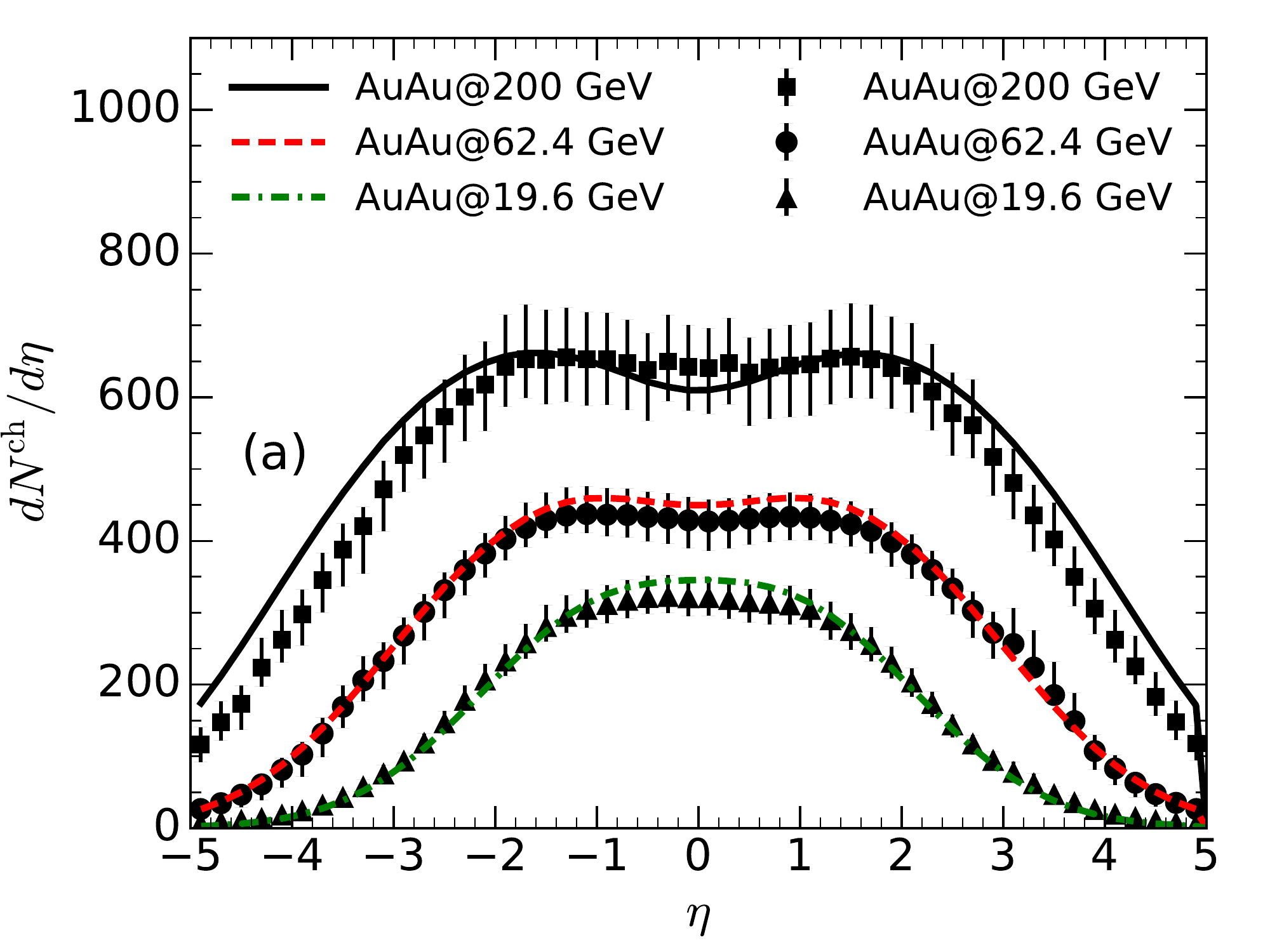}
  \includegraphics[width=0.9\linewidth]{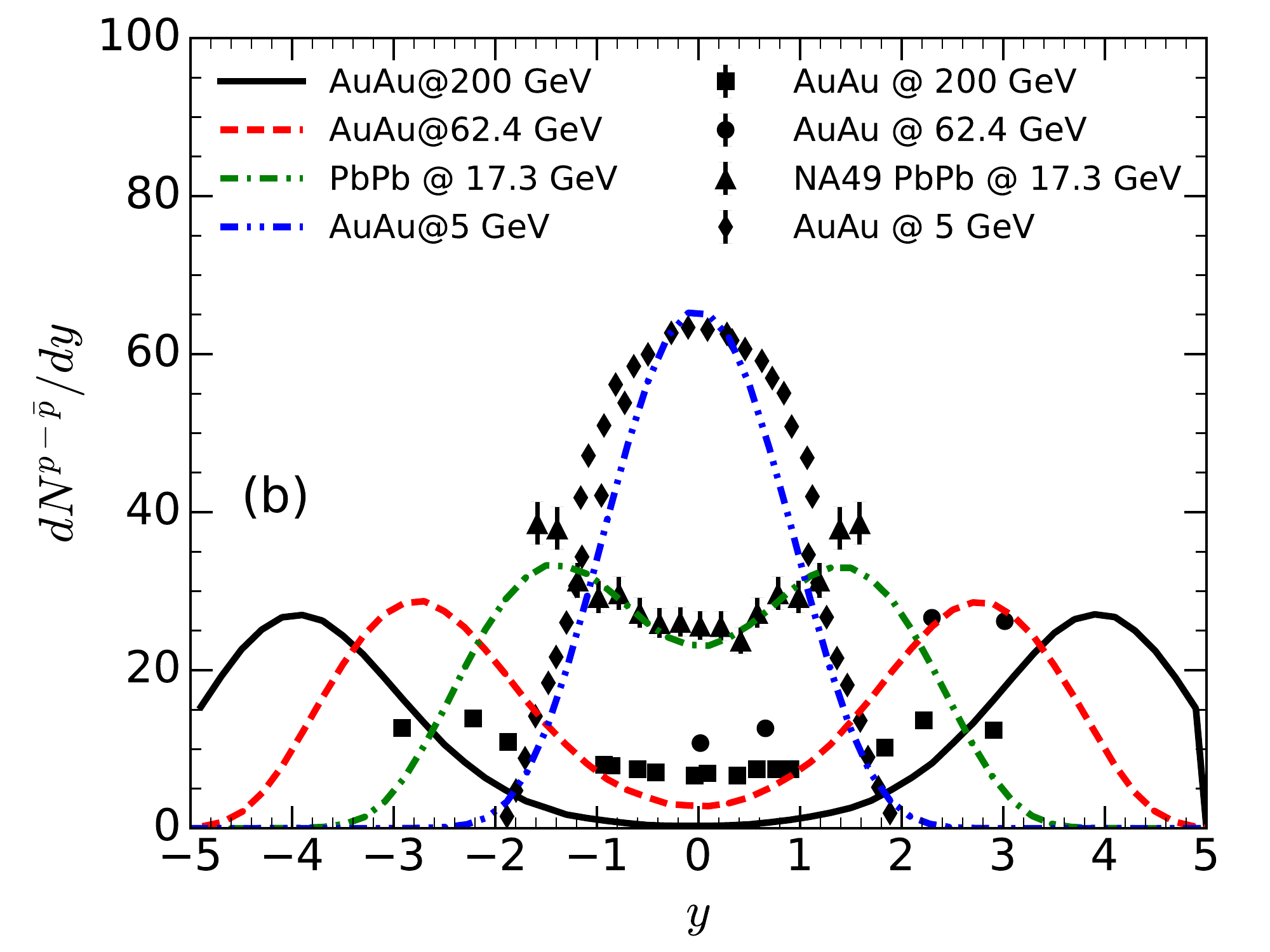}
  \caption{The rapidity distributions of charged hadrons (a) and net protons (b) at four different collision energies from our hybrid simulations with the model using valence quarks + rapidity loss fluctuations.}
  \label{fig11}
\end{figure}
%
%
In order to present first comparisons to experimental measurements we initialize hybrid (ideal hydrodynamics\footnote{The numerical code is publicly available at \url{http://www.physics.mcgill.ca/music}.} \cite{Schenke:2010nt} + hadronic cascade \cite{Bass:1998ca,Bleicher:1999xi}) simulations using source terms obtained from the four variants of our initial state model.
Fig.~\ref{fig11.1} compares the shape of the net proton rapidity distribution among the four initial state models for 0-5\% Pb+Pb collisions at the top SPS energy \cite{Anticic:2010mp}. The hybrid model evolves the system's energy and net baryon density distributions that are dynamically injected by the source terms discussed in Sections \ref{sec:sources} and \ref{sec:hydro} until freeze-out. 

The net proton rapidity distribution maintains a similar shape as the initial net baryon distribution shown in Fig.~\ref{fig9}. The nucleon model with a constant deceleration time produces the smallest number of net protons in the mid-rapidity region after the dynamical evolution. Additional fluctuations, included by using quark degrees of freedom or fluctuations of the final rapidities, allow more baryons to be transported from the forward to central rapidities and lead to larger net-proton numbers around mid-rapidity. For the used string tension $\sigma = 1$ GeV/fm, the valence quark + final rapidity fluctuations model provides a reasonable description of the experimental data \cite{Anticic:2010mp}, while other models underestimates the net proton yield at $y=0$.

Fig.~\ref{fig11} further studies the charged hadron $dN^\mathrm{ch}/d\eta$ and net proton $dN^{p-\bar{p}}/dy$ at different collision energies using the valence quark + final rapidity fluctuation model. At every collision energy, we adjust an overall normalization factor for the system's total entropy such that the measured charged hadron multiplicity is reproduced. This amounts to modeling the energy dependence of particle production. We found that this normalization factor is the same from $\sqrt{s_{NN}} = 62.4$ GeV to $\sqrt{s_{NN}} = 19.6$ GeV. But it is about 25\% larger at the top 200 GeV. The shape of the charged hadron pseudo-rapidity distribution $dN^\mathrm{ch}/d\eta$ is reasonably predicted by the valence quark + final rapidity fluctuation model. The $dN^\mathrm{ch}/d\eta$ is slightly wider compared to the RHIC measurements at 200 GeV \cite{Alver:2010ck}. For the net proton rapidity distribution, a reasonable agreement with experimental data is found for Au+Au collisions at $\sqrt{s_\mathrm{NN}} = 5$ GeV \cite{Ahle:1998jc,Ahle:1999in,Barrette:1999ry} and Pb+Pb collisions at $\sqrt{s_\mathrm{NN}} = 17.3$ GeV \cite{Anticic:2010mp}. The net proton yield is underestimated in the central rapidity region at the higher RHIC energies \cite{Arsene:2009aa}. Considering finite net baryon diffusion effects in the hydrodynamic simulations will likely improve the agreement with experimental data \cite{Shen:2017ruz}. 

\subsection{Longitudinal fluctuations}

The degree of longitudinal fluctuations can be quantified by studying two-particle multiplicity rapidity correlations \cite{Aaboud:2016jnr} and the event-plane decorrelation ratio $r_n$ as a function of rapidity \cite{Khachatryan:2015oea,Aaboud:2017tql}.   
\begin{figure}[h!]
  \centering
  \includegraphics[width=0.9\linewidth]{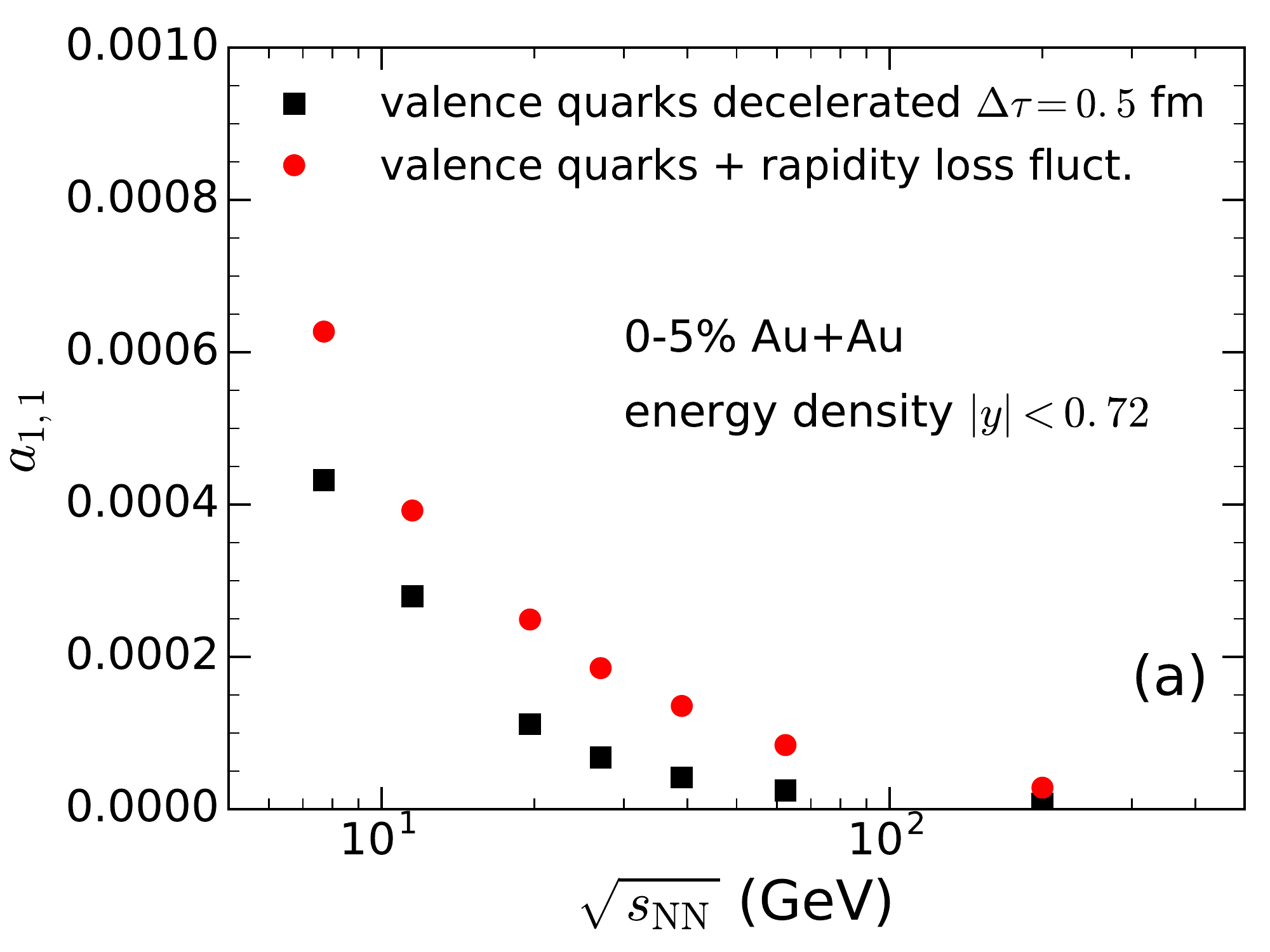} \\
  \includegraphics[width=0.9\linewidth]{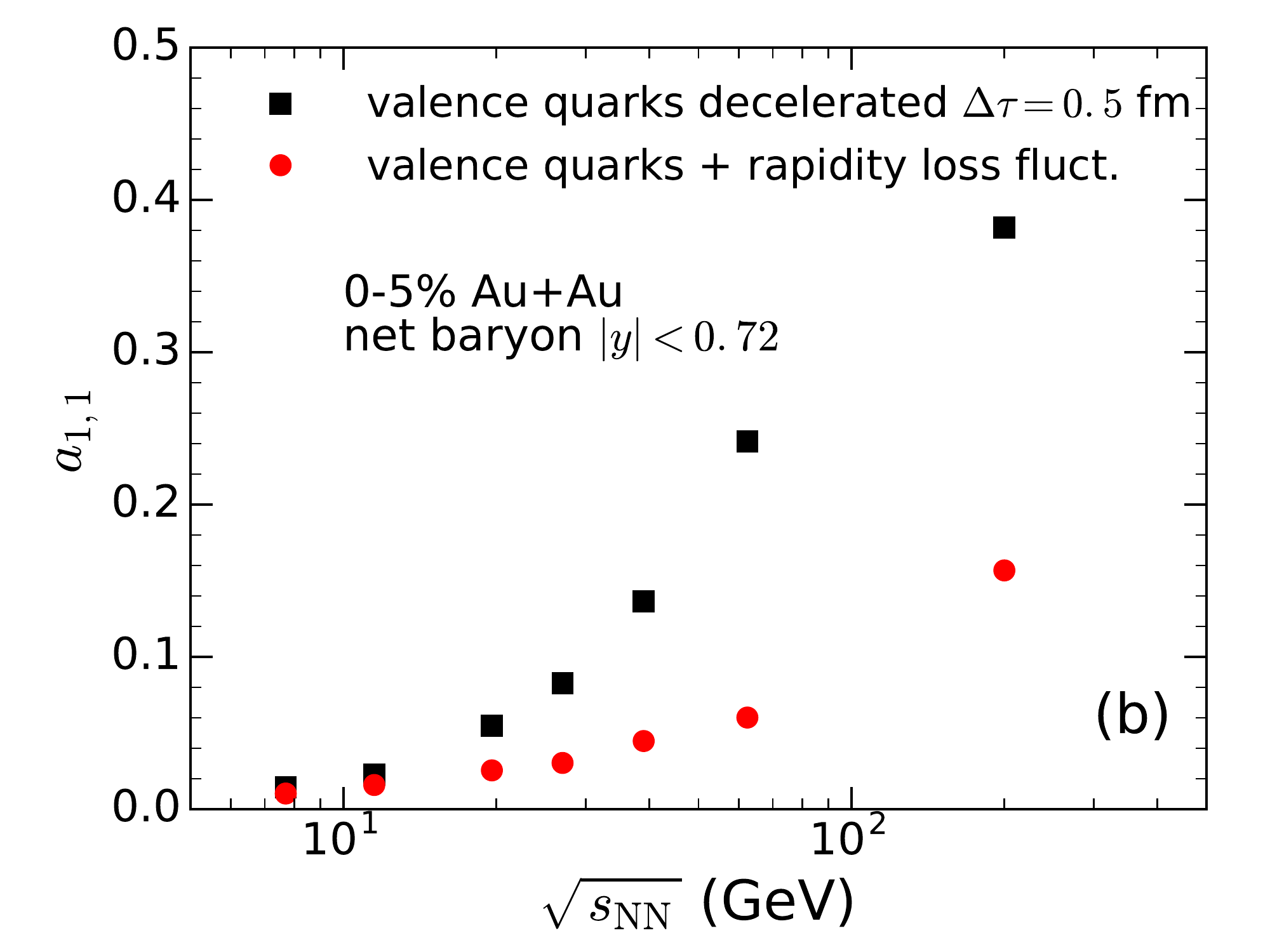}
  \caption{The longitudinal fluctuation coefficients $\{a_{1,1}\}$ for initial energy (a) and net baryon number (b) density profiles as functions of collision energy in 0-5\% Au+Au collisions. The rapidity window is chosen to be $\vert y \vert < 0.72$ to be consistent with the current STAR measurements.}
  \label{fig6}
\end{figure}
%

The two particle rapidity multiplicity correlation function is defined as \cite{Jia:2015jga} 
\begin{equation}
C(\eta_1, \eta_2) = \frac{\langle \frac{dN}{d\eta}(\eta_1) \frac{dN}{d\eta}(\eta_2) \rangle}{\langle \frac{dN}{d\eta}(\eta_1) \rangle \langle \frac{dN}{d\eta}(\eta_2) \rangle}\,,
\end{equation}
and the corresponding normalized multiplicity correlation function is
\begin{equation}
C_N(\eta_1, \eta_2) = \frac{C(\eta_1, \eta_2)}{C_\rho(\eta_1) C_\rho(\eta_2)},
\end{equation}
where the denominator $C_\rho(\eta_1) = \frac{1}{2 Y} \int_{-Y}^{Y} C(\eta_1, \eta_2) d \eta_2$ is the marginal distribution. The analyzed rapidity window is chosen to be from $-Y$ to $Y$. The normalized multiplicity correlation function can be expanded into a Legendre series with coefficients
\begin{eqnarray}
a_{n,m}&= & \int \frac{d\eta_1}{Y} \frac{d\eta_2}{Y} C_N(\eta_1, \eta_2) \notag\\
&& \qquad \times \frac{T_n(\eta_1)T_m(\eta_2) + T_n(\eta_2)T_m(\eta_1)}{2},
\end{eqnarray}
where $T_n(\eta) = \sqrt{n + \frac{1}{2}} P_n(\frac{\eta}{Y})$ and $P_n(x)$ are the standard Legendre polynomials.

In Fig.~\ref{fig6}, we present the first Legendre coefficient in the series, $a_{1,1}$, for initial energy and net baryon profiles. The coefficient $a_{1,1}$ is the least affected by short range correlations \cite{Aaboud:2016jnr} and is only little affected by the late stage hydrodynamical evolution \cite{Pang:2015zrq}.

We find that the $a_{1,1}$ coefficient of the system's initial energy density decreases as a function of collision energy. This is because the produced strings stretch over a longer range in rapidity for higher collision energies, which reduces longitudinal fluctuations in the mid-rapidity region. In contrast, the collision energy dependence of $a_{1,1}$ for the initial net baryon density is opposite. This is because there are fewer net baryons in the mid-rapidity region for higher collision energies, which increases the fluctuations.

The model that contains both initial valence quarks and rapidity loss fluctuations produces a larger $a_{1,1}$ coefficient for the energy density distribution. The collision-by-collision rapidity loss fluctuation introduces additional fluctuations to particle production along the longitudinal direction. For the net baryon density, the valence quarks + final rapidity fluctuation model transports more net baryons to the mid-rapidity region and hence reduces the net baryon density fluctuation. In this case, the $a_{1,1}$ coefficient is smaller compared to the model with valence quarks and fixed rapidity loss.

The longitudinal fluctuation can also be constrained by studying the event-plane rapidity decorrelation ratio $r_n$ defined as
\begin{equation}
r_n (\eta_a, \eta_b) = \frac{\langle \Re\{ \mathcal{E}_n(-\eta_a) \cdot \mathcal{E}^*_n(\eta_b) \} \rangle_\mathrm{ev} }{\langle \Re\{ \mathcal{E}_n(\eta_a) \cdot \mathcal{E}^*_n(\eta_b) \} \rangle_\mathrm{ev}}.
\end{equation}
Here the ratio $r_n$ captures the decorrelation of the initial eccentricity of the energy density profile between $\eta_a$ and $-\eta_a$. The reference (space-time) rapidity is chosen to be $\eta_b = 2$.  

\begin{figure}[ht!]
  \centering
  \includegraphics[width=0.9\linewidth]{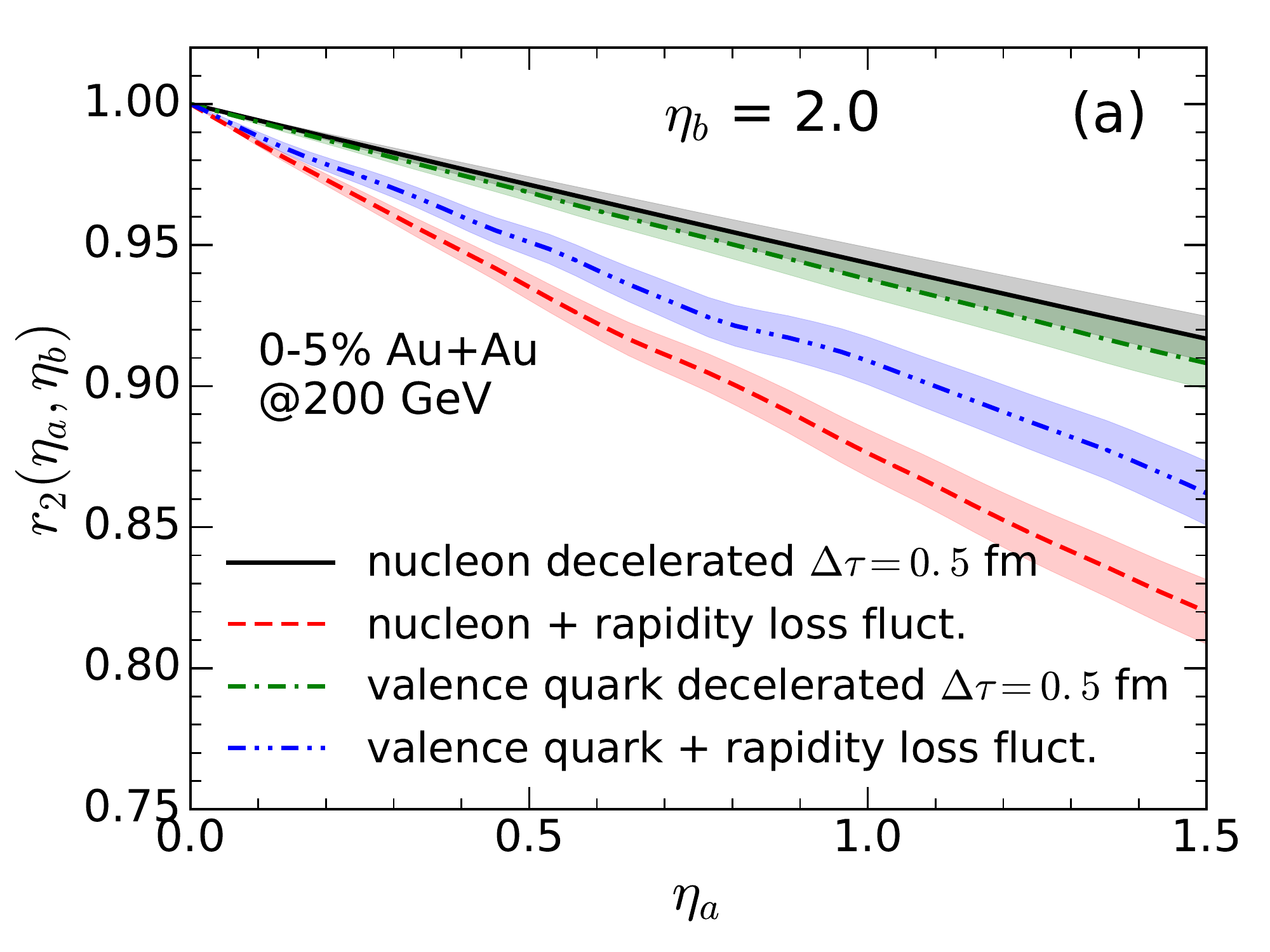} \\ 
  \includegraphics[width=0.9\linewidth]{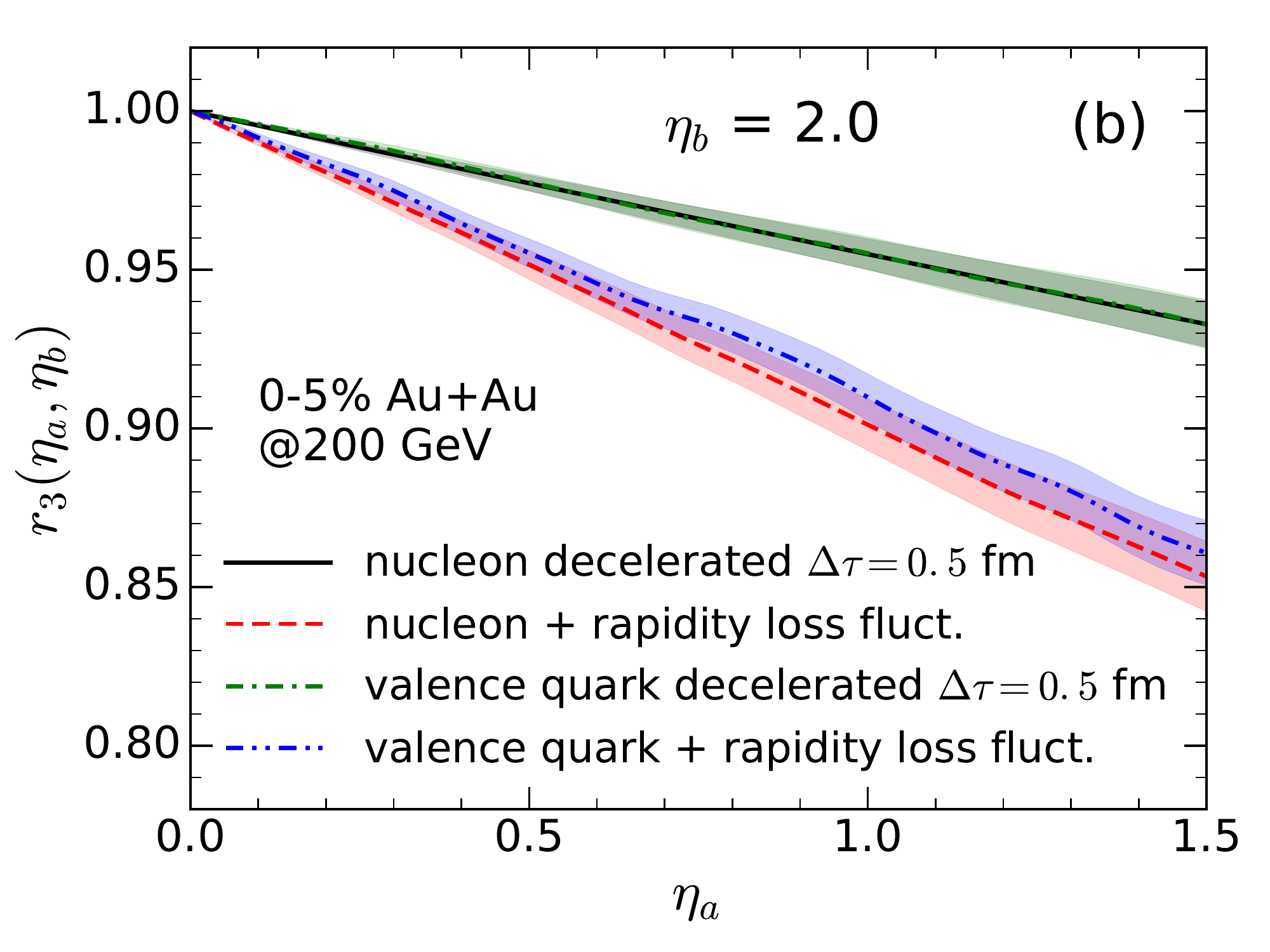}
  \caption{The participant plane decorrelation coefficients $r_{2,3} (\eta_a, \eta_b = 2.0)$ in 0-5\% Au+Au collisions at 200 GeV for four different models.}
  \label{fig7}
\end{figure}
%
In Fig.~\ref{fig7}, we compare the initial eccentricity $r_n$ ratio for the four initial state models. The nucleon + constant deceleration model gives the smallest event-plane decorrelation. The fluctuation introduced by sampling the valence quarks' rapidities results in a larger decorrelation. Allowing for a fluctuating rapidity loss introduces extra fluctuations that further reduce the event-plane correlation at large $\eta$ difference. We find that the fluctuations of the rapidity loss have a larger effect than the valence quark sampling in these $r_n$ observables.

%
%
\begin{figure}[ht!]
  \centering
  \includegraphics[width=0.9\linewidth]{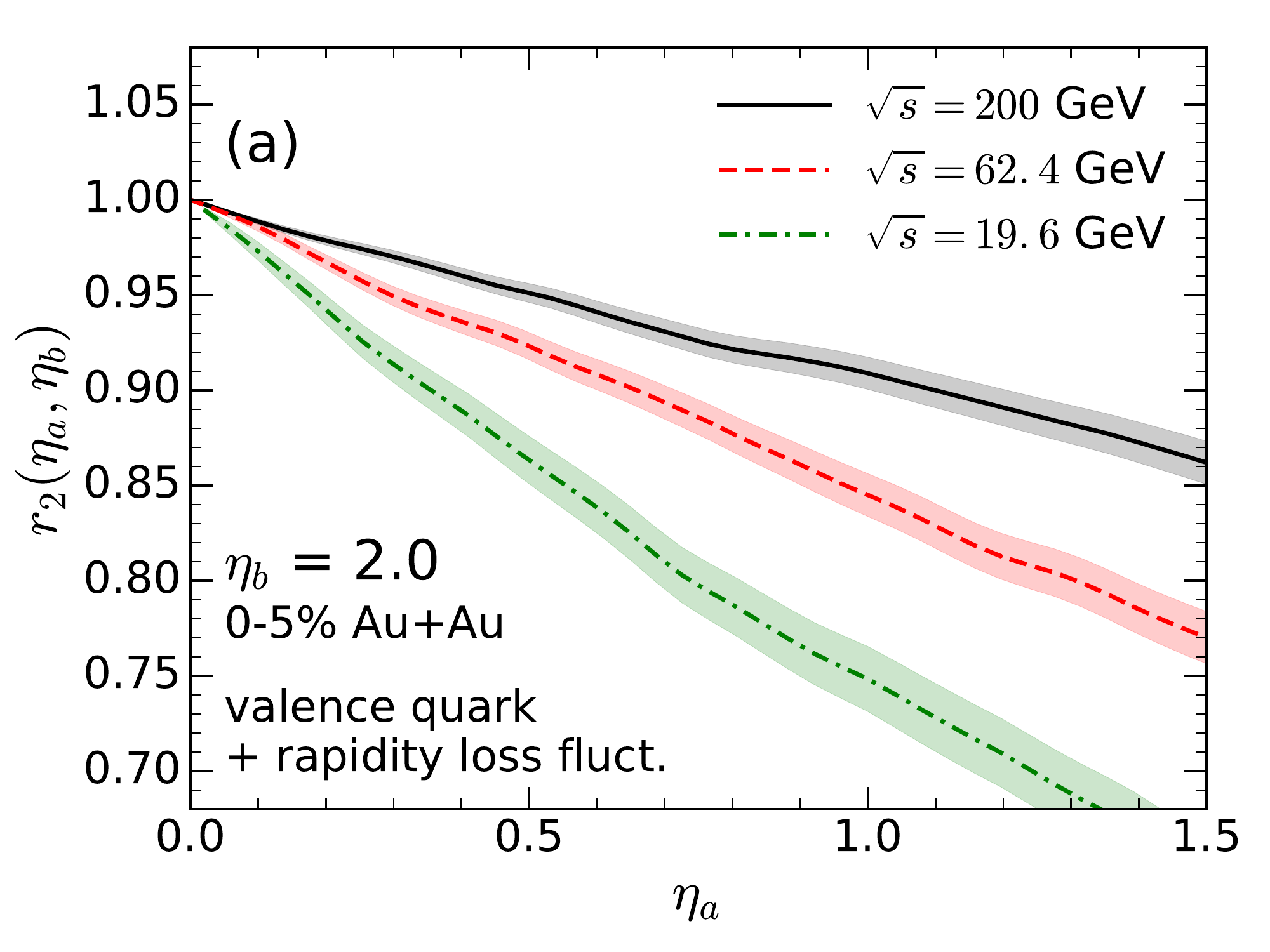} \\
  \includegraphics[width=0.9\linewidth]{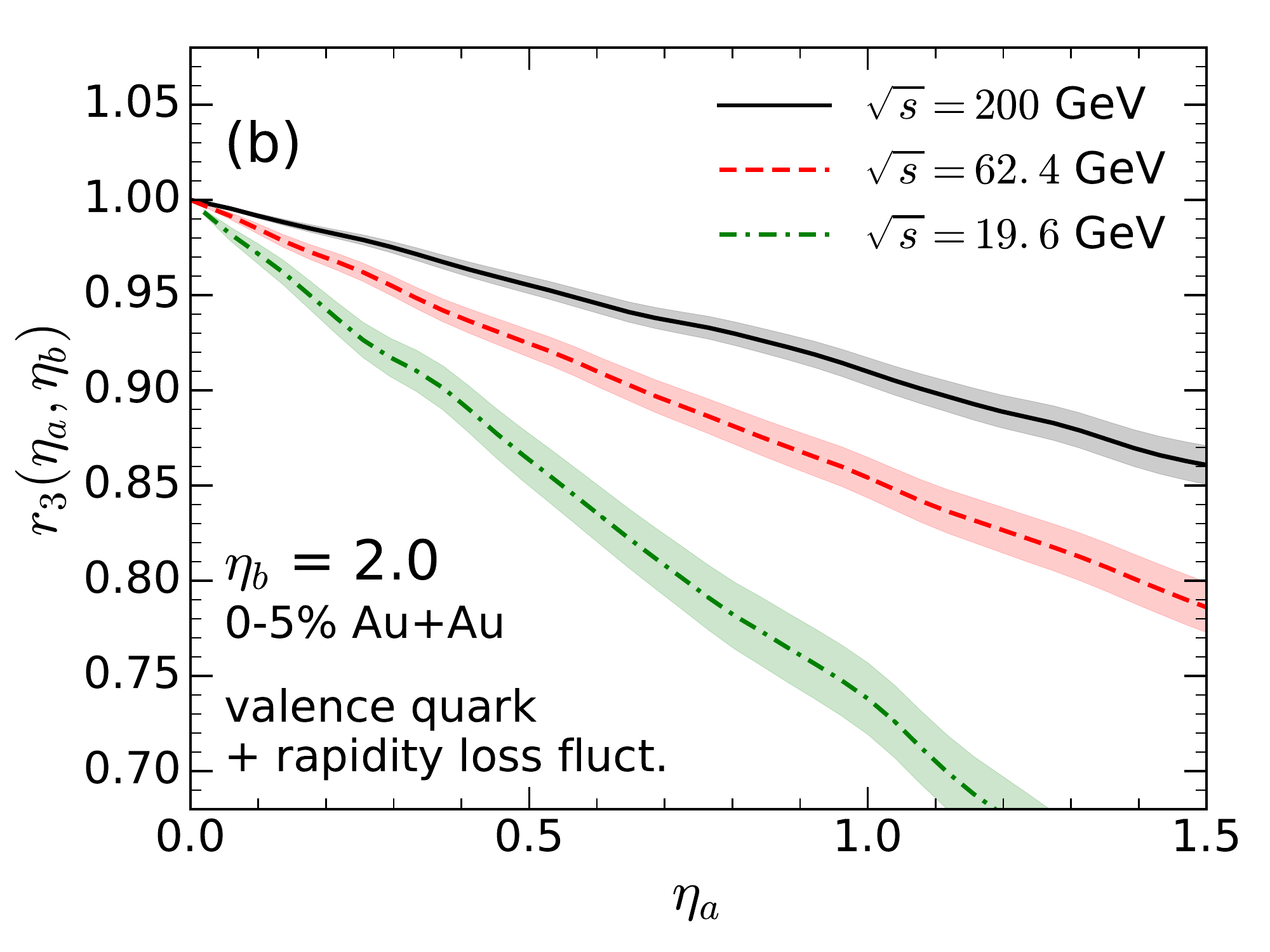}
  \caption{The participant plane decorrelation coefficients $r_{2,3} (\eta_a, \eta_b = 2.0)$ in 0-5\% Au+Au collisions at different collision energies.}
  \label{fig8}
\end{figure}
%
%
In Fig.\ref{fig8}, we study the collision energy dependence of the $r_n$ ratio at various RHIC BES energies. The decorrelation effect becomes stronger at lower collision energy, which is driven predominantly by the reduction of $Y_\mathrm{beam}$ with decreasing collision energy. To remove this effect we can study $r_n$ as a function of the scaled variable $\tilde{\eta}_a = \eta_a/Y_\mathrm{beam}$. We find that the decorrelation is still stronger at lower collision energy in our models, even after such a scaling.  

\subsection{Moments of the net proton distribution}

Longitudinal fluctuations in the initial state models can lead to non-trivial net baryon number fluctuations within a certain rapidity interval. In this section, we study the moments of the decelerated proton distribution in our models. Because only protons are measured in the experiments, we converted the decelerated baryons to protons by randomly assigning the proton or neutron identification to the nucleons in our model. In every $^{197}_{79}$Au+$^{197}_{79}$Au collision, we require the total proton number to be 158 and total neutron number to be 236. This identification procedure introduces additional fluctuations but does not overwhelm the initial state fluctuations. Here we study the moments of the decelerated proton distributions up to $\sqrt{s} = 27$ GeV. At these low collision energies, the decelerated proton yield dominates over the produced proton anti-proton pairs and is a good proxy for the total proton yield.

The central moments of the decelerated proton distribution can be computed as
\begin{eqnarray}
C_1 &=& \langle N \rangle = M \\
C_2 &=& \langle (\Delta N)^2 \rangle = \sigma^2 \\
C_3 &=& \langle (\Delta N)^3 \rangle = S \sigma^3 \\
C_4 &=& \langle (\Delta N)^4 \rangle - 3 C_2^2 = \kappa \sigma^4.
\end{eqnarray}
In order to reduce finite volume effects, one usually computes ratios of these moments
\begin{eqnarray}\label{eq:momentratios}
\frac{\sigma^2}{M} = \frac{C_2}{C_1}, S \sigma = \frac{C_3}{C_2}, \kappa \sigma^2 = \frac{C_4}{C_2}.
\end{eqnarray}
The Poisson limit of these ratios are 1. Here we study these fluctuation moments in 0-5\% central Au+Au collisions. Centrality is determined by the number of participants. 

%
\begin{figure}[ht!]
  \centering
   \includegraphics[width=1.0\linewidth]{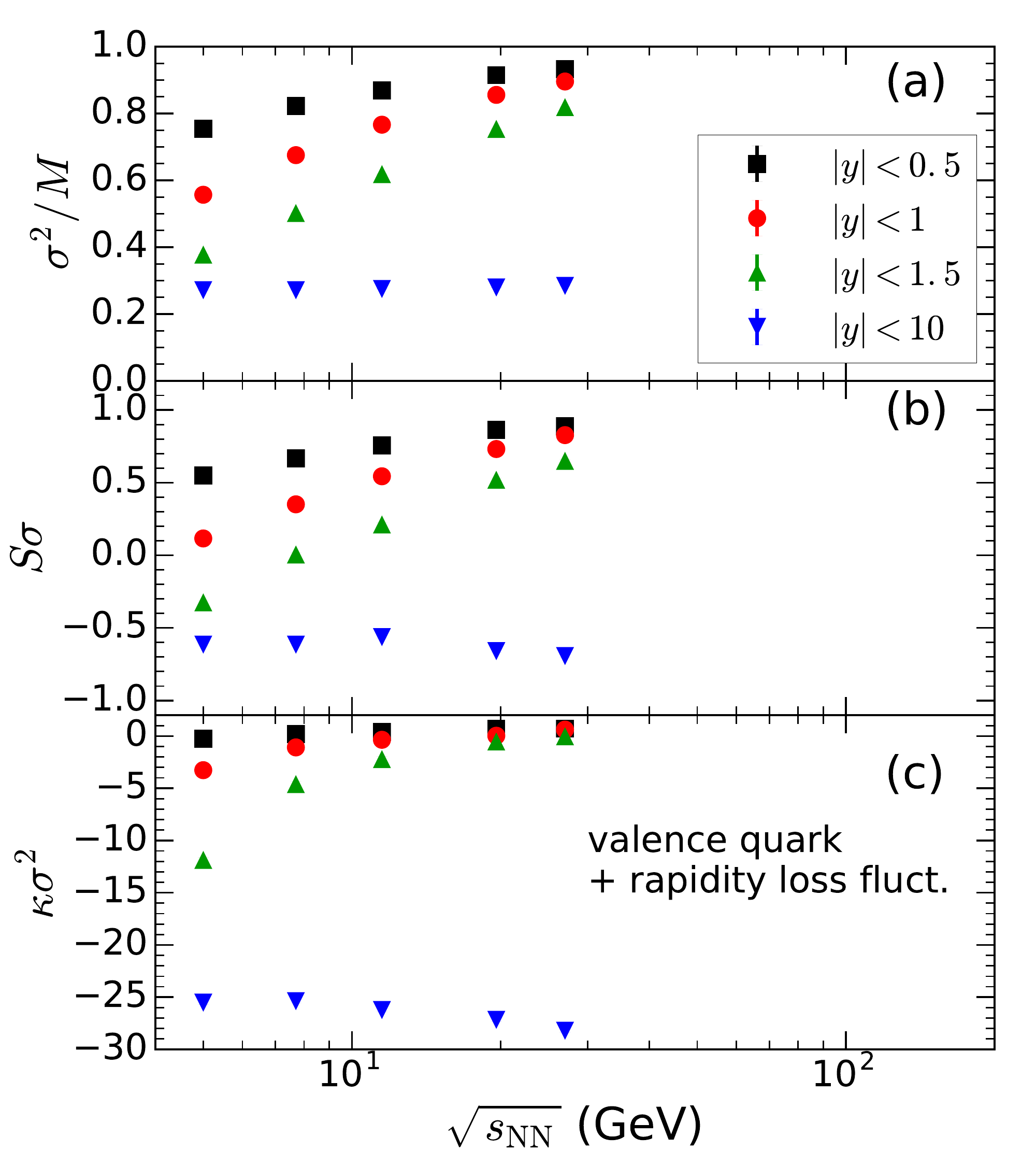}
  \caption{The ratios of proton fluctuation moments, $C_2/C_1 = \sigma^2/M$, $C_3/C_2 = S \sigma$, $C_4/C_2 = \kappa \sigma^2$, as functions of the collision energy from the valence quark + rapidity loss fluctuation model. Their dependence on the rapidity cut window are shown. }
  \label{fig:protonMoments}
\end{figure}
%
Fig.~\ref{fig:protonMoments} shows the ratios of decelerated proton moments defined in Eqs.~(\ref{eq:momentratios}) as a function of collision energy using the valence quark + rapidity loss fluctuation model. With a small rapidity window $\vert y \vert < 0.5$, these ratios increase with the collision energy and approach the Poisson limit at high collision energy. This is because fewer and fewer baryons are decelerated to the mid-rapidity region when the energy increases.
As the rapidity acceptance increases, more baryons are included in the analysis and the ratios of proton fluctuation moments decrease and deviate away from the Poisson limit. In the $y \rightarrow \infty$ limit, these ratios approach those of the fluctuation moments of the number of participants $N_\mathrm{part}$ in the simulation. Interestingly, we find negative skewness and kurtosis of the $N_\mathrm{part}$ distribution in our models.

%
\begin{figure}[ht!]
  \centering
   \includegraphics[width=1.0\linewidth]{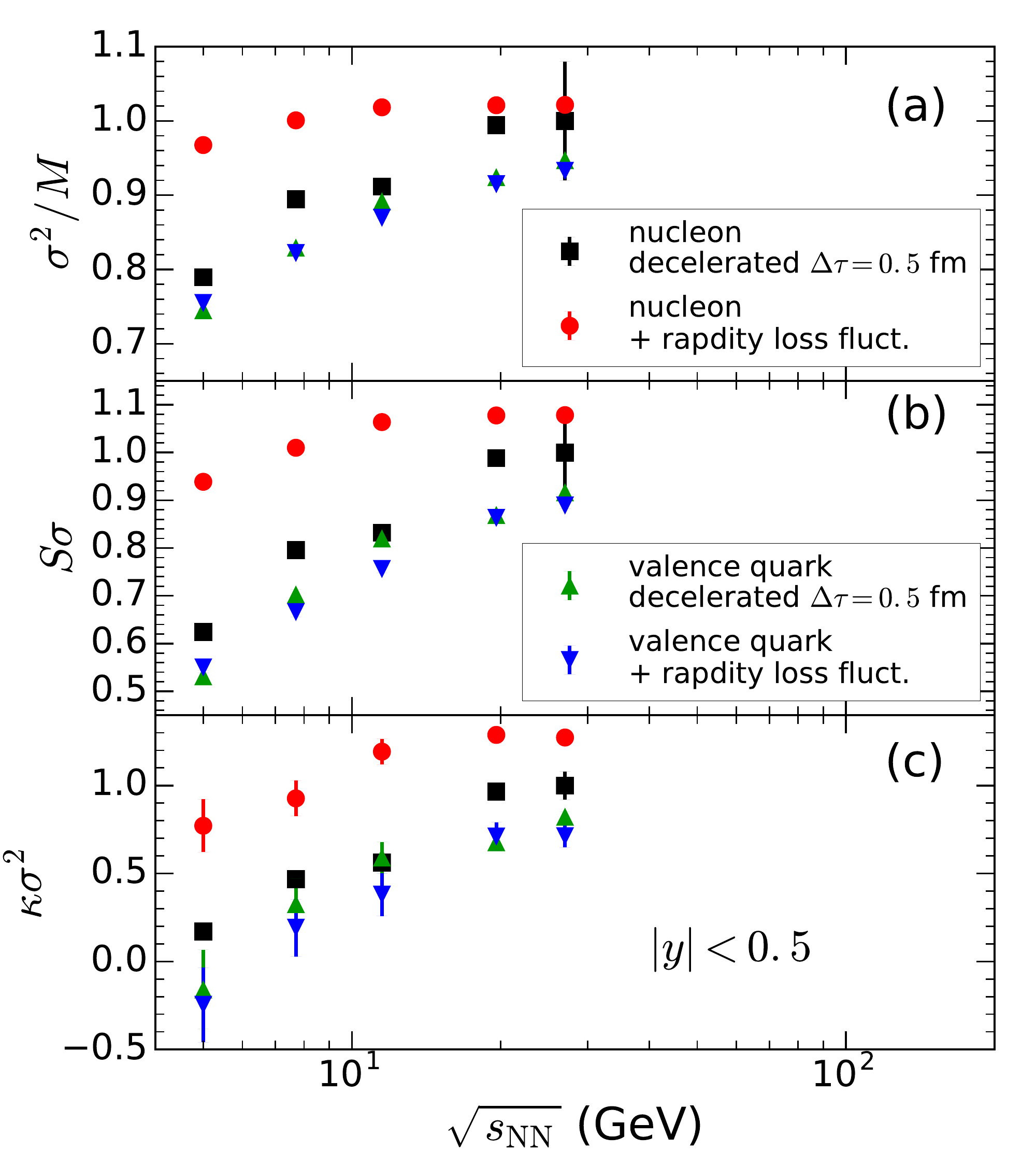}
  \caption{Same as Fig.~\ref{fig:protonMoments} but for comparison among different models at a same rapidity window, $\vert y \vert < 0.5$.}
  \label{fig:protonMoments_comp}
\end{figure}
%

Fig.~\ref{fig:protonMoments_comp} compares the moments of decelerated protons among the four initial state models within a same rapidity window. The nucleon + rapidity loss fluctuation model gives the largest ratios. The ratios are larger than the Poisson limit for $\sqrt{s} \ge 11.5$ GeV. We checked that their values approach the Poisson result in the high energy limit. Compared to the nucleon with constant time deceleration case, the additional fluctuations lead to an increase of all the central moments $C_n$ of the net proton distribution. The increase is larger for higher order of $n$.
In contrast, the models using valence quark participants produce less fluctuations near the mid-rapidity region. In this case all the moment ratios are smaller than the Poisson limit and rapidity loss fluctuations do not modify the moments of the stopped protons.

Independent of the details of the model, we find that initial state fluctuations of the proton number within a given rapidity window by themselves generate large cumulant ratios. If one were to fold our result with another Poisson distribution, which is similar to what would occur in the grand-canonical picture, where after hydrodynamic evolution Poisson distributions are sampled for a given freeze-out temperature and baryon chemical potential \cite{Li:2017via}, then the cumulant ratios would increase by 1. That would lead to an overestimation of the experimental data \cite{Adamczyk:2013dal} for most studied energies.

\section{Conclusions and Outlook}\label{sec:conc}
Heavy ion collisions with center of mass energies as realized in the RHIC beam energy scan or the NA61/SHINE program have a complex early time behavior, owing to the fact that the colliding nuclei have a finite extend in the beam direction. Furthermore, the assumption of boost invariance is not valid for most of the BES energies, and the net-baryon density can reach values much greater than at top RHIC energies, where it is usually ignored.

All these complications demand the development of a sophisticated initial state model to complement state-of-the-art hydrodynamic simulations.
In this work we have presented a three-dimensional dynamical initial state model that integrates with 3+1D hydrodynamic simulations. This model addresses all issues mentioned above: It provides fluctuating three dimensional distributions of both energy and net-baryon density, which dynamically enter the hydrodynamic simulation via source terms. 

The model is based on the formation of strings between (nucleon or quark) participants and the gradual deceleration of the string ends. After a certain time the strings are assumed to thermalize and become part of the hydrodynamic medium, leading to complex structures of energy (and baryon) deposition in space-time. 

Various sources of fluctuations are present in this initial state description. First we have the usual fluctuations of nucleon positions in the incoming nuclei. When using quark degrees of freedom, the quarks' positions in the nucleon also fluctuate. This leads to the usual fluctuations in the transverse plane of the collision, but fluctuations in the longitudinal coordinate of participating nucleons also generate longitudinal fluctuations. Additional fluctuations in the longitudinal structure occur in the quark based model because of fluctuations in the momentum fraction $x$ of the participating quarks. Finally, we allow for random fluctuations of the amount of rapidity lost by each participant similar to the LEXUS model. 

Apart from presenting average quantities such as the initial (and final) distributions of net baryons (protons), we concentrate on measures of fluctuations and how various realizations of our initial state model affect them. We presented Legendre coefficients of net baryon and energy density rapidity fluctuations, measures of decorrelations of the transverse geometry with rapidity, and cumulant ratios for net proton distributions obtained directly from the initial state.

We find significant sensitivity to the details of the initial state model for all of these fluctuation measures. We conclude that a certain subset of experimental observables will have to be used to constrain model parameters, so that the effect of initial state fluctuations can be estimated reliably in the experimental program aimed at revealing critical fluctuations. 

In particular we find significant contributions to the cumulants of net-proton distributions solely from the initial state - these will have to be folded with additional fluctuations potentially occurring during the hydrodynamic evolution and at freeze-out. Full scale event-by-event viscous hydrodynamic simulations using the presented initial states will then allow for direct comparison with experimental data and hopefully provide insight into whether critical behavior is present on top of the background provided by our model. 

\appendix*
\section{Algorithm for string generation}

We assume that strings are produced between all colliding nucleons. To reduce multiple strings overlapping with each other, we want to reduce the number of strings that connect to the same colliding nucleon. This can be achieved by adopting the following algorithm. We introduce a cost function for the production probability of a string from a given binary collision,
\begin{equation}
f(N_\mathrm{conn}) = \exp(- F_\mathrm{asy} N_\mathrm{conn}).
\end{equation}
Here $N_\mathrm{conn}$ is the total number of connections already attached to the colliding nucleons. The probability of generating one new string is exponentially suppressed with $N_\mathrm{conn}$. To increase the numerical efficiency in asymmetric collisions, we introduce an asymmetry factor $F_\mathrm{asy}$ defined as
\begin{equation}
F_\mathrm{asy} = \frac{1}{\mathrm{max}\{N_A, N_B\}} \frac{N_A N_B}{N_A + N_B},
\end{equation}
where $N_A$ and $N_B$ are the atomic numbers of the two colliding nuclei. For symmetric collision systems $F_\mathrm{asy} = 1/2$, but for asymmetric systems such as p+A collisions, $F_\mathrm{asy} = 1/(1 + N_A) \ll 1/2$. This asymmetry factor $F_\mathrm{asy}$ allows a high acceptance probability for a p+A collisions when $N_\mathrm{conn}$ is large. We apply this algorithm when looping over the binary collision list until all the colliding nucleons are connected with at least one string. Once all string production points are determined, they are sorted according their production time.

When valence quarks are used as constitutes for the string ends, a similar cost function $f (N^q_\mathrm{conn}) = \exp(- N^q_\mathrm{conn})$ is applied to sample the colliding valence quark pair for the binary collision.

\section*{Acknowledgments}
The authors thank Sangyong Jeon for useful discussions. BPS and CS are supported under DOE Contract No. DE-SC0012704. This research used resources of the National Energy Research Scientific Computing Center, which is supported by the Office of Science of the U.S. Department of Energy under Contract No. DE-AC02-05CH11231. BPS acknowledges a DOE Office of Science Early Career Award. CS gratefully acknowledges a Goldhaber Distinguished Fellowship from Brookhaven Science Associates.
This work is supported in part by the U.S. Department of Energy, Office of Science, Office of Nuclear Physics, within the framework of the Beam Energy Scan Theory (BEST) Topical Collaboration.

\bibliography{ref}

\begin{thebibliography}{48}%
\makeatletter
\providecommand \@ifxundefined [1]{%
 \@ifx{#1\undefined}
}%
\providecommand \@ifnum [1]{%
 \ifnum #1\expandafter \@firstoftwo
 \else \expandafter \@secondoftwo
 \fi
}%
\providecommand \@ifx [1]{%
 \ifx #1\expandafter \@firstoftwo
 \else \expandafter \@secondoftwo
 \fi
}%
\providecommand \natexlab [1]{#1}%
\providecommand \enquote  [1]{``#1''}%
\providecommand \bibnamefont  [1]{#1}%
\providecommand \bibfnamefont [1]{#1}%
\providecommand \citenamefont [1]{#1}%
\providecommand \href@noop [0]{\@secondoftwo}%
\providecommand \href [0]{\begingroup \@sanitize@url \@href}%
\providecommand \@href[1]{\@@startlink{#1}\@@href}%
\providecommand \@@href[1]{\endgroup#1\@@endlink}%
\providecommand \@sanitize@url [0]{\catcode `\\12\catcode `\$12\catcode
  `\&12\catcode `\#12\catcode `\^12\catcode `\_12\catcode `\%12\relax}%
\providecommand \@@startlink[1]{}%
\providecommand \@@endlink[0]{}%
\providecommand \url  [0]{\begingroup\@sanitize@url \@url }%
\providecommand \@url [1]{\endgroup\@href {#1}{\urlprefix }}%
\providecommand \urlprefix  [0]{URL }%
\providecommand \Eprint [0]{\href }%
\providecommand \doibase [0]{http://dx.doi.org/}%
\providecommand \selectlanguage [0]{\@gobble}%
\providecommand \bibinfo  [0]{\@secondoftwo}%
\providecommand \bibfield  [0]{\@secondoftwo}%
\providecommand \translation [1]{[#1]}%
\providecommand \BibitemOpen [0]{}%
\providecommand \bibitemStop [0]{}%
\providecommand \bibitemNoStop [0]{.\EOS\space}%
\providecommand \EOS [0]{\spacefactor3000\relax}%
\providecommand \BibitemShut  [1]{\csname bibitem#1\endcsname}%
\let\auto@bib@innerbib\@empty
\bibitem [{\citenamefont {Adamczyk}\ \emph
  {et~al.}(2014{\natexlab{a}})\citenamefont {Adamczyk} \emph
  {et~al.}}]{Adamczyk:2013dal}%
  \BibitemOpen
  \bibfield  {author} {\bibinfo {author} {\bibfnamefont {L.}~\bibnamefont
  {Adamczyk}} \emph {et~al.} (\bibinfo {collaboration} {STAR}),\ }\bibfield
  {title} {\enquote {\bibinfo {title} {{Energy Dependence of Moments of
  Net-proton Multiplicity Distributions at RHIC}},}\ }\href {\doibase
  10.1103/PhysRevLett.112.032302} {\bibfield  {journal} {\bibinfo  {journal}
  {Phys. Rev. Lett.}\ }\textbf {\bibinfo {volume} {112}},\ \bibinfo {pages}
  {032302} (\bibinfo {year} {2014}{\natexlab{a}})},\ \Eprint
  {http://arxiv.org/abs/1309.5681} {arXiv:1309.5681 [nucl-ex]} \BibitemShut
  {NoStop}%
\bibitem [{\citenamefont {Adamczyk}\ \emph
  {et~al.}(2014{\natexlab{b}})\citenamefont {Adamczyk} \emph
  {et~al.}}]{Adamczyk:2014fia}%
  \BibitemOpen
  \bibfield  {author} {\bibinfo {author} {\bibfnamefont {L.}~\bibnamefont
  {Adamczyk}} \emph {et~al.} (\bibinfo {collaboration} {STAR}),\ }\bibfield
  {title} {\enquote {\bibinfo {title} {{Beam energy dependence of moments of
  the net-charge multiplicity distributions in Au+Au collisions at RHIC}},}\
  }\href {\doibase 10.1103/PhysRevLett.113.092301} {\bibfield  {journal}
  {\bibinfo  {journal} {Phys. Rev. Lett.}\ }\textbf {\bibinfo {volume} {113}},\
  \bibinfo {pages} {092301} (\bibinfo {year} {2014}{\natexlab{b}})},\ \Eprint
  {http://arxiv.org/abs/1402.1558} {arXiv:1402.1558 [nucl-ex]} \BibitemShut
  {NoStop}%
\bibitem [{\citenamefont {Adare}\ \emph {et~al.}(2016)\citenamefont {Adare}
  \emph {et~al.}}]{Adare:2015aqk}%
  \BibitemOpen
  \bibfield  {author} {\bibinfo {author} {\bibfnamefont {A.}~\bibnamefont
  {Adare}} \emph {et~al.} (\bibinfo {collaboration} {PHENIX}),\ }\bibfield
  {title} {\enquote {\bibinfo {title} {{Measurement of higher cumulants of
  net-charge multiplicity distributions in Au$+$Au collisions at
  $\sqrt{s_{_{NN}}}=7.7-200$ GeV}},}\ }\href {\doibase
  10.1103/PhysRevC.93.011901} {\bibfield  {journal} {\bibinfo  {journal} {Phys.
  Rev.}\ }\textbf {\bibinfo {volume} {C93}},\ \bibinfo {pages} {011901}
  (\bibinfo {year} {2016})},\ \Eprint {http://arxiv.org/abs/1506.07834}
  {arXiv:1506.07834 [nucl-ex]} \BibitemShut {NoStop}%
\bibitem [{\citenamefont {Adamczyk}\ \emph {et~al.}(2017)\citenamefont
  {Adamczyk} \emph {et~al.}}]{Adamczyk:2017iwn}%
  \BibitemOpen
  \bibfield  {author} {\bibinfo {author} {\bibfnamefont {L.}~\bibnamefont
  {Adamczyk}} \emph {et~al.} (\bibinfo {collaboration} {STAR}),\ }\bibfield
  {title} {\enquote {\bibinfo {title} {{Bulk Properties of the Medium Produced
  in Relativistic Heavy-Ion Collisions from the Beam Energy Scan Program, }},}\
  }\href@noop {} {\  (\bibinfo {year} {2017})},\ \Eprint
  {http://arxiv.org/abs/1701.07065} {arXiv:1701.07065 [nucl-ex]} \BibitemShut
  {NoStop}%
\bibitem [{\citenamefont
  {Mackowiak-Pawlowska}(2017)}]{Mackowiak-Pawlowska:2017rcx}%
  \BibitemOpen
  \bibfield  {author} {\bibinfo {author} {\bibfnamefont {Maja}\ \bibnamefont
  {Mackowiak-Pawlowska}} (\bibinfo {collaboration} {NA61/SHINE}),\ }\bibfield
  {title} {\enquote {\bibinfo {title} {{Recent results from NA61/SHINE}},}\
  }in\ \href {http://inspirehep.net/record/1610347/files/arXiv:1707.04735.pdf}
  {\emph {\bibinfo {booktitle} {{9th Workshop "Excited QCD" 2017 Sintra,
  Portugal, May 7-13, 2017}}}}\ (\bibinfo {year} {2017})\ \Eprint
  {http://arxiv.org/abs/1707.04735} {arXiv:1707.04735 [nucl-ex]} \BibitemShut
  {NoStop}%
\bibitem [{\citenamefont {Stephanov}\ \emph {et~al.}(1998)\citenamefont
  {Stephanov}, \citenamefont {Rajagopal},\ and\ \citenamefont
  {Shuryak}}]{Stephanov:1998dy}%
  \BibitemOpen
  \bibfield  {author} {\bibinfo {author} {\bibfnamefont {Misha~A.}\
  \bibnamefont {Stephanov}}, \bibinfo {author} {\bibfnamefont {K.}~\bibnamefont
  {Rajagopal}}, \ and\ \bibinfo {author} {\bibfnamefont {Edward~V.}\
  \bibnamefont {Shuryak}},\ }\bibfield  {title} {\enquote {\bibinfo {title}
  {{Signatures of the tricritical point in QCD}},}\ }\href {\doibase
  10.1103/PhysRevLett.81.4816} {\bibfield  {journal} {\bibinfo  {journal}
  {Phys. Rev. Lett.}\ }\textbf {\bibinfo {volume} {81}},\ \bibinfo {pages}
  {4816--4819} (\bibinfo {year} {1998})},\ \Eprint
  {http://arxiv.org/abs/hep-ph/9806219} {arXiv:hep-ph/9806219 [hep-ph]}
  \BibitemShut {NoStop}%
\bibitem [{\citenamefont {Stephanov}\ \emph {et~al.}(1999)\citenamefont
  {Stephanov}, \citenamefont {Rajagopal},\ and\ \citenamefont
  {Shuryak}}]{Stephanov:1999zu}%
  \BibitemOpen
  \bibfield  {author} {\bibinfo {author} {\bibfnamefont {Misha~A.}\
  \bibnamefont {Stephanov}}, \bibinfo {author} {\bibfnamefont {K.}~\bibnamefont
  {Rajagopal}}, \ and\ \bibinfo {author} {\bibfnamefont {Edward~V.}\
  \bibnamefont {Shuryak}},\ }\bibfield  {title} {\enquote {\bibinfo {title}
  {{Event-by-event fluctuations in heavy ion collisions and the QCD critical
  point}},}\ }\href {\doibase 10.1103/PhysRevD.60.114028} {\bibfield  {journal}
  {\bibinfo  {journal} {Phys. Rev.}\ }\textbf {\bibinfo {volume} {D60}},\
  \bibinfo {pages} {114028} (\bibinfo {year} {1999})},\ \Eprint
  {http://arxiv.org/abs/hep-ph/9903292} {arXiv:hep-ph/9903292 [hep-ph]}
  \BibitemShut {NoStop}%
\bibitem [{\citenamefont {Stephanov}(2004)}]{Stephanov:2004wx}%
  \BibitemOpen
  \bibfield  {author} {\bibinfo {author} {\bibfnamefont {Mikhail~A.}\
  \bibnamefont {Stephanov}},\ }\bibfield  {title} {\enquote {\bibinfo {title}
  {{QCD phase diagram and the critical point}},}\ }\bibfield  {booktitle}
  {\emph {\bibinfo {booktitle} {{Non-perturbative quantum chromodynamics.
  Proceedings, 8th Workshop, Paris, France, June 7-11, 2004}}},\ }\href
  {\doibase 10.1142/S0217751X05027965} {\bibfield  {journal} {\bibinfo
  {journal} {Prog. Theor. Phys. Suppl.}\ }\textbf {\bibinfo {volume} {153}},\
  \bibinfo {pages} {139--156} (\bibinfo {year} {2004})},\ \bibinfo {note}
  {[Int. J. Mod. Phys.A20,4387(2005)]},\ \Eprint
  {http://arxiv.org/abs/hep-ph/0402115} {arXiv:hep-ph/0402115 [hep-ph]}
  \BibitemShut {NoStop}%
\bibitem [{\citenamefont {Heinz}\ and\ \citenamefont
  {Snellings}(2013)}]{Heinz:2013th}%
  \BibitemOpen
  \bibfield  {author} {\bibinfo {author} {\bibfnamefont {Ulrich}\ \bibnamefont
  {Heinz}}\ and\ \bibinfo {author} {\bibfnamefont {Raimond}\ \bibnamefont
  {Snellings}},\ }\bibfield  {title} {\enquote {\bibinfo {title} {{Collective
  flow and viscosity in relativistic heavy-ion collisions}},}\ }\href@noop {}
  {\bibfield  {journal} {\bibinfo  {journal} {Ann. Rev. Nucl. Part. Sci.}\
  }\textbf {\bibinfo {volume} {63}},\ \bibinfo {pages} {123--151} (\bibinfo
  {year} {2013})}\BibitemShut {NoStop}%
\bibitem [{\citenamefont {Gale}\ \emph {et~al.}(2013)\citenamefont {Gale},
  \citenamefont {Jeon},\ and\ \citenamefont {Schenke}}]{Gale:2013da}%
  \BibitemOpen
  \bibfield  {author} {\bibinfo {author} {\bibfnamefont {Charles}\ \bibnamefont
  {Gale}}, \bibinfo {author} {\bibfnamefont {Sangyong}\ \bibnamefont {Jeon}}, \
  and\ \bibinfo {author} {\bibfnamefont {Bjoern}\ \bibnamefont {Schenke}},\
  }\bibfield  {title} {\enquote {\bibinfo {title} {{Hydrodynamic Modeling of
  Heavy-Ion Collisions}},}\ }\href@noop {} {\bibfield  {journal} {\bibinfo
  {journal} {Int. J. Mod. Phys.}\ }\textbf {\bibinfo {volume} {A28}},\ \bibinfo
  {pages} {1340011} (\bibinfo {year} {2013})}\BibitemShut {NoStop}%
\bibitem [{\citenamefont {Petersen}(2014)}]{Petersen:2014yqa}%
  \BibitemOpen
  \bibfield  {author} {\bibinfo {author} {\bibfnamefont {Hannah}\ \bibnamefont
  {Petersen}},\ }\bibfield  {title} {\enquote {\bibinfo {title} {{Anisotropic
  flow in transport + hydrodynamics hybrid approaches}},}\ }\href {\doibase
  10.1088/0954-3899/41/12/124005} {\bibfield  {journal} {\bibinfo  {journal}
  {J. Phys.}\ }\textbf {\bibinfo {volume} {G41}},\ \bibinfo {pages} {124005}
  (\bibinfo {year} {2014})},\ \Eprint {http://arxiv.org/abs/1404.1763}
  {arXiv:1404.1763 [nucl-th]} \BibitemShut {NoStop}%
\bibitem [{\citenamefont {Karpenko}\ \emph {et~al.}(2015)\citenamefont
  {Karpenko}, \citenamefont {Huovinen}, \citenamefont {Petersen},\ and\
  \citenamefont {Bleicher}}]{Karpenko:2015xea}%
  \BibitemOpen
  \bibfield  {author} {\bibinfo {author} {\bibfnamefont {Iu.~A.}\ \bibnamefont
  {Karpenko}}, \bibinfo {author} {\bibfnamefont {P.}~\bibnamefont {Huovinen}},
  \bibinfo {author} {\bibfnamefont {H.}~\bibnamefont {Petersen}}, \ and\
  \bibinfo {author} {\bibfnamefont {M.}~\bibnamefont {Bleicher}},\ }\bibfield
  {title} {\enquote {\bibinfo {title} {{Estimation of the shear viscosity at
  finite net-baryon density from $A+A$ collision data at $\sqrt{s_\mathrm{NN}}
  = 7.7-200$ GeV}},}\ }\href {\doibase 10.1103/PhysRevC.91.064901} {\bibfield
  {journal} {\bibinfo  {journal} {Phys. Rev.}\ }\textbf {\bibinfo {volume}
  {C91}},\ \bibinfo {pages} {064901} (\bibinfo {year} {2015})},\ \Eprint
  {http://arxiv.org/abs/1502.01978} {arXiv:1502.01978 [nucl-th]} \BibitemShut
  {NoStop}%
\bibitem [{\citenamefont {Bass}\ \emph {et~al.}(1998)\citenamefont {Bass} \emph
  {et~al.}}]{Bass:1998ca}%
  \BibitemOpen
  \bibfield  {author} {\bibinfo {author} {\bibfnamefont {S.~A.}\ \bibnamefont
  {Bass}} \emph {et~al.},\ }\bibfield  {title} {\enquote {\bibinfo {title}
  {{Microscopic models for ultrarelativistic heavy ion collisions}},}\ }\href
  {\doibase 10.1016/S0146-6410(98)00058-1} {\bibfield  {journal} {\bibinfo
  {journal} {Prog. Part. Nucl. Phys.}\ }\textbf {\bibinfo {volume} {41}},\
  \bibinfo {pages} {255--369} (\bibinfo {year} {1998})},\ \bibinfo {note}
  {[Prog. Part. Nucl. Phys.41,225(1998)]},\ \Eprint
  {http://arxiv.org/abs/nucl-th/9803035} {arXiv:nucl-th/9803035 [nucl-th]}
  \BibitemShut {NoStop}%
\bibitem [{\citenamefont {Bleicher}\ \emph {et~al.}(1999)\citenamefont
  {Bleicher} \emph {et~al.}}]{Bleicher:1999xi}%
  \BibitemOpen
  \bibfield  {author} {\bibinfo {author} {\bibfnamefont {M.}~\bibnamefont
  {Bleicher}} \emph {et~al.},\ }\bibfield  {title} {\enquote {\bibinfo {title}
  {{Relativistic hadron hadron collisions in the ultrarelativistic quantum
  molecular dynamics model}},}\ }\href {\doibase 10.1088/0954-3899/25/9/308}
  {\bibfield  {journal} {\bibinfo  {journal} {J. Phys.}\ }\textbf {\bibinfo
  {volume} {G25}},\ \bibinfo {pages} {1859--1896} (\bibinfo {year} {1999})},\
  \Eprint {http://arxiv.org/abs/hep-ph/9909407} {arXiv:hep-ph/9909407 [hep-ph]}
  \BibitemShut {NoStop}%
\bibitem [{\citenamefont {Pang}\ \emph {et~al.}(2012)\citenamefont {Pang},
  \citenamefont {Wang},\ and\ \citenamefont {Wang}}]{Pang:2012he}%
  \BibitemOpen
  \bibfield  {author} {\bibinfo {author} {\bibfnamefont {Longgang}\
  \bibnamefont {Pang}}, \bibinfo {author} {\bibfnamefont {Qun}\ \bibnamefont
  {Wang}}, \ and\ \bibinfo {author} {\bibfnamefont {Xin-Nian}\ \bibnamefont
  {Wang}},\ }\bibfield  {title} {\enquote {\bibinfo {title} {{Effects of
  initial flow velocity fluctuation in event-by-event (3+1)D hydrodynamics}},}\
  }\href {\doibase 10.1103/PhysRevC.86.024911} {\bibfield  {journal} {\bibinfo
  {journal} {Phys. Rev.}\ }\textbf {\bibinfo {volume} {C86}},\ \bibinfo {pages}
  {024911} (\bibinfo {year} {2012})},\ \Eprint {http://arxiv.org/abs/1205.5019}
  {arXiv:1205.5019 [nucl-th]} \BibitemShut {NoStop}%
\bibitem [{\citenamefont {Pang}\ \emph {et~al.}(2016)\citenamefont {Pang},
  \citenamefont {Petersen}, \citenamefont {Qin}, \citenamefont {Roy},\ and\
  \citenamefont {Wang}}]{Pang:2015zrq}%
  \BibitemOpen
  \bibfield  {author} {\bibinfo {author} {\bibfnamefont {Long-Gang}\
  \bibnamefont {Pang}}, \bibinfo {author} {\bibfnamefont {Hannah}\ \bibnamefont
  {Petersen}}, \bibinfo {author} {\bibfnamefont {Guang-You}\ \bibnamefont
  {Qin}}, \bibinfo {author} {\bibfnamefont {Victor}\ \bibnamefont {Roy}}, \
  and\ \bibinfo {author} {\bibfnamefont {Xin-Nian}\ \bibnamefont {Wang}},\
  }\bibfield  {title} {\enquote {\bibinfo {title} {{Decorrelation of
  anisotropic flow along the longitudinal direction}},}\ }\href {\doibase
  10.1140/epja/i2016-16097-x} {\bibfield  {journal} {\bibinfo  {journal} {Eur.
  Phys. J.}\ }\textbf {\bibinfo {volume} {A52}},\ \bibinfo {pages} {97}
  (\bibinfo {year} {2016})},\ \Eprint {http://arxiv.org/abs/1511.04131}
  {arXiv:1511.04131 [nucl-th]} \BibitemShut {NoStop}%
\bibitem [{\citenamefont {Okai}\ \emph {et~al.}(2017)\citenamefont {Okai},
  \citenamefont {Kawaguchi}, \citenamefont {Tachibana},\ and\ \citenamefont
  {Hirano}}]{Okai:2017ofp}%
  \BibitemOpen
  \bibfield  {author} {\bibinfo {author} {\bibfnamefont {Michito}\ \bibnamefont
  {Okai}}, \bibinfo {author} {\bibfnamefont {Koji}\ \bibnamefont {Kawaguchi}},
  \bibinfo {author} {\bibfnamefont {Yasuki}\ \bibnamefont {Tachibana}}, \ and\
  \bibinfo {author} {\bibfnamefont {Tetsufumi}\ \bibnamefont {Hirano}},\
  }\bibfield  {title} {\enquote {\bibinfo {title} {{New approach to
  initializing hydrodynamic fields and mini-jet propagation in quark-gluon
  fluids}},}\ }\href {\doibase 10.1103/PhysRevC.95.054914} {\bibfield
  {journal} {\bibinfo  {journal} {Phys. Rev.}\ }\textbf {\bibinfo {volume}
  {C95}},\ \bibinfo {pages} {054914} (\bibinfo {year} {2017})},\ \Eprint
  {http://arxiv.org/abs/1702.07541} {arXiv:1702.07541 [nucl-th]} \BibitemShut
  {NoStop}%
\bibitem [{\citenamefont {Jeon}\ and\ \citenamefont
  {Kapusta}(1997)}]{Jeon:1997bp}%
  \BibitemOpen
  \bibfield  {author} {\bibinfo {author} {\bibfnamefont {Sangyong}\
  \bibnamefont {Jeon}}\ and\ \bibinfo {author} {\bibfnamefont {Joseph~I.}\
  \bibnamefont {Kapusta}},\ }\bibfield  {title} {\enquote {\bibinfo {title}
  {{Linear extrapolation of ultrarelativistic nucleon-nucleon scattering to
  nucleus-nucleus collisions}},}\ }\href {\doibase 10.1103/PhysRevC.56.468}
  {\bibfield  {journal} {\bibinfo  {journal} {Phys. Rev.}\ }\textbf {\bibinfo
  {volume} {C56}},\ \bibinfo {pages} {468--480} (\bibinfo {year} {1997})},\
  \Eprint {http://arxiv.org/abs/nucl-th/9703033} {arXiv:nucl-th/9703033
  [nucl-th]} \BibitemShut {NoStop}%
\bibitem [{\citenamefont {Monnai}\ and\ \citenamefont
  {Schenke}(2016)}]{Monnai:2015sca}%
  \BibitemOpen
  \bibfield  {author} {\bibinfo {author} {\bibfnamefont {Akihiko}\ \bibnamefont
  {Monnai}}\ and\ \bibinfo {author} {\bibfnamefont {Bjoern}\ \bibnamefont
  {Schenke}},\ }\bibfield  {title} {\enquote {\bibinfo {title} {{Pseudorapidity
  correlations in heavy ion collisions from viscous fluid dynamics}},}\
  }\href@noop {} {\bibfield  {journal} {\bibinfo  {journal} {Phys. Lett.}\
  }\textbf {\bibinfo {volume} {B752}},\ \bibinfo {pages} {317--321} (\bibinfo
  {year} {2016})}\BibitemShut {NoStop}%
\bibitem [{\citenamefont {Pang}\ \emph {et~al.}(2013)\citenamefont {Pang},
  \citenamefont {Wang},\ and\ \citenamefont {Wang}}]{Pang:2012uw}%
  \BibitemOpen
  \bibfield  {author} {\bibinfo {author} {\bibfnamefont {Longgang}\
  \bibnamefont {Pang}}, \bibinfo {author} {\bibfnamefont {Qun}\ \bibnamefont
  {Wang}}, \ and\ \bibinfo {author} {\bibfnamefont {Xin-Nian}\ \bibnamefont
  {Wang}},\ }\bibfield  {title} {\enquote {\bibinfo {title} {{Effect of
  longitudinal fluctuation in event-by-event (3+1)D hydrodynamics}},}\
  }\bibfield  {booktitle} {\emph {\bibinfo {booktitle} {{Proceedings, 23rd
  International Conference on Ultrarelativistic Nucleus-Nucleus Collisions :
  Quark Matter 2012 (QM 2012): Washington, DC, USA, August 13-18, 2012}}},\
  }\href {\doibase 10.1016/j.nuclphysa.2013.02.140} {\bibfield  {journal}
  {\bibinfo  {journal} {Nucl. Phys.}\ }\textbf {\bibinfo {volume} {A904-905}},\
  \bibinfo {pages} {811c--814c} (\bibinfo {year} {2013})},\ \Eprint
  {http://arxiv.org/abs/1211.1570} {arXiv:1211.1570 [nucl-th]} \BibitemShut
  {NoStop}%
\bibitem [{\citenamefont {Bozek}\ and\ \citenamefont
  {Broniowski}(2016)}]{Bozek:2015bna}%
  \BibitemOpen
  \bibfield  {author} {\bibinfo {author} {\bibfnamefont {Piotr}\ \bibnamefont
  {Bozek}}\ and\ \bibinfo {author} {\bibfnamefont {Wojciech}\ \bibnamefont
  {Broniowski}},\ }\bibfield  {title} {\enquote {\bibinfo {title} {{The torque
  effect and fluctuations of entropy deposition in rapidity in
  ultra-relativistic nuclear collisions}},}\ }\href {\doibase
  10.1016/j.physletb.2015.11.054} {\bibfield  {journal} {\bibinfo  {journal}
  {Phys. Lett.}\ }\textbf {\bibinfo {volume} {B752}},\ \bibinfo {pages}
  {206--211} (\bibinfo {year} {2016})},\ \Eprint
  {http://arxiv.org/abs/1506.02817} {arXiv:1506.02817 [nucl-th]} \BibitemShut
  {NoStop}%
\bibitem [{\citenamefont {Broniowski}\ and\ \citenamefont
  {Bożek}(2016)}]{Broniowski:2015oif}%
  \BibitemOpen
  \bibfield  {author} {\bibinfo {author} {\bibfnamefont {Wojciech}\
  \bibnamefont {Broniowski}}\ and\ \bibinfo {author} {\bibfnamefont {Piotr}\
  \bibnamefont {Bożek}},\ }\bibfield  {title} {\enquote {\bibinfo {title}
  {{Simple model for rapidity fluctuations in the initial state of
  ultrarelativistic heavy-ion collisions}},}\ }\href {\doibase
  10.1103/PhysRevC.93.064910} {\bibfield  {journal} {\bibinfo  {journal} {Phys.
  Rev.}\ }\textbf {\bibinfo {volume} {C93}},\ \bibinfo {pages} {064910}
  (\bibinfo {year} {2016})},\ \Eprint {http://arxiv.org/abs/1512.01945}
  {arXiv:1512.01945 [nucl-th]} \BibitemShut {NoStop}%
\bibitem [{\citenamefont {Schenke}\ and\ \citenamefont
  {Schlichting}(2016)}]{Schenke:2016ksl}%
  \BibitemOpen
  \bibfield  {author} {\bibinfo {author} {\bibfnamefont {Bjoern}\ \bibnamefont
  {Schenke}}\ and\ \bibinfo {author} {\bibfnamefont {Soeren}\ \bibnamefont
  {Schlichting}},\ }\bibfield  {title} {\enquote {\bibinfo {title} {{3D glasma
  initial state for relativistic heavy ion collisions}},}\ }\href {\doibase
  10.1103/PhysRevC.94.044907} {\bibfield  {journal} {\bibinfo  {journal} {Phys.
  Rev.}\ }\textbf {\bibinfo {volume} {C94}},\ \bibinfo {pages} {044907}
  (\bibinfo {year} {2016})},\ \Eprint {http://arxiv.org/abs/1605.07158}
  {arXiv:1605.07158 [hep-ph]} \BibitemShut {NoStop}%
\bibitem [{\citenamefont {Miller}\ \emph {et~al.}(2007)\citenamefont {Miller},
  \citenamefont {Reygers}, \citenamefont {Sanders},\ and\ \citenamefont
  {Steinberg}}]{Miller:2007ri}%
  \BibitemOpen
  \bibfield  {author} {\bibinfo {author} {\bibfnamefont {Michael~L.}\
  \bibnamefont {Miller}}, \bibinfo {author} {\bibfnamefont {Klaus}\
  \bibnamefont {Reygers}}, \bibinfo {author} {\bibfnamefont {Stephen~J.}\
  \bibnamefont {Sanders}}, \ and\ \bibinfo {author} {\bibfnamefont {Peter}\
  \bibnamefont {Steinberg}},\ }\bibfield  {title} {\enquote {\bibinfo {title}
  {{Glauber modeling in high energy nuclear collisions}},}\ }\href {\doibase
  10.1146/annurev.nucl.57.090506.123020} {\bibfield  {journal} {\bibinfo
  {journal} {Ann. Rev. Nucl. Part. Sci.}\ }\textbf {\bibinfo {volume} {57}},\
  \bibinfo {pages} {205--243} (\bibinfo {year} {2007})},\ \Eprint
  {http://arxiv.org/abs/nucl-ex/0701025} {arXiv:nucl-ex/0701025 [nucl-ex]}
  \BibitemShut {NoStop}%
\bibitem [{\citenamefont {Bialas}\ \emph {et~al.}(2016)\citenamefont {Bialas},
  \citenamefont {Bzdak},\ and\ \citenamefont {Koch}}]{Bialas:2016epd}%
  \BibitemOpen
  \bibfield  {author} {\bibinfo {author} {\bibfnamefont {Andrzej}\ \bibnamefont
  {Bialas}}, \bibinfo {author} {\bibfnamefont {Adam}\ \bibnamefont {Bzdak}}, \
  and\ \bibinfo {author} {\bibfnamefont {Volker}\ \bibnamefont {Koch}},\
  }\bibfield  {title} {\enquote {\bibinfo {title} {{Stopped nucleons in
  configuration space}},}\ }\href@noop {} {\  (\bibinfo {year} {2016})},\
  \Eprint {http://arxiv.org/abs/1608.07041} {arXiv:1608.07041 [hep-ph]}
  \BibitemShut {NoStop}%
\bibitem [{\citenamefont {Li}\ and\ \citenamefont
  {Kapusta}(2017)}]{Li:2016wzh}%
  \BibitemOpen
  \bibfield  {author} {\bibinfo {author} {\bibfnamefont {Ming}\ \bibnamefont
  {Li}}\ and\ \bibinfo {author} {\bibfnamefont {Joseph~I.}\ \bibnamefont
  {Kapusta}},\ }\bibfield  {title} {\enquote {\bibinfo {title} {{High Baryon
  Densities in Heavy Ion Collisions at Energies Attainable at the BNL
  Relativistic Heavy Ion Collider and the CERN Large Hadron Collider}},}\
  }\href {\doibase 10.1103/PhysRevC.95.011901} {\bibfield  {journal} {\bibinfo
  {journal} {Phys. Rev.}\ }\textbf {\bibinfo {volume} {C95}},\ \bibinfo {pages}
  {011901} (\bibinfo {year} {2017})},\ \Eprint
  {http://arxiv.org/abs/1604.08525} {arXiv:1604.08525 [nucl-th]} \BibitemShut
  {NoStop}%
\bibitem [{\citenamefont {Gyulassy}\ and\ \citenamefont
  {Wang}(1994)}]{Gyulassy:1994ew}%
  \BibitemOpen
  \bibfield  {author} {\bibinfo {author} {\bibfnamefont {Miklos}\ \bibnamefont
  {Gyulassy}}\ and\ \bibinfo {author} {\bibfnamefont {Xin-Nian}\ \bibnamefont
  {Wang}},\ }\bibfield  {title} {\enquote {\bibinfo {title} {{HIJING 1.0: A
  Monte Carlo program for parton and particle production in high-energy
  hadronic and nuclear collisions}},}\ }\href {\doibase
  10.1016/0010-4655(94)90057-4} {\bibfield  {journal} {\bibinfo  {journal}
  {Comput. Phys. Commun.}\ }\textbf {\bibinfo {volume} {83}},\ \bibinfo {pages}
  {307} (\bibinfo {year} {1994})},\ \Eprint
  {http://arxiv.org/abs/nucl-th/9502021} {arXiv:nucl-th/9502021 [nucl-th]}
  \BibitemShut {NoStop}%
\bibitem [{\citenamefont {Capella}\ \emph {et~al.}(1994)\citenamefont
  {Capella}, \citenamefont {Sukhatme}, \citenamefont {Tan},\ and\ \citenamefont
  {Tran Thanh~Van}}]{Capella:1992yb}%
  \BibitemOpen
  \bibfield  {author} {\bibinfo {author} {\bibfnamefont {A.}~\bibnamefont
  {Capella}}, \bibinfo {author} {\bibfnamefont {U.}~\bibnamefont {Sukhatme}},
  \bibinfo {author} {\bibfnamefont {C-I}\ \bibnamefont {Tan}}, \ and\ \bibinfo
  {author} {\bibfnamefont {J.}~\bibnamefont {Tran Thanh~Van}},\ }\bibfield
  {title} {\enquote {\bibinfo {title} {{Dual parton model}},}\ }\href {\doibase
  10.1016/0370-1573(94)90064-7} {\bibfield  {journal} {\bibinfo  {journal}
  {Phys. Rept.}\ }\textbf {\bibinfo {volume} {236}},\ \bibinfo {pages}
  {225--329} (\bibinfo {year} {1994})}\BibitemShut {NoStop}%
\bibitem [{\citenamefont {Shen}\ \emph {et~al.}(2017)\citenamefont {Shen},
  \citenamefont {Denicol}, \citenamefont {Gale}, \citenamefont {Jeon},
  \citenamefont {Monnai},\ and\ \citenamefont {Schenke}}]{Shen:2017ruz}%
  \BibitemOpen
  \bibfield  {author} {\bibinfo {author} {\bibfnamefont {Chun}\ \bibnamefont
  {Shen}}, \bibinfo {author} {\bibfnamefont {Gabriel}\ \bibnamefont {Denicol}},
  \bibinfo {author} {\bibfnamefont {Charles}\ \bibnamefont {Gale}}, \bibinfo
  {author} {\bibfnamefont {Sangyong}\ \bibnamefont {Jeon}}, \bibinfo {author}
  {\bibfnamefont {Akihiko}\ \bibnamefont {Monnai}}, \ and\ \bibinfo {author}
  {\bibfnamefont {Bjoern}\ \bibnamefont {Schenke}},\ }\bibfield  {title}
  {\enquote {\bibinfo {title} {{A hybrid approach to relativistic heavy-ion
  collisions at the RHIC BES energies}},}\ }in\ \href
  {http://inspirehep.net/record/1591546/files/arXiv:1704.04109.pdf} {\emph
  {\bibinfo {booktitle} {{26th International Conference on Ultrarelativistic
  Nucleus-Nucleus Collisions (Quark Matter 2017) Chicago,Illinois, USA,
  February 6-11, 2017}}}}\ (\bibinfo {year} {2017})\ \Eprint
  {http://arxiv.org/abs/1704.04109} {arXiv:1704.04109 [nucl-th]} \BibitemShut
  {NoStop}%
\bibitem [{\citenamefont {Andersson}\ \emph {et~al.}(1983)\citenamefont
  {Andersson}, \citenamefont {Gustafson}, \citenamefont {Ingelman},\ and\
  \citenamefont {Sjostrand}}]{Andersson:1983ia}%
  \BibitemOpen
  \bibfield  {author} {\bibinfo {author} {\bibfnamefont {Bo}~\bibnamefont
  {Andersson}}, \bibinfo {author} {\bibfnamefont {G.}~\bibnamefont
  {Gustafson}}, \bibinfo {author} {\bibfnamefont {G.}~\bibnamefont {Ingelman}},
  \ and\ \bibinfo {author} {\bibfnamefont {T.}~\bibnamefont {Sjostrand}},\
  }\bibfield  {title} {\enquote {\bibinfo {title} {{Parton Fragmentation and
  String Dynamics}},}\ }\href {\doibase 10.1016/0370-1573(83)90080-7}
  {\bibfield  {journal} {\bibinfo  {journal} {Phys. Rept.}\ }\textbf {\bibinfo
  {volume} {97}},\ \bibinfo {pages} {31--145} (\bibinfo {year}
  {1983})}\BibitemShut {NoStop}%
\bibitem [{\citenamefont {Werner}(1995)}]{Werner:1994tw}%
  \BibitemOpen
  \bibfield  {author} {\bibinfo {author} {\bibfnamefont {K.}~\bibnamefont
  {Werner}},\ }\bibfield  {title} {\enquote {\bibinfo {title} {{The String
  model for ultrarelativistic nuclear scattering}},}\ }\bibfield  {booktitle}
  {\emph {\bibinfo {booktitle} {{Relativistic heavy ion physics. Proceedings,
  International School, Workshop for Young Physicists, Prague, Czech Republic,
  September 19-23, 1994}}},\ }\href {\doibase 10.1007/BF01688547} {\bibfield
  {journal} {\bibinfo  {journal} {Czech. J. Phys.}\ }\textbf {\bibinfo {volume}
  {45}},\ \bibinfo {pages} {591--609} (\bibinfo {year} {1995})},\ \Eprint
  {http://arxiv.org/abs/nucl-th/9412006} {arXiv:nucl-th/9412006 [nucl-th]}
  \BibitemShut {NoStop}%
\bibitem [{\citenamefont {Ivanyi}\ \emph {et~al.}(2000)\citenamefont {Ivanyi},
  \citenamefont {Schram}, \citenamefont {Sailer},\ and\ \citenamefont
  {Soff}}]{Ivanyi:1999bv}%
  \BibitemOpen
  \bibfield  {author} {\bibinfo {author} {\bibfnamefont {B.}~\bibnamefont
  {Ivanyi}}, \bibinfo {author} {\bibfnamefont {Z.}~\bibnamefont {Schram}},
  \bibinfo {author} {\bibfnamefont {K.}~\bibnamefont {Sailer}}, \ and\ \bibinfo
  {author} {\bibfnamefont {G.}~\bibnamefont {Soff}},\ }\bibfield  {title}
  {\enquote {\bibinfo {title} {{Nucleus-nucleus collisions in the dynamical
  string model}},}\ }\href {\doibase 10.1103/PhysRevC.61.024908} {\bibfield
  {journal} {\bibinfo  {journal} {Phys. Rev.}\ }\textbf {\bibinfo {volume}
  {C61}},\ \bibinfo {pages} {024908} (\bibinfo {year} {2000})},\ \Eprint
  {http://arxiv.org/abs/hep-ph/9901371} {arXiv:hep-ph/9901371 [hep-ph]}
  \BibitemShut {NoStop}%
\bibitem [{\citenamefont {Kajantie}\ and\ \citenamefont
  {McLerran}(1982)}]{Kajantie:1982jt}%
  \BibitemOpen
  \bibfield  {author} {\bibinfo {author} {\bibfnamefont {K.}~\bibnamefont
  {Kajantie}}\ and\ \bibinfo {author} {\bibfnamefont {Larry~D.}\ \bibnamefont
  {McLerran}},\ }\bibfield  {title} {\enquote {\bibinfo {title} {{Initial
  Conditions for Hydrodynamical Calculations of Ultrarelativistic Nuclear
  Collisions}},}\ }\href {\doibase 10.1016/0370-2693(82)90277-5} {\bibfield
  {journal} {\bibinfo  {journal} {Phys. Lett.}\ }\textbf {\bibinfo {volume}
  {119B}},\ \bibinfo {pages} {203--206} (\bibinfo {year} {1982})}\BibitemShut
  {NoStop}%
\bibitem [{\citenamefont {Kajantie}\ and\ \citenamefont
  {McLerran}(1983)}]{Kajantie:1982nh}%
  \BibitemOpen
  \bibfield  {author} {\bibinfo {author} {\bibfnamefont {K.}~\bibnamefont
  {Kajantie}}\ and\ \bibinfo {author} {\bibfnamefont {Larry~D.}\ \bibnamefont
  {McLerran}},\ }\bibfield  {title} {\enquote {\bibinfo {title} {{Energy
  Densities, Initial Conditions and Hydrodynamic Equations for
  Ultrarelativistic Nucleus-nucleus Collisions}},}\ }\href {\doibase
  10.1016/0550-3213(83)90662-4} {\bibfield  {journal} {\bibinfo  {journal}
  {Nucl. Phys.}\ }\textbf {\bibinfo {volume} {B214}},\ \bibinfo {pages}
  {261--284} (\bibinfo {year} {1983})}\BibitemShut {NoStop}%
\bibitem [{\citenamefont {Gao}\ \emph {et~al.}(2014)\citenamefont {Gao},
  \citenamefont {Guzzi}, \citenamefont {Huston}, \citenamefont {Lai},
  \citenamefont {Li}, \citenamefont {Nadolsky}, \citenamefont {Pumplin},
  \citenamefont {Stump},\ and\ \citenamefont {Yuan}}]{Gao:2013xoa}%
  \BibitemOpen
  \bibfield  {author} {\bibinfo {author} {\bibfnamefont {Jun}\ \bibnamefont
  {Gao}}, \bibinfo {author} {\bibfnamefont {Marco}\ \bibnamefont {Guzzi}},
  \bibinfo {author} {\bibfnamefont {Joey}\ \bibnamefont {Huston}}, \bibinfo
  {author} {\bibfnamefont {Hung-Liang}\ \bibnamefont {Lai}}, \bibinfo {author}
  {\bibfnamefont {Zhao}\ \bibnamefont {Li}}, \bibinfo {author} {\bibfnamefont
  {Pavel}\ \bibnamefont {Nadolsky}}, \bibinfo {author} {\bibfnamefont {Jon}\
  \bibnamefont {Pumplin}}, \bibinfo {author} {\bibfnamefont {Daniel}\
  \bibnamefont {Stump}}, \ and\ \bibinfo {author} {\bibfnamefont {C.~P.}\
  \bibnamefont {Yuan}},\ }\bibfield  {title} {\enquote {\bibinfo {title} {{CT10
  next-to-next-to-leading order global analysis of QCD}},}\ }\href {\doibase
  10.1103/PhysRevD.89.033009} {\bibfield  {journal} {\bibinfo  {journal} {Phys.
  Rev.}\ }\textbf {\bibinfo {volume} {D89}},\ \bibinfo {pages} {033009}
  (\bibinfo {year} {2014})},\ \Eprint {http://arxiv.org/abs/1302.6246}
  {arXiv:1302.6246 [hep-ph]} \BibitemShut {NoStop}%
\bibitem [{\citenamefont {Eskola}\ \emph {et~al.}(2009)\citenamefont {Eskola},
  \citenamefont {Paukkunen},\ and\ \citenamefont {Salgado}}]{Eskola:2009uj}%
  \BibitemOpen
  \bibfield  {author} {\bibinfo {author} {\bibfnamefont {K.~J.}\ \bibnamefont
  {Eskola}}, \bibinfo {author} {\bibfnamefont {H.}~\bibnamefont {Paukkunen}}, \
  and\ \bibinfo {author} {\bibfnamefont {C.~A.}\ \bibnamefont {Salgado}},\
  }\bibfield  {title} {\enquote {\bibinfo {title} {{EPS09: A New Generation of
  NLO and LO Nuclear Parton Distribution Functions}},}\ }\href {\doibase
  10.1088/1126-6708/2009/04/065} {\bibfield  {journal} {\bibinfo  {journal}
  {JHEP}\ }\textbf {\bibinfo {volume} {04}},\ \bibinfo {pages} {065} (\bibinfo
  {year} {2009})},\ \Eprint {http://arxiv.org/abs/0902.4154} {arXiv:0902.4154
  [hep-ph]} \BibitemShut {NoStop}%
\bibitem [{\citenamefont {Anticic}\ \emph {et~al.}(2011)\citenamefont {Anticic}
  \emph {et~al.}}]{Anticic:2010mp}%
  \BibitemOpen
  \bibfield  {author} {\bibinfo {author} {\bibfnamefont {T.}~\bibnamefont
  {Anticic}} \emph {et~al.} (\bibinfo {collaboration} {NA49}),\ }\bibfield
  {title} {\enquote {\bibinfo {title} {{Centrality dependence of proton and
  antiproton spectra in Pb+Pb collisions at 40A GeV and 158A GeV measured at
  the CERN SPS}},}\ }\href {\doibase 10.1103/PhysRevC.83.014901} {\bibfield
  {journal} {\bibinfo  {journal} {Phys. Rev.}\ }\textbf {\bibinfo {volume}
  {C83}},\ \bibinfo {pages} {014901} (\bibinfo {year} {2011})},\ \Eprint
  {http://arxiv.org/abs/1009.1747} {arXiv:1009.1747 [nucl-ex]} \BibitemShut
  {NoStop}%
\bibitem [{\citenamefont {Schenke}\ \emph {et~al.}(2010)\citenamefont
  {Schenke}, \citenamefont {Jeon},\ and\ \citenamefont
  {Gale}}]{Schenke:2010nt}%
  \BibitemOpen
  \bibfield  {author} {\bibinfo {author} {\bibfnamefont {Bjoern}\ \bibnamefont
  {Schenke}}, \bibinfo {author} {\bibfnamefont {Sangyong}\ \bibnamefont
  {Jeon}}, \ and\ \bibinfo {author} {\bibfnamefont {Charles}\ \bibnamefont
  {Gale}},\ }\bibfield  {title} {\enquote {\bibinfo {title} {{(3+1)D
  hydrodynamic simulation of relativistic heavy-ion collisions}},}\ }\href
  {\doibase 10.1103/PhysRevC.82.014903} {\bibfield  {journal} {\bibinfo
  {journal} {Phys. Rev.}\ }\textbf {\bibinfo {volume} {C82}},\ \bibinfo {pages}
  {014903} (\bibinfo {year} {2010})},\ \Eprint {http://arxiv.org/abs/1004.1408}
  {arXiv:1004.1408 [hep-ph]} \BibitemShut {NoStop}%
\bibitem [{\citenamefont {Alver}\ \emph {et~al.}(2011)\citenamefont {Alver}
  \emph {et~al.}}]{Alver:2010ck}%
  \BibitemOpen
  \bibfield  {author} {\bibinfo {author} {\bibfnamefont {B.}~\bibnamefont
  {Alver}} \emph {et~al.} (\bibinfo {collaboration} {PHOBOS}),\ }\bibfield
  {title} {\enquote {\bibinfo {title} {{Phobos results on charged particle
  multiplicity and pseudorapidity distributions in Au+Au, Cu+Cu, d+Au, and p+p
  collisions at ultra-relativistic energies}},}\ }\href {\doibase
  10.1103/PhysRevC.83.024913} {\bibfield  {journal} {\bibinfo  {journal} {Phys.
  Rev.}\ }\textbf {\bibinfo {volume} {C83}},\ \bibinfo {pages} {024913}
  (\bibinfo {year} {2011})},\ \Eprint {http://arxiv.org/abs/1011.1940}
  {arXiv:1011.1940 [nucl-ex]} \BibitemShut {NoStop}%
\bibitem [{\citenamefont {Ahle}\ \emph {et~al.}(1998)\citenamefont {Ahle} \emph
  {et~al.}}]{Ahle:1998jc}%
  \BibitemOpen
  \bibfield  {author} {\bibinfo {author} {\bibfnamefont {L.}~\bibnamefont
  {Ahle}} \emph {et~al.} (\bibinfo {collaboration} {E-802}),\ }\bibfield
  {title} {\enquote {\bibinfo {title} {{Particle production at high baryon
  density in central Au+Au reactions at 11.6A GeV/c}},}\ }\href {\doibase
  10.1103/PhysRevC.57.R466} {\bibfield  {journal} {\bibinfo  {journal} {Phys.
  Rev.}\ }\textbf {\bibinfo {volume} {C57}},\ \bibinfo {pages} {R466--R470}
  (\bibinfo {year} {1998})}\BibitemShut {NoStop}%
\bibitem [{\citenamefont {Ahle}\ \emph {et~al.}(1999)\citenamefont {Ahle} \emph
  {et~al.}}]{Ahle:1999in}%
  \BibitemOpen
  \bibfield  {author} {\bibinfo {author} {\bibfnamefont {L.}~\bibnamefont
  {Ahle}} \emph {et~al.} (\bibinfo {collaboration} {E802}),\ }\bibfield
  {title} {\enquote {\bibinfo {title} {{Proton and deuteron production in Au +
  Au reactions at 11.6/A-GeV/c}},}\ }\href {\doibase
  10.1103/PhysRevC.60.064901} {\bibfield  {journal} {\bibinfo  {journal} {Phys.
  Rev.}\ }\textbf {\bibinfo {volume} {C60}},\ \bibinfo {pages} {064901}
  (\bibinfo {year} {1999})}\BibitemShut {NoStop}%
\bibitem [{\citenamefont {Barrette}\ \emph {et~al.}(2000)\citenamefont
  {Barrette} \emph {et~al.}}]{Barrette:1999ry}%
  \BibitemOpen
  \bibfield  {author} {\bibinfo {author} {\bibfnamefont {J.}~\bibnamefont
  {Barrette}} \emph {et~al.} (\bibinfo {collaboration} {E877}),\ }\bibfield
  {title} {\enquote {\bibinfo {title} {{Proton and pion production in Au + Au
  collisions at 10.8A-GeV/c}},}\ }\href {\doibase 10.1103/PhysRevC.62.024901}
  {\bibfield  {journal} {\bibinfo  {journal} {Phys. Rev.}\ }\textbf {\bibinfo
  {volume} {C62}},\ \bibinfo {pages} {024901} (\bibinfo {year} {2000})},\
  \Eprint {http://arxiv.org/abs/nucl-ex/9910004} {arXiv:nucl-ex/9910004
  [nucl-ex]} \BibitemShut {NoStop}%
\bibitem [{\citenamefont {Arsene}\ \emph {et~al.}(2009)\citenamefont {Arsene}
  \emph {et~al.}}]{Arsene:2009aa}%
  \BibitemOpen
  \bibfield  {author} {\bibinfo {author} {\bibfnamefont {I.~C.}\ \bibnamefont
  {Arsene}} \emph {et~al.} (\bibinfo {collaboration} {BRAHMS}),\ }\bibfield
  {title} {\enquote {\bibinfo {title} {{Nuclear stopping and rapidity loss in
  Au+Au collisions at s(NN)**(1/2) = 62.4-GeV}},}\ }\href {\doibase
  10.1016/j.physletb.2009.05.049} {\bibfield  {journal} {\bibinfo  {journal}
  {Phys. Lett.}\ }\textbf {\bibinfo {volume} {B677}},\ \bibinfo {pages}
  {267--271} (\bibinfo {year} {2009})},\ \Eprint
  {http://arxiv.org/abs/0901.0872} {arXiv:0901.0872 [nucl-ex]} \BibitemShut
  {NoStop}%
\bibitem [{\citenamefont {Aaboud}\ \emph
  {et~al.}(2017{\natexlab{a}})\citenamefont {Aaboud} \emph
  {et~al.}}]{Aaboud:2016jnr}%
  \BibitemOpen
  \bibfield  {author} {\bibinfo {author} {\bibfnamefont {Morad}\ \bibnamefont
  {Aaboud}} \emph {et~al.} (\bibinfo {collaboration} {ATLAS}),\ }\bibfield
  {title} {\enquote {\bibinfo {title} {{Measurement of forward-backward
  multiplicity correlations in lead-lead, proton-lead, and proton-proton
  collisions with the ATLAS detector}},}\ }\href {\doibase
  10.1103/PhysRevC.95.064914} {\bibfield  {journal} {\bibinfo  {journal} {Phys.
  Rev.}\ }\textbf {\bibinfo {volume} {C95}},\ \bibinfo {pages} {064914}
  (\bibinfo {year} {2017}{\natexlab{a}})},\ \Eprint
  {http://arxiv.org/abs/1606.08170} {arXiv:1606.08170 [hep-ex]} \BibitemShut
  {NoStop}%
\bibitem [{\citenamefont {Khachatryan}\ \emph {et~al.}(2015)\citenamefont
  {Khachatryan} \emph {et~al.}}]{Khachatryan:2015oea}%
  \BibitemOpen
  \bibfield  {author} {\bibinfo {author} {\bibfnamefont {Vardan}\ \bibnamefont
  {Khachatryan}} \emph {et~al.} (\bibinfo {collaboration} {CMS}),\ }\bibfield
  {title} {\enquote {\bibinfo {title} {{Evidence for transverse momentum and
  pseudorapidity dependent event plane fluctuations in PbPb and pPb
  collisions}},}\ }\href {\doibase 10.1103/PhysRevC.92.034911} {\bibfield
  {journal} {\bibinfo  {journal} {Phys. Rev.}\ }\textbf {\bibinfo {volume}
  {C92}},\ \bibinfo {pages} {034911} (\bibinfo {year} {2015})},\ \Eprint
  {http://arxiv.org/abs/1503.01692} {arXiv:1503.01692 [nucl-ex]} \BibitemShut
  {NoStop}%
\bibitem [{\citenamefont {Aaboud}\ \emph
  {et~al.}(2017{\natexlab{b}})\citenamefont {Aaboud} \emph
  {et~al.}}]{Aaboud:2017tql}%
  \BibitemOpen
  \bibfield  {author} {\bibinfo {author} {\bibfnamefont {Morad}\ \bibnamefont
  {Aaboud}} \emph {et~al.} (\bibinfo {collaboration} {ATLAS}),\ }\bibfield
  {title} {\enquote {\bibinfo {title} {{Measurement of longitudinal flow
  de-correlations in Pb+Pb collisions at $\sqrt{s_\mathrm{NN}} = 2.76$ and 5.02
  TeV with the ATLAS detector}},}\ }\href@noop {} {\  (\bibinfo {year}
  {2017}{\natexlab{b}})},\ \Eprint {http://arxiv.org/abs/1709.02301}
  {arXiv:1709.02301 [nucl-ex]} \BibitemShut {NoStop}%
\bibitem [{\citenamefont {Jia}\ \emph {et~al.}(2016)\citenamefont {Jia},
  \citenamefont {Radhakrishnan},\ and\ \citenamefont {Zhou}}]{Jia:2015jga}%
  \BibitemOpen
  \bibfield  {author} {\bibinfo {author} {\bibfnamefont {Jiangyong}\
  \bibnamefont {Jia}}, \bibinfo {author} {\bibfnamefont {Sooraj}\ \bibnamefont
  {Radhakrishnan}}, \ and\ \bibinfo {author} {\bibfnamefont {Mingliang}\
  \bibnamefont {Zhou}},\ }\bibfield  {title} {\enquote {\bibinfo {title}
  {{Forward-backward multiplicity fluctuation and longitudinal harmonics in
  high-energy nuclear collisions}},}\ }\href {\doibase
  10.1103/PhysRevC.93.044905} {\bibfield  {journal} {\bibinfo  {journal} {Phys.
  Rev.}\ }\textbf {\bibinfo {volume} {C93}},\ \bibinfo {pages} {044905}
  (\bibinfo {year} {2016})},\ \Eprint {http://arxiv.org/abs/1506.03496}
  {arXiv:1506.03496 [nucl-th]} \BibitemShut {NoStop}%
\bibitem [{\citenamefont {Li}\ \emph {et~al.}(2017)\citenamefont {Li},
  \citenamefont {Xu},\ and\ \citenamefont {Song}}]{Li:2017via}%
  \BibitemOpen
  \bibfield  {author} {\bibinfo {author} {\bibfnamefont {Jixing}\ \bibnamefont
  {Li}}, \bibinfo {author} {\bibfnamefont {Hao-jie}\ \bibnamefont {Xu}}, \ and\
  \bibinfo {author} {\bibfnamefont {Huichao}\ \bibnamefont {Song}},\ }\bibfield
   {title} {\enquote {\bibinfo {title} {{Non-critical fluctuations of (net)
  charges and (net) protons from iEBE-VISHNU hybrid model}},}\ }\href@noop {}
  {\  (\bibinfo {year} {2017})},\ \Eprint {http://arxiv.org/abs/1707.09742}
  {arXiv:1707.09742 [nucl-th]} \BibitemShut {NoStop}%
\end{thebibliography}%

\end{document}